\documentclass[graybox]{svmult}

\usepackage{mathptmx}       % selects Times Roman as basic font
\usepackage{helvet}         % selects Helvetica as sans-serif font
\usepackage{courier}        % selects Courier as typewriter font
\usepackage{type1cm}        % activate if the above 3 fonts are
\usepackage{makeidx}         % allows index generation
\usepackage{graphicx}        % standard LaTeX graphics tool
\usepackage{multicol}        % used for the two-column index
\usepackage[bottom]{footmisc}% places footnotes at page bottom

\makeindex             % used for the subject index
\begin{document}

\title*{A New Approach To The Treatment Of Separatrix Chaos And Its Applications}
\author{S.M. Soskin, R. Mannella, O.M. Yevtushenko, I.A. Khovanov,
%and
P.V.E. McClintock}
%\author{Stanislav Soskin, Riccardo Mannella, Oleg Yevtushenko, Igor Khovanov,
%and Peter McClintock}
\institute{S.M. Soskin \at Institute of Semiconductor Physics,
%National Academy of Sciences of Ukraine,
03028 Kiev, Ukraine, \\ \email{ssoskin@ictp.it}
\and R. Mannella \at Dipartimento di Fisica, Universit\`{a} di Pisa, 56127
Pisa, Italy, \\ \email{mannella@df.unipi.it}
\and O.M. Yevtushenko \at Physics Department, Ludwig-Maximilians-Universit{\"a}t M{\"u}nchen,
   D-80333 M{\"u}nchen, Germany, \\ \email{bom@ictp.it}
\and I.A. Khovanov \at School of Engineering, University of Warwick, Coventry CV4 7AL,
UK,\\ \email{i.khovanov@warwick.ac.uk}
\and P.V.E. McClintock \at Physics Department, Lancaster University, Lancaster LA1 4YB,
UK,\\ \email{p.v.e.mcclintock@lancaster.ac.uk}
}
\maketitle

\abstract*{We consider time-periodically perturbed 1D Hamiltonian systems
possessing one or more separatrices. If the perturbation is weak, then the
separatrix chaos is most developed when the perturbation frequency lies in the
%logarithmically small or moderate
characteristic range of frequencies of oscillations of the unperturbed system:
this corresponds to the involvement
of resonance dynamics into the separatrix chaos. We develop a method matching
the discrete chaotic dynamics of the separatrix map and the continuous regular
dynamics of the resonance Hamiltonian. The method has allowed us to solve the
long-standing problem of an accurate analytic description of the maximum of the
separatrix chaotic layer width as a function of the perturbation frequency. Moreover,
it has also allowed us to predict and describe new phenomena including, in
particular: (i) a drastic facilitation of the onset of global chaos between
neighbouring separatrices, and (ii) a huge increase in the size of the
low-dimensional stochastic web.}

\abstract{We consider time-periodically perturbed 1D Hamiltonian systems
possessing one or more separatrices. If the perturbation is weak, then the
separatrix chaos is most developed when the perturbation frequency lies in the
logarithmically small or moderate ranges: this corresponds to the involvement
of resonance dynamics into the separatrix chaos. We develop a method matching
the discrete chaotic dynamics of the separatrix map and the continuous regular
dynamics of the resonance Hamiltonian. The method has allowed us to solve the
long-standing problem of an accurate description of the maximum of the
separatrix chaotic layer width as a function of the perturbation frequency. It
has also allowed us to predict and describe new phenomena including, in
particular: (i) a drastic facilitation of the onset of global chaos between
neighbouring separatrices, and (ii) a huge increase in the size of the
low-dimensional stochastic web.}

\section{Introduction}
\label{sec:1}

Separatrix chaos is the germ of Hamiltonian chaos \cite{zaslavsky:1998}.
Consider an integrable Hamiltonian system possessing a saddle, i.e.\ a
hyperbolic point in the one-dimensional case, or a hyperbolic invariant torus,
in higher-dimensional cases. The stable (incoming) and unstable (outgoing)
manifolds of the saddle are called {\it separatrices} \cite{gelfreich}: they
separate trajectories that have different phase space topologies. If a weak
time-periodic perturbation is added, then the separatrix is destroyed; it is
replaced by a {\it separatrix chaotic layer} (SCL)
\cite{zaslavsky:1998,gelfreich,lichtenberg_lieberman,treschev}. Even if the
unperturbed system does not possess a separatrix, the resonant part of the
perturbation generates a separatrix in the auxiliary resonance phase space
while the non-resonant part of the perturbation destroys this separatrix,
replacing it with a chaotic layer
\cite{zaslavsky:1998,gelfreich,lichtenberg_lieberman,Chirikov:79}. Thus
separatrix chaos is of a fundamental importance for Hamiltonian chaos.

One of the most important characteristics of SCL is its width in energy (or
expressed in related quantities). It can be easily found {\it numerically} by
integration of the Hamiltonian equations with a set of initial conditions in
the vicinity of the separatrix: the space occupied by the chaotic trajectory in
the Poincar\'{e} section has a higher dimension than that for a regular
trajectory, e.g.\ in the 3/2D case the regular trajectories lie on lines i.e.\
1D objects while the chaotic trajectory lies within the SCL i.e.\ the object outer boundaries
of which limit a 2D area.

On the other hand, it is important to be able to describe {\it theoretically}
both the outer boundaries of the SCL and its width. There is a long and rich
history of the such studies. The results may be classified as follows.

\subsection{Heuristic results}

Consider a 1D Hamiltonian system perturbed by a weak time-periodic
perturbation:

\begin{eqnarray}
&& H=H_0(p,q)+hV(p,q,t),\\
&& V(p,q,t+2\pi/\omega_f)=V(p,q,t),\quad\quad h\ll 1,
\nonumber
\end{eqnarray}

\noindent where $H_0(p,q)$ possesses a separatrix and, for the sake of
notational compactness, all relevant parameters of $H_0$ and $V$, except
possibly for $\omega_f$, are assumed to be $\sim 1$.

Physicists proposed a number of different heuristic criteria
\cite{ZF:1968,Chirikov:79,lichtenberg_lieberman,Zaslavsky:1991,zaslavsky:1998,zaslavsky:2005}
for the SCL width $\Delta E$ in terms of energy $E\equiv H_0(p,q)$
which
gave qualitatively similar results:
\begin{eqnarray}
&&\Delta E \equiv \Delta E(\omega_f)\sim  \omega_f\delta,\quad\quad
\\
&&\delta \equiv h|\epsilon|,
\nonumber
\\
&&
%\quad\quad
|\epsilon|\stackrel{<}{\sim} 1\quad\quad\quad\quad\quad\quad\quad\quad\quad \quad\quad\quad\quad  {\rm for}\quad
\omega_f\stackrel{<}{\sim} 1,
\nonumber
\\
&&
|\epsilon|\propto \exp(-a\omega_f)\ll 1 \quad \quad (a\sim 1) \quad \quad {\rm for}\quad
\omega_f\gg 1.
\nonumber
\end{eqnarray}

\noindent The quantity $\delta \equiv h|\epsilon|$ is called the {\it
separatrix split} \cite{zaslavsky:1998} (see also Eq.\ (4) below): it
determines the maximum distance between the perturbed incoming and outgoing
separatrices
\cite{ZF:1968,Chirikov:79,lichtenberg_lieberman,Zaslavsky:1991,zaslavsky:1998,zaslavsky:2005,abdullaev,gelfreich,treschev}.

It follows from (2) that the maximum of $\Delta E$ should lie in the frequency
range $\omega_f\sim 1$ while the maximum itself should be $\sim h$:

\begin{equation}
\Delta E_{\max}\equiv\max_{\omega_f}\{\Delta E(\omega_f)\}\sim h,
%\quad\quad\quad
\quad\quad\quad \omega_f^{(\max)}\sim 1.
\end{equation}

\subsection{Mathematical and accurate physical results}

Many papers studied the SCL by  mathematical or accurate physical methods.

For the range $\omega_f\gg 1$, many works studied the separatrix splitting (see
the review \cite{gelfreich} and references therein) and the SCL width in terms
of normal coordinates (see the review \cite{treschev} and references therein).
Though quantities studied in these works differ from those typically studied by
physicists
\cite{ZF:1968,Chirikov:79,lichtenberg_lieberman,Zaslavsky:1991,zaslavsky:1998,zaslavsky:2005},
they implicitly confirm the main qualitative conclusion from the heuristic
formula (2) in the high frequency range: provided that $\omega_f\gg 1$ the SCL
width is exponentially small.

There were also several works studying the SCL in the opposite (i.e.\
adiabatic) limit $\omega_f\rightarrow 0$: see e.g.\
\cite{Neishtadt:1986,E&E:1991,Neishtadt:1997,prl2005,13_prime} and references
therein. In the context of the SCL width, it is most important that $\Delta
E(\omega_f\rightarrow 0)\sim h$ for most of the systems
\cite{Neishtadt:1986,E&E:1991,Neishtadt:1997}. For a particular class of
systems, namely for ac-driven spatially periodic systems (e.g.\ the ac-driven
pendulum), the width of the SCL part above the separatrix diverges in the
adiabatic limit \cite{prl2005,13_prime}: the divergence develops for
$\omega_f\ll 1/\ln(1/h)$.

Finally, there is a qualitative estimation of the SCL width for the range
$\omega_f\sim 1$ within the Kolmogorov-Arnold-Moser (KAM) theory
\cite{treschev}. The quantitative estimate within the KAM theory is
lacking,
apparently being very difficult for
this frequency range \cite{vasya}. It follows from the results in
\cite{treschev} that the width in this range is of the order of the separatrix
split, which itself is of the order of $h$.

Thus it could seem to follow that, for all systems except ac-driven spatially
periodic systems, the maximum in the SCL width is $\sim h$ and occurs in the
range $\omega_f\sim 1$, very much in agreement with the heuristic result (3).
Even for ac-driven spatially periodic systems, this conclusion could seem to
apply to the width of the SCL part below the separatrix over the whole
frequency range, and to the width of the SCL part above the separatrix for
$\omega_f \stackrel{>}{\sim} 1/\ln(1/h)$.

\subsection{Numerical evidence for high peaks in $\Delta E(\omega_f)$ and their
rough estimation}

The above conclusion disagrees with several numerical studies carried out
during the last decade (see e.g.\
\cite{prl2005,13_prime,shevchenko:1998,luo1,soskin2000,luo2,vecheslavov,shevchenko})
which have revealed the existence of sharp peaks in $\Delta E(\omega_f)$ in the
frequency range $1/\ln(1/h)\stackrel{<}{\sim}\omega_f\stackrel{<}{\sim} 1$ the
heights of which greatly exceed $h$ (see also Figs.\ 2, 3, 5, 6 below). Thus,
the peaks represent the general {\it dominant feature} of the function $\Delta
E(\omega_f)$. They were related by the authors of
\cite{shevchenko:1998,luo1,soskin2000,luo2,vecheslavov,shevchenko} to the
absorption of nonlinear resonances by the SCL. For some partial case, rough
heuristic estimates for the position and magnitude of the peaks were made in
\cite{shevchenko:1998,shevchenko}.

\subsection{Accurate description of the peaks and of the related phenomena}

Until recently, accurate analytic estimates for the peaks were lacking. It is
explicitly stated in \cite{luo2} that the search for the mechanism through
which resonances are involved in separatrix chaos, and for an accurate analytic
description of the peaks in the SCL width as function of the perturbation
frequency, are being among the most important and challenging tasks in the
study of separatrix chaos. The first step towards accomplishing them was taken
through the proposal \cite{pre2008,proceedings} of a new approach to the
theoretical treatment of the separatrix chaos in the relevant frequency range.
It was developed and applied to the onset of global chaos between two close
separatrices. The application of the approach \cite{pre2008,proceedings} to the
commoner single-separatrix case was also discussed. The approach has been
further developed \cite{icnf_approach,pre_submitted}, including an explicit
theory for the single-separatrix case \cite{pre_submitted}.

The present paper reviews the new approach
\cite{pre2008,proceedings,icnf_approach,pre_submitted} and its applications to
the single-separatrix \cite{pre_submitted} and double-separatrix
\cite{pre2008,proceedings} cases. We also briefly review application to the
enlargement of the low-dimensional stochastic web
\cite{icnf_enlargement} and discuss other promising
applications.

Though the form of our treatment differs from typical forms of mathematical
theorems in this subject (cf.\ \cite{gelfreich,treschev}), it yields the {\it
exact} expressions for the leading term in the relevant asymptotic
expansions (the parameter of
smallness is $\alpha\equiv1/\ln(1/h)$) and, in some case, even for the next-order
term. Our theory is in excellent agreement with
results obtained by numerical integration of the equations of motion.

Sec.\ 2 describes the basic ideas underlying the approach. Sec.\ 3 is
devoted to the leading-order asymptotic description of the single-separatrix
chaotic layers. Sec.\ 4 presents an asymptotic description of the onset of
global chaos in between two close separatrices. Sec.\ 5 describes the increase
in sizes of a stochastic web. Conclusions are
drawn in Sec.\ 6. Sec. 7 presents the Appendix, which explicitly
matches the separatrix map and the
resonance Hamiltonian descriptions for the double-separatrix case.

\section{Basic ideas of the approach}
\label{sec:2}

The new approach \cite {pre2008,proceedings,icnf_approach,pre_submitted} may be
formulated briefly as a matching between the discrete chaotic dynamics of the
separatrix map in the immediate vicinity of the separatrix and the continuous
regular dynamics of the resonance Hamiltonian beyond that region. The present
section describes the general features of the approach in more detail.

Motion near the separatrix may be approximated by the {\it separatrix map} (SM)
\cite{ZF:1968,Chirikov:79,lichtenberg_lieberman,Zaslavsky:1991,zaslavsky:1998,zaslavsky:2005,abdullaev,treschev,shevchenko:1998,shevchenko,pre2008,proceedings,vered}.
This was introduced in \cite{ZF:1968} and its various modifications were
subsequently used in many studies. It is sometimes known as the {\it whisker
map}. It was re-derived rigorously in \cite {vered} as the leading-order
approximation of motion near the separatrix in the asymptotic limit
$h\rightarrow 0$, and an estimate of the error was also carried out
in \cite {vered} (see also the review
\cite{treschev} and references therein).

The main ideas that allow one to introduce the SM
\cite{ZF:1968,Chirikov:79,lichtenberg_lieberman,Zaslavsky:1991,zaslavsky:1998,zaslavsky:2005,abdullaev,treschev,pre2008,proceedings,vered}
are as follows. For the sake of simplicity, let us consider a perturbation $V$
that does not depend on the momentum: $V\equiv V(q,t)$. A system with energy
close to the separatrix value spends most of its time in the vicinity of the
saddle(s), where the velocity $\dot{q}$ is exponentially small. Differentiating
$E\equiv H_0(p,q)$ with respect to time and allowing for the equations of
motion of the system (1), we can show that ${\rm d}E/{\rm d}t\equiv
\partial V/\partial q \dot{q}\propto\dot{q}$. Thus, the perturbation can
significantly change the energy only when the velocity is not small i.e.\
during the relatively short intervals while the system is away from the
saddle(s): these intervals correspond to {\it pulses} of velocity as a function
of time (cf. Fig. 20 in the Appendix below). Consequently, it is possible to
approximate the continuous Hamiltonian dynamics by a discrete dynamics which
maps the energy $E$, the perturbation angle $\varphi\equiv \omega_f t$, and the
velocity sign $\sigma\equiv{\rm sgn}(\dot{q})$, from pulse to pulse.

The actual form of the SM may vary, depending on the system under study, but
its features relevant in the present context are similar for all systems. For
the sake of clarity, consider the explicit case when the separatrix of
$H_0(p,q)$ possesses a single saddle and two symmetric loops while
$V=q\cos(\omega_ft)$. Then the SM reads \cite{pre2008} (cf.\ Appendix):

\begin{eqnarray}
&&E_{i+1}=E_i+\sigma_ih\epsilon\sin(\varphi_i),
\\
&&\varphi_{i+1}=\varphi_i+\frac{\omega_f\pi(3- {\rm
sgn}(E_{i+1}-E_s)) }{2\omega(E_{i+1})}, \nonumber
\\
&&\sigma_{i+1}=\sigma_i \, {\rm sgn}(E_s-E_{i+1}), \quad\quad |\sigma_i|=1, \nonumber
\\
&&\quad\quad\quad\epsilon \equiv \epsilon(\omega_f)=
{\rm sgn}\left(\left.\frac{\partial
H_0}{\partial p}\right|_{t\rightarrow -\infty}\right)
%\sigma_i
\int_{-\infty}^{\infty}
%_{i{\rm th}\quad{\rm pulse}}
{\rm d}t\;\left.\frac{\partial H_0}{\partial p}\right|_{E_s}\sin(\omega_ft)
,
\nonumber
\\
&&\quad\quad\quad E_i\equiv \left.H_0(p,q)\right|_{t_i-\Delta},
\quad\quad\varphi_i\equiv
\omega_ft_i,\quad\quad\sigma_i\equiv{\rm sgn}\left(\left.\frac{\partial
H_0}{\partial p}\right|_{t_i}\right), \nonumber
\end{eqnarray}

\noindent where $E_s$ is the separatrix energy, $\omega(E)$ is the frequency of
oscillation with energy $E$ in the unperturbed case (i.e.\ for $h=0$), $t_i$ is
the instant corresponding to the $i$-th turning point in the trajectory $q(t)$
(cf. Fig. 20 in the Appendix below),
and $\Delta$ is an arbitrary value from the range of time intervals which
greatly exceed the characteristic duration of the velocity pulse while being
much smaller than the interval between the subsequent pulses
\cite{ZF:1968,Chirikov:79,lichtenberg_lieberman,Zaslavsky:1991,zaslavsky:1998,zaslavsky:2005,abdullaev,treschev,vered}.
Consider the two most general ideas of our approach.

\begin{enumerate}

\item If a trajectory of the SM includes a state with $E=E_s$ and an
    arbitrary $\varphi$ and $\sigma$, then this trajectory is chaotic.
    Indeed, the angle $\varphi$ of such a state is not correlated with the
    angle of the state at the previous step of the SM, due to the
    divergence of $\omega^{-1}(E\rightarrow E_s)$. Therefore, the angle at
    the previous step may deviate from a multiple of $2\pi$ by an arbitrary
    value. Hence the energy of the state at the previous step may deviate
    from $E_s$ by an arbitrary value within the interval
    $[-h|\epsilon|,h|\epsilon|]$. The velocity sign $\sigma$ is not
    correlated with that at the previous step either\footnote{Formally,
    ${\rm sgn}(E-E_s)$ is not defined for $E=E_s$ but, if to shift $E$ from
    $E_s$ for an infinitesemal value, ${\rm sgn}(E-E_s)$ acquires a value
    equal to either $+1$ or $-1$, depending on the sign of the shift.
    Given that $\sigma_{i+1}$ is proportional to
    ${\rm sgn}(E_s-E_{i+1})$ while the latter is random-like (as
    it has been shown above),
    $\sigma_{i+1}$ is not correlated with
    $\sigma_{i}$ if $E_{i+1}=E_s\pm 0$.}. Given that a regular trajectory
    of the SM cannot include a step where all three variables change
    random-like, we conclude that such a trajectory must be chaotic.

Though the above arguments may appear to be obvious, they cannot be
considered a mathematically rigorous proof, so that the statement about the
chaotic nature of the SM trajectory which includes any state with $E=E_s$
should be considered as a conjecture supported by the above arguments and
by numerical iteration of the SM. Possibly, a mathematically rigorous proof
should involve an analysis of the Lyapunov exponents for the SM (cf.\
\cite{lichtenberg_lieberman}) but this appears to be a technically
difficult problem. We emphasize however that a rigorous proof of the
conjecture is not crucial for the validity of the main results of the
present paper, namely for the {\it leading} terms in the asymptotic
expressions describing (i) the peaks of the SCL width as a function of the
perturbation frequency in the single-separatrix case, and (ii) the related
quantities for the double-separatrix case. It will become obvious from the
next item that, to derive the leading term, it is sufficient to know that
the chaotic trajectory does visit areas of the phase space where the energy
deviates from the separatrix by values of the order of the separatrix split
$\delta\equiv h|\epsilon|$, which is a widely accepted fact
\cite{ZF:1968,Chirikov:79,lichtenberg_lieberman,Zaslavsky:1991,zaslavsky:1998,zaslavsky:2005,abdullaev,gelfreich,treschev}.

\item It is well known
    \cite{ZF:1968,Chirikov:79,lichtenberg_lieberman,Zaslavsky:1991,zaslavsky:1998,zaslavsky:2005,abdullaev,gelfreich,treschev,shevchenko:1998,shevchenko,pre2008,proceedings},
    that, at the leading-order approximation, the frequency of
    eigenoscillation as function of the energy near the separatrix is
    proportional to the reciprocal of the logarithmic factor

\begin{eqnarray}
&&\omega(E)=\frac{b\pi\omega_0}{\ln\left(\frac{\Delta H}{|E-E_s|}\right)},
\quad  b=\frac{3- {\rm sgn}(E-E_s)}{2},
\\
%\quad\quad
&&|E-E_s|\ll\Delta H\equiv E_s-E_{st},
\nonumber
\end{eqnarray}

\noindent where $E_{st}$ is the energy of the stable states.

Given that the argument of the logarithm is large in the relevant range of
$E$, the function $\omega (E)$ is nearly constant for a substantial
variation of the argument. Therefore, as the SM maps the state
$(E_0=E_s,\varphi_0,\sigma_0)$ onto the state with $E=E_1\equiv
E_s+\sigma_0 h\epsilon\sin(\varphi_0)$, the value of $\omega(E)$ for the
given ${\rm sgn}(\sigma_0\epsilon\sin(\varphi_0))$ is nearly the same for
most of the angles $\varphi_0$ (except in the vicinity of multiples of
$\pi$),

\begin{eqnarray}
&&\omega(E)\approx\omega_r^{(\pm)},
\\
&&\omega_r^{(\pm)}
\equiv\omega(E_s\pm
h),
%\epsilon(\omega_f=\omega(E_s+h))
%\quad\quad {\rm for}
\quad\quad {\rm
sgn}(\sigma_0\epsilon\sin(\varphi_0))=\pm 1.
\nonumber
\end{eqnarray}

Moreover, if the deviation of the SM trajectory from the separatrix
increases further, $\omega(E)$ remains close to $\omega_r^{(\pm)}$ provided
the deviation is not too large, namely if $\ln(|E-E_s|/h)\ll\ln(\Delta
H/h)$. If $\omega_f\stackrel{<}{\sim}\omega_r^{(\pm)}$, then the evolution
of the SM (4) may be regular-like for a long time until the energy returns
to the close vicinity of the separatrix, where the trajectory becomes
chaotic. Such behavior is especially pronounced if the perturbation
frequency is close to $\omega_r^{(+)}$ or $\omega_r^{(-)}$ or to one of
their multiples of relatively low order: the resonance between the
perturbation and the eigenoscillation gives rise to an accumulation of
energy changes for many steps of the SM, which results in a deviation of
$E$ from $E_s$ that greatly exceeds the separatrix split $h|\epsilon|$.
Consider a state at the boundary of the SCL. The deviation of energy of
such a state from $E_s$ depends on its position at the boundary. In turn,
the maximum deviation is a function of $\omega_f$. The latter function
possesses the absolute maximum at $\omega_f$ close to $\omega_r^{(+)}$ or
$\omega_r^{(-)}$ typically\footnote{For the SM relating to ac-driven
spatially periodic systems, the time during which the SM undergoes a
regular-like evolution above the separatrix diverges in the adiabatic limit
$\omega_f\rightarrow 0$ \cite{13_prime}, and the width of the part of the
SM layer above the separatrix diverges too. However, we do not consider
this case here since it is irrelevant to the main subject of the present
paper i.e.\ to the involvement of the resonance dynamics into the
separatrix chaotic motion.}, for the upper or lower boundary of the SCL
respectively. This corresponds to the absorption of the, respectively upper
and lower, 1st-order nonlinear resonance by the SCL.

\end{enumerate}

\noindent The second of these intuitive ideas has been explicitly confirmed
\cite{pre2008} (see Appendix): in the relevant range of energies, the
separatrix map has been shown to reduce to two differential equations
which are identical
to the equations of motion of the auxiliary resonance Hamiltonian describing
the resonance dynamics in terms of the conventional canonically conjugate slow
variables, action $I$ and slow angle $\tilde{\psi}\equiv n \psi-\omega_ft$
where $\psi$ is the angle variable
\cite{Chirikov:79,lichtenberg_lieberman,Zaslavsky:1991,zaslavsky:1998,zaslavsky:2005,abdullaev}
(see Eq.\ (16) below) and $n$ is the relevant resonance number i.e.\ the integer
closest to the ratio $\omega_f/\omega_r^{(\pm)}$.

Thus the matching between the discrete chaotic dynamics of the SM and the
continuous regular-like dynamics of the resonance Hamiltonian arises in the
following way \cite{pre2008}. After the chaotic trajectory of the SM visits any
state on the separatrix, the system transits in one step of the SM to a given
upper or lower curve in the $I-\tilde{\psi}$ plane which has been called
\cite{pre2008} the upper or lower {\it generalized separatrix split} (GSS)
curve\footnote{The GSS curve corresponds to the step of the SM which follows
the state with $E=E_s$, as described above.} respectively:

\begin{equation}
E=E_{GSS}^{(\pm)}(\tilde{\psi})\equiv
E_s\pm\delta|\sin(\tilde{\psi})|,\quad\quad \delta\equiv
h|\epsilon|,
\end{equation}

\noindent where $\delta$ is the conventional separatrix split
\cite{zaslavsky:1998}, $\epsilon$ is the amplitude of the Melnikov-like
integral defined in Eq.\ (4) above (cf.\
\cite{ZF:1968,Chirikov:79,lichtenberg_lieberman,Zaslavsky:1991,zaslavsky:1998,zaslavsky:2005,abdullaev,gelfreich,treschev,shevchenko:1998,vecheslavov,shevchenko,pre2008,proceedings}),
and the angle $\tilde{\psi}$ may take any value either from the range
$[0,\pi]$ or from the range $[\pi, 2\pi]$\footnote{Of these two intervals, the relevant one is
that in which the derivative ${\rm d}E/{\rm d}t$ in the nonlinear resonance
equations (see Eq.\ (16) below) is positive or negative, for the case of the
upper or lower GSS curve respectively.}.

After that, because of the closeness of $\omega_f$ to the $n$-th harmonic of
$\omega(E)$ in the relevant range of $E$\footnote{I.e.\ $E$ determined by Eq.\
(7) for any $\tilde{\psi}$ except from the vicinity of multiples of
$\pi$. As shown in \cite{pre2008}, Eq.\ (7) is irrelevant to the boundary of
the chaotic layer in the range of $\tilde{\psi}$ close to multiples of $\pi$
while the boundary in this range of $\tilde{\psi}$ still lies in the resonance
range of energies, where $\omega(E)\approx \omega^{(\pm)}$.}, for a relatively
long time the system follows the {\it nonlinear resonance} (NR) dynamics (see
Eq.\ (16) below), during which the deviation of the energy from the separatrix
value grows, greatly exceeding $\delta$ for most of the trajectory. As time
passes, $\tilde{\psi}$ is moving and, at some point, the growth of the
deviation changes for the decrease. This decrease lasts until the
system hits the GSS curve, after which it returns to the separatrix just for
one step of the separatrix map. At the separatrix, the slow angle
$\tilde{\psi}$ changes random-like, so that a new stage of evolution similar to
the one just described occurs, i.e. the nonlinear resonance
dynamics starting from the GSS curve
with a new (random) value of $\tilde{\psi}$.

Of course, the SM cannot describe the variation of the energy during the
velocity pulses (i.e.\ in between instants relevant to the SM): in some cases
this variation can be comparable to the change within the SM dynamics. This
additional variation will be taken into account below, where relevant.

One might argue that, even for the instants relevant to the SM, the SM
describes the original Hamiltonian dynamics only approximately \cite{vered} and
may therefore miss some fine details of the motion: for example, the above
picture does not include small windows of stability on the separatrix itself.
However these fine details are irrelevant in the present context, in particular
the relative portion of the windows of stability on the separatrix apparently
vanishes in the asymptotic limit $h\rightarrow 0$.

\begin{figure}[t]
\sidecaption
\includegraphics*[scale=.22]{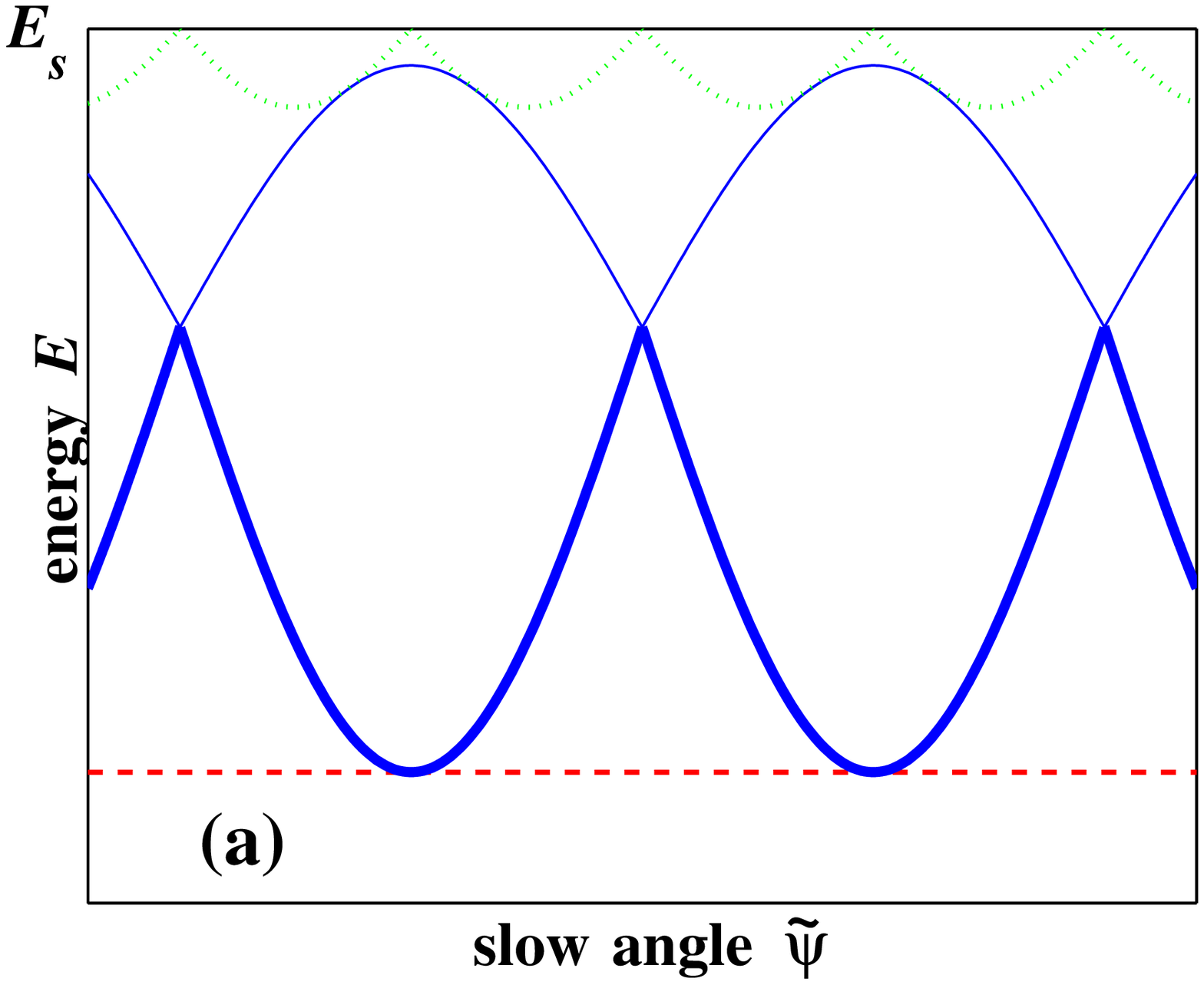}
%\vskip 0.2 cm
\includegraphics*[scale=.22]{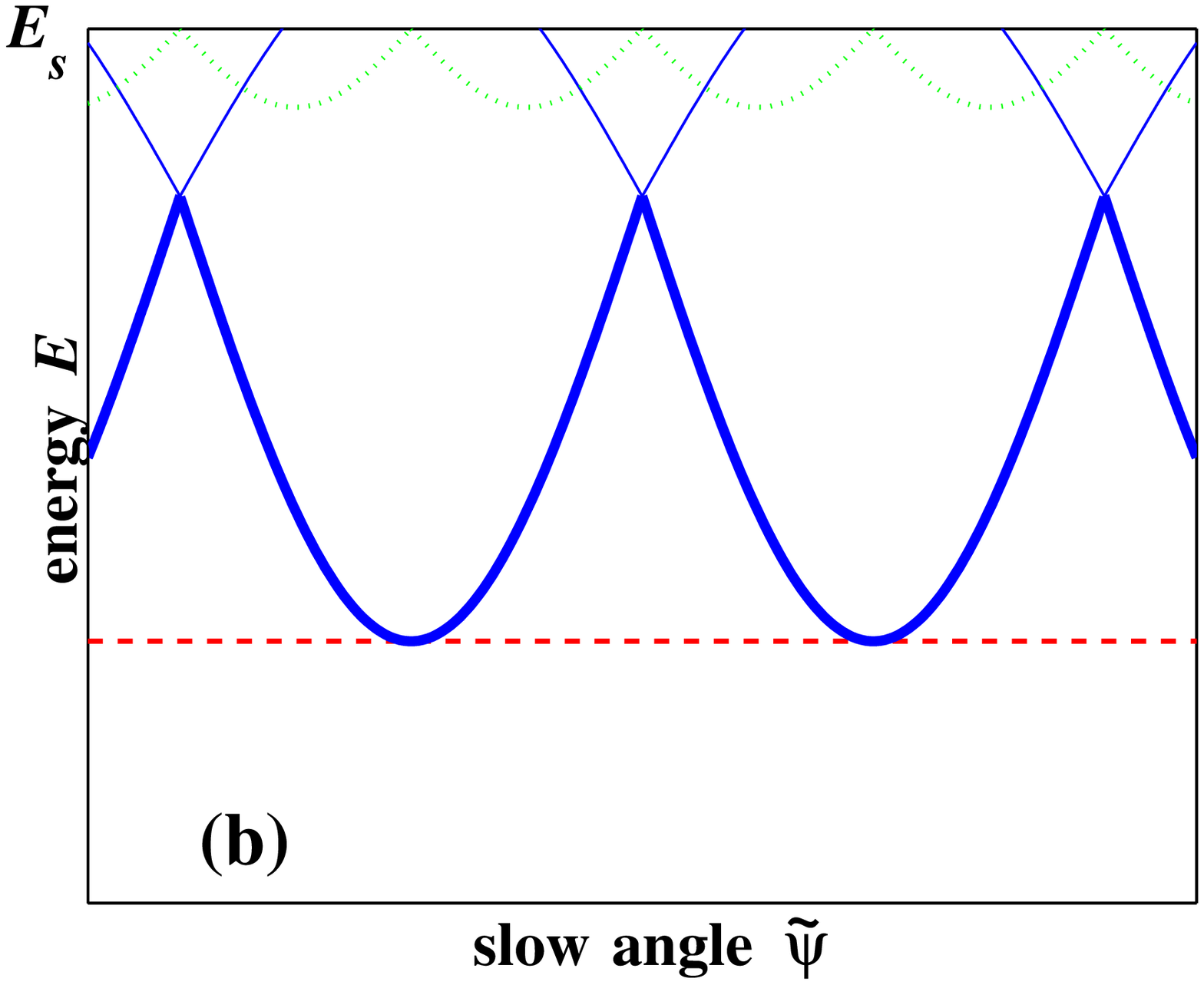}
%\vskip 0.2 cm
\includegraphics*[scale=.22]{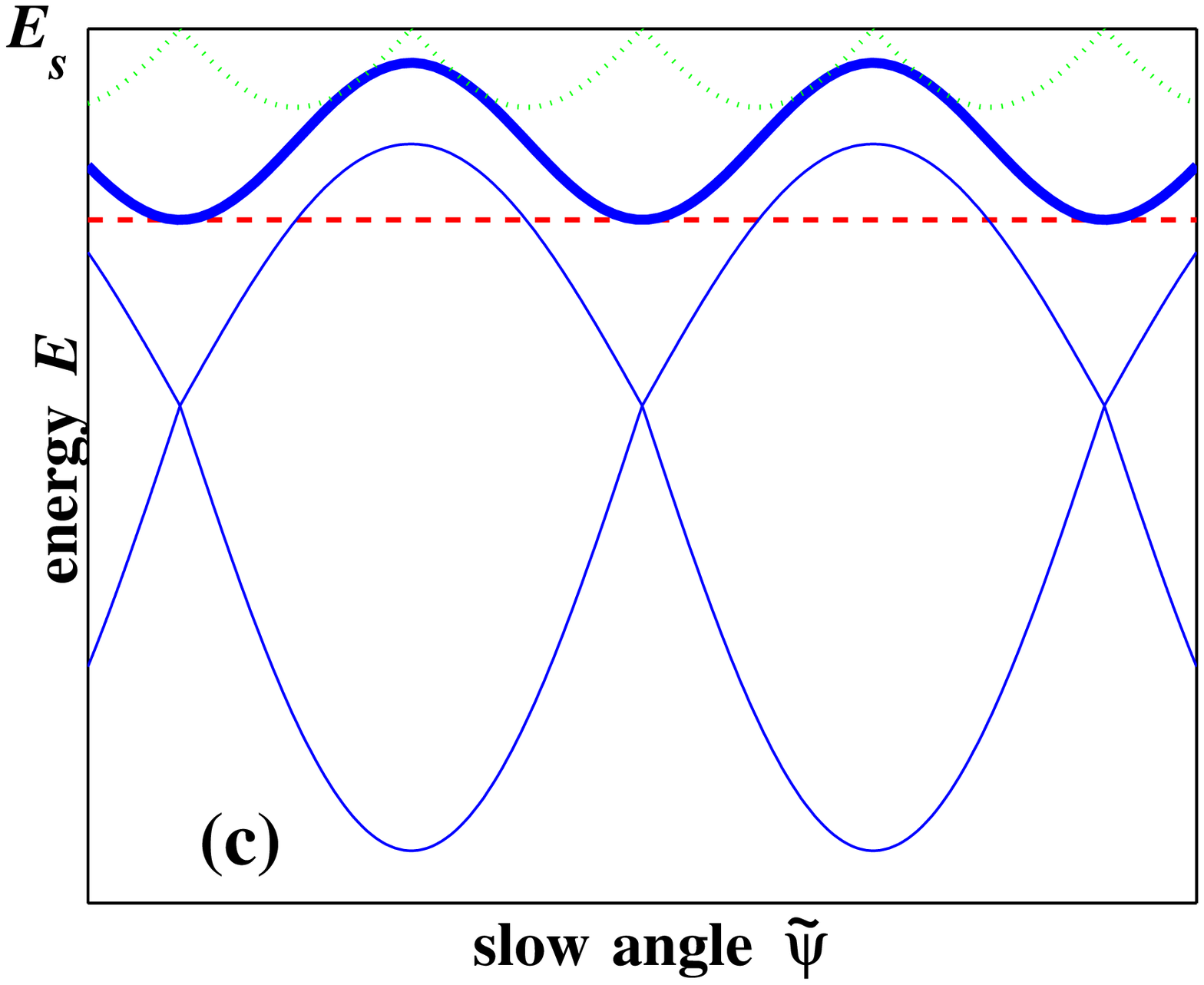}
%\vskip 0.2 cm
\caption {Schematic figure illustrating the formation of the peak of the
function $\Delta E^{(-)}_{sm}(\omega_f)$: (a) $\omega_f=\omega_{\max}$; (b)
$\omega_f<\omega_{\max}$; (c) $\omega_f>\omega_{\max}$. The relevant (lower)
GSS curve is shown by the dotted line. The relevant trajectories of the
resonance Hamiltonian are shown by solid lines. The lower boundary of the layer
is marked by a thick solid line: in (a) and (b) the lower boundary is formed by
the lower part of the resonance separatrix while, in (c) it is formed by the
resonance trajectory tangent to the GSS curve. The dashed line marks, for a
given $\omega_f$, the maximal deviation of the lower boundary from the
separatrix energy $E_s$. } \label{aba:fig1}
\end{figure}

The boundary of the SM chaotic layer is formed by those parts of the SM chaotic
trajectory which deviate from the separatrix more than others. It follows from
the structure of the chaotic trajectory described above that the upper/lower
boundary of the SM chaotic layer is formed in one of the two following ways
\cite{pre2008,proceedings}: (i) if there exists a {\it self-intersecting}
resonance trajectory (in other words, the resonance separatrix) the lower/upper
part of which (i.e.\ the part situated below/above the self-intersection)
touches or intersects the upper/lower GSS curve while the upper/lower part does
not, then the upper/lower boundary of the layer is formed by the upper/lower
part of this self-intersecting trajectory (Figs.\ 1(a) and 1(b)); (ii)
otherwise the boundary is formed by the resonance trajectory {\it tangent} to
the GSS curve (Fig.\ 1(c)). It is shown below that, in both cases, the
variation of the energy along the resonance trajectory is larger than the
separatrix split $\delta$ by a logarithmically large factor $\propto \ln(1/h)$.
Therefore, over the boundary of the SM chaotic layer the largest deviation of
the energy from the separatrix value, $\Delta E^{(\pm)}_{sm}$, may be taken, in
the leading-order approximation, to be equal to the largest variation of the
energy along the resonance trajectory forming the boundary, while the latter
trajectory can be entirely described within the resonance Hamiltonian
formalism.

Finally, we mention that, as is obvious from the above description of the
boundary, $\Delta E^{(\pm)}_{sm}\equiv \Delta E^{(\pm)}_{sm}(\omega_f)$
possesses a local maximum $\Delta E^{(\pm)}_{\max,sm}$ at $\omega_f$ for which
the resonance separatrix just {\it touches} the corresponding GSS curve (see
Fig.\ 1(a)).

\section{Single-Separatrix Chaotic Layer}
\label{sec:3}

It is clear from Sec.\ 2 above that $\Delta E^{(\pm)}_{\max,sm}$ is equal in
leading order to the width $\Delta E_{NR}$ of the nonlinear resonance which
touches the separatrix. In Sec.\ 3.1 below, we {\it roughly} estimate $\Delta E_{NR}$
in order to classify two different types of systems. Secs.\ 3.2 and 3.3 present
the {\it accurate} leading-order asymptotic theory for the two types of systems. The
{\it next-order} correction is estimated in Sec.\ 3.4, while a {\it discussion}
is presented in Sec.\ 3.5.

\subsection
{Rough estimates. Classification of systems.}
\label{subsec:2}

Let us roughly estimate $\Delta E_{NR}$: it will turn out that it is thus possible to
classify all systems into two different types. With this aim, we expand the
perturbation $V$ into two Fourier series in $t$ and in $\psi$ respectively:

\begin{equation}
V\equiv  \frac{1}{2}\sum_l V^{(l)}(E,\psi){\rm e}^{-{\rm
i}l\omega_ft}+{\rm c.c.}\equiv\frac{1}{2}\sum_{l,k}
V^{(l)}_{k}(E){\rm e}^{{\rm i}(k\psi-l\omega_ft)}+{\rm c.c.}
\end{equation}

As in standard nonlinear resonance theory
\cite{Chirikov:79,lichtenberg_lieberman,Zaslavsky:1991,zaslavsky:1998,zaslavsky:2005},
we single out the relevant (for a given peak) numbers $K$ and $L$ for the blind
indices
$k$ and $l$ respectively, and denote the
absolute value of $V^{(L)}_{K}$ as $V_0$:

\begin{equation}
V_0(E)\equiv |V^{(L)}_{K}(E)|.
\end{equation}

To estimate the width of the resonance roughly, we use the pendulum approximation of
resonance dynamics
\cite{Chirikov:79,lichtenberg_lieberman,Zaslavsky:1991,zaslavsky:1998,zaslavsky:2005,abdullaev}:

\begin{equation}
%\Delta E_{NR}\sim \sqrt{\frac{8hV_0\omega_f}{|{\rm d}\omega/{\rm d}E|}}.
\Delta E_{NR}\sim \sqrt{8hV_0\omega_f/|{\rm d}\omega/{\rm d}E|}.
\end{equation}

This approximation assumes constancy of ${\rm d}\omega/{\rm d}E$ in the
resonance range of energies, which is not the case here: in reality,
$\omega(E)\propto 1/\ln(1/|E-E_s|)$ in the vicinity of the separatrix
\cite{ZF:1968,Chirikov:79,lichtenberg_lieberman,Zaslavsky:1991,zaslavsky:1998,zaslavsky:2005,abdullaev,treschev,shevchenko:1998,vecheslavov,shevchenko,pre2008,proceedings},
so that the relevant derivative $|{\rm d}\omega/{\rm d}E|\sim
(\omega_r^{(\pm)})^2/(\omega_0|E-E_s|)$ varies strongly within the resonance range.
However, for our rough estimate we may use the maximal value of $|E-E_s|$,
which is approximately equal to $\Delta E_{NR}$. If $\omega_f$ is of the order
of $\omega_r^{(\pm)}\sim \omega_0/\ln(1/h)$, then Eq.\ (10) reduces to the following
approximate asymptotic equation for $\Delta E_{NR}$:

\begin{equation}
\Delta E_{NR}\sim V_0(E=E_s\pm\Delta E_{NR})h\ln(1/h),
\quad\quad
h\rightarrow 0 .
\end{equation}

The asymptotic solution of Eq.\ (11) depends on $V_0(E_s\pm\Delta E_{NR})$ as a
function of $\Delta E_{NR}$. In this context, all systems can be divided in two
types.

\begin{itemize}

\item[{\bf I}] The separatrix of the unperturbed system has {\it two or
    more} saddles while the relevant Fourier coefficient $V^{(L)}\equiv
    V^{(L)}(E,\psi)$ possesses {\it different} values on adjacent saddles.
    Given that, for $E\rightarrow E_s$, the system stays most of time near
    one of the saddles, the coefficient $V^{(L)}(E\rightarrow E_s,\psi)$ as
    a function of $\psi$ is nearly a \lq\lq square wave\rq\rq: it
    oscillates between the values at the different saddles. The relevant
    $K$ is typically odd and, therefore, $V_0(E\rightarrow E_s)$ approaches
    a well defined non-zero value. Thus, the quantity $V_0(E=E_s\pm\Delta E_{NR})$
    in Eq.\ (11) may be approximated by this non-zero limit, and we
    conclude therefore that

\begin{equation}
\Delta E_{NR}\propto h\ln(1/h), \quad\quad h\rightarrow 0 .
\end{equation}

\item[{\bf II}] Either (i) the separatrix of the unperturbed system has a
    {\it single saddle}, or (ii) it has more than one saddle but the
    perturbation coefficient $V^{(L)}$ is {\it identical} for all saddles.
    Then $V^{(L)}(E\rightarrow E_s,\psi)$, as a periodic function of
    $\psi$, significantly differs from its value at the saddle(s) only
    during a small part of the period in $\psi$: this part is $\sim
    \omega(E)/\omega_0\sim 1/\ln(1/|E_s-E|)$. Hence, $V_0(E_s\pm\Delta
    E_{NR})\propto 1/\ln(1/\Delta E_{NR})$. Substituting this value in Eq.\
    (11), we conclude that

\begin{equation}
\Delta E_{NR}\propto h, \quad\quad h\rightarrow 0 .
\end{equation}

\end{itemize}

\noindent Thus, for systems of type I, the maximum width of the SM chaotic
layer is proportional to $h$ times a logarithmically large factor
$\propto\ln(1/h)$ while, for systems of type II, it is proportional to $h$
times a numerical factor.

As shown below, the variation of energy in between the instants relevant to the
SM is $\sim h$, i.e.\ much less than $\Delta E_{NR}$ (12) for systems of the
type I, and of the same order as $\Delta E_{NR}$ (13) for systems of type II.
Therefore, one may expect that the maximum width of the layer for the original
Hamiltonian system (1), $\Delta E^{(\pm)}$, is at least roughly approximated by
that for the SM, $\Delta E_{sm}^{(\pm)}$, so that the above classification of
systems is relevant to $\Delta E^{(\pm)}$ too. This is confirmed both by
numerical integration of the equations of motion of the original Hamiltonian
system and by the accurate theory presented in the next two sub-sections.

\subsection{Asymptotic theory for systems of type I.}

For the sake of clarity, we consider a particular example of a type I system;
its generalization is straightforward.

We choose an archetypal example: the ac-driven pendulum (sometimes referred to
as a pendulum subject to a dipole time-periodic perturbation)
\cite{Zaslavsky:1991,prl2005,13_prime}:

\begin{eqnarray}
&& H=H_0+hV,
\\
&&  H_0=\frac{p^2}{2}-\cos(q), \quad\quad V=-q\cos(\omega_ft),
\quad\quad h\ll 1. \nonumber
\end{eqnarray}

Fig.\ 2 presents the results of numerical simulations for a few values of $h$ and several
values of $\omega_f$. It shows that: (i) that the function $\Delta
E^{(-)}(\omega_f)$ indeed possesses sharp peaks whose heights greatly exceed
the estimates by the heuristic \cite{Zaslavsky:1991}, adiabatic \cite{E&E:1991}
and moderate-frequency \cite{treschev}  theories (see inset); (ii)
as predicted by our rough estimates of Sec.\ 3.1, the 1st peak of $\Delta
E^{(-)}(\omega_f)$ shifts to smaller values of $\omega_f$ while its magnitude
grows, as $h$ decreases. Below, we develop a leading-order asymptotic
theory, in which
the
parameter of smallness is $1/\ln(1/h)$,
and compare it with results of the simulations.

\begin{figure}[b]
%\sidecaption
\includegraphics*[scale=.34]{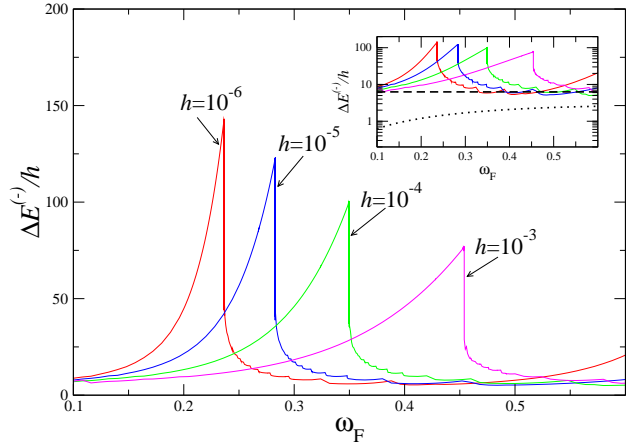}
%[tb]
%\includegraphics*[width = 8.5 cm]{Fig2.eps}
%\vskip 0.2 cm
\caption {Computer simulations for the ac driven pendulum (14) (an archetypal
example of type I): the deviation $\Delta E^{(-)}$ of the lower boundary of the
chaotic layer from the separatrix, normalized by the perturbation amplitude
$h$, is plotted as a function of the perturbation frequency $\omega_f$, for
various $h$. The inset presents the same data but with a logarithmic ordinate
and with the estimates by the heuristic \cite{Zaslavsky:1991}, adiabatic
\cite{E&E:1991} and moderate-frequency \cite{treschev} theories.
The heuristic estimate is shown by the dotted line: as an example of the
heuristic estimate, we use the formula from \cite{Zaslavsky:1991}: $\Delta
E^{(-)}/h=2\pi\omega_f/\cosh(\pi\omega_f/2)$. The adiabatic and
moderate-frequency estimates are shown by the dashed line: the adiabatic
estimate for $\Delta E^{(-)}(\omega_f)$ is equal approximately to $2\pi$; the
estimate following from the results of the work \cite{treschev}
for $\omega_f\sim 1$ is of the same order, so that it is schematically
represented in the inset in Fig.\ 2 by the same line as for the adiabatic
estimate (dashed line). The inset shows explicitly that the simulation results
exceed the estimates of the former theories by 1 or 2 orders of magnitude, over
a wide range of frequencies.}
\end{figure}

Before moving on, we note that the SM (approximated in the relevant case by
nonlinear resonance dynamics) considers states of the system only at discrete
instants. Apart from the variation of energy within the SM dynamics, a
variation of energy in the Hamiltonian system also occurs in between the
instants relevant to the SM. Given that $\omega_f\ll 1$, this latter variation
may be considered in adiabatic approximation and it is of the order of $h$
\cite{E&E:1991,shevchenko}. It follows from the above rough estimates, and from
the more accurate consideration below, that the variation of energy within the
SM dynamics for systems of type I is logarithmically larger i.e.\ larger by the
factor $\ln(1/h)$. The variation of energy in between the instants relevant to
the SM may therefore be neglected to leading-order for systems of type I: $\Delta
E^{(-)}\simeq \Delta E^{(-)}_{sm}$. For the sake of notational compactness, we
shall henceforth omit the subscript ``$sm$" in this subsection.

For the system (14), the separatrix energy is equal to 1, while the
asymptotic (for $E\rightarrow E_s$) dependence $\omega(E)$ is
\cite{Zaslavsky:1991}:

\begin{eqnarray}
&& \omega(E)\simeq \frac{\pi}{\ln(32/|E_s-E|)},
\\
&&  E_s=1, \quad\quad |E_s-E|\ll 1. \nonumber
\end{eqnarray}

\noindent Let us consider the range of energies below $E_s$ (the range above
$E_s$ may be considered in an analogous way) and assume that $\omega_f$ is
close to an odd multiple of $\omega_r^{(-)}$. The nonlinear resonance dynamics
of the slow variables in the range of approximately resonant energies may be
described as follows \cite{pre2008,PR} (cf.\ also
\cite{Chirikov:79,lichtenberg_lieberman,Zaslavsky:1991,zaslavsky:1998,zaslavsky:2005,abdullaev}):

\begin{eqnarray}
&& \frac{{\rm d}I}{{\rm
d}t}=-\frac{\partial{\tilde{H}}(I,{\tilde{\psi}})}{\partial{\tilde{\psi}}},
\quad\quad \frac{{\rm d}{\tilde{\psi}}}{{\rm
d}t}=\frac{\partial{\tilde{H}}(I,{\tilde{\psi}})}{\partial I},
\\
&& \tilde{H}(I,\tilde{\psi})=\int_{I(E_s)}^{I}{\rm d}\tilde{I}\;
(n\omega-\omega_f)\;-\; nhq_n\cos(\tilde{\psi}) \nonumber
\\
&& \quad\; \equiv\; n(E-E_s)-\omega_f(I-I(E_s))\;-\;
nhq_n\cos(\tilde{\psi})\;, \nonumber
\\
&& I \equiv I(E) = \int_{E_{\rm min}}^E \frac{{\rm
d}\tilde{E}}{\omega(\tilde{E})}, \quad\quad  E \equiv H_0(p,q),
\nonumber
\\
&& \tilde{\psi}=n\psi-\omega_ft, \quad\quad \nonumber
\\
&& \psi= \pi+{\rm sign}(p)\omega(E)\int^q_{q_{\rm min}(E)}\frac{{\rm
d}\tilde{q}}{\sqrt{2(E-U(\tilde{q}))}}+2\pi l, \nonumber
\\
&& q_n\equiv q_n(E)= \frac{1}{2\pi}\int_0^{2\pi} \!\!\!\!\! {\rm
d}\psi \; q(E,\psi)\cos(n\psi) ,
%\quad i\equiv \sqrt{-1},
\nonumber
\\
&& |n\omega-\omega_f|\ll\omega,\quad\quad n\equiv 2j-1, \quad\quad
j=1,2,3,\ldots, \nonumber
\end{eqnarray}

\noindent where $I$ and $\psi$ are the canonical variables action and angle
respectively
\cite{Chirikov:79,lichtenberg_lieberman,Zaslavsky:1991,zaslavsky:1998,zaslavsky:2005,abdullaev};
$E_{\rm min}$ is the minimal energy over all $q,p$, $ E \equiv H_0(p,q)$;
$q_{\rm min}(E)$ is the minimum coordinate of the conservative motion with a
given value of energy $E$; $l$ is the number of right turning points in the
trajectory $[q(\tau)]$ of the conservative motion with energy $E$ and given
initial state $(q_0,p_0)$.

The resonance Hamiltonian $\tilde{H}(I,\tilde{\psi})$ is obtained in the following
way. First, the original Hamiltonian $H$ is transformed to action-angle variables
$I-\psi$. Then it is multiplied by $n$ and the term $\omega_fI$ is extracted
(the latter two operations correspond to
the transformation $\psi\rightarrow\tilde{\psi}\equiv
n\psi-\omega_ft$). Finally, the result is being averaged over time i.e.\
only the resonance term in the
double Fourier expansion of the perturbation is kept (it may be done
since the
effect of the fast-oscillating terms
on the dynamics of slow variables is small: see the estimate of the
corrections in Sec.\ 3.4 below).

Let us derive asymptotic expression for $I(E)$, substituting the asymptotic
expression (15) for $\omega(E)$ into the definition of $I(E)$ (16) and carrying
out the integration:

\begin{equation}
I(E)\simeq I(E_s)-\frac{E_s-E}{\pi}\left ( \ln \left
(\frac{32}{E_s-E} \right )+1 \right ).
\end{equation}

As for the asymptotic value $q_n(E\rightarrow E_s)$, it can be
seen that $q(E\rightarrow E_s,\psi)$, as a function of $\psi$,
asymptotically approaches a \lq\lq square wave\rq\rq, oscillating
between $-\pi$ and $\pi$, so that, for sufficiently small $j$,

\begin{eqnarray}
&& q_{2j-1}(E\rightarrow E_s)\simeq (-1)^{j+1}\frac{2}{2j-1},\\
%\quad\quad
&& q_{2j}=0,\nonumber\\
&& j=1,2,...\ll \frac{\pi}{2\omega(E)}.\nonumber
\end{eqnarray}

The next issue is the analysis of the phase space of the resonant
Hamiltonian (16). Substituting $\tilde{H}$ (16) into the equations
of motion (16), it can be seen that their stationary points have
the following values of the slow angle

\begin{equation}
\tilde{\psi}_+=\pi, \quad\quad \tilde{\psi}_-=0,
\end{equation}

\noindent while the corresponding action is determined by the
equation

\begin{equation}
n\omega-\omega_f\mp nh\frac{{\rm d}q_n}{{\rm d}I}=0, \quad\quad
n\equiv 2j-1,
\end{equation}

\noindent where the sign \lq\lq $\mp$\rq\rq corresponds to
$\tilde{\psi}_{\mp}$ (19).

The term $\propto h$ in (20) may be neglected to leading-order (cf.\
\cite{Chirikov:79,lichtenberg_lieberman,Zaslavsky:1991,zaslavsky:1998,zaslavsky:2005,abdullaev,pre2008,PR}), and Eq.\ (20)
reduces to the resonance condition

\begin{equation}
(2j-1)\omega(E_r^{(j)})=\omega_f,
\end{equation}

\noindent the lowest-order solution of which is

\begin{equation}
E_s-E_r^{(j)}\simeq 32\exp\left(-\frac{(2j-1)\pi}{\omega_f}\right).
\end{equation}

Eqs.\ (19) and (22) together with (17) explicitly determine the elliptic and
hyperbolic points of the Hamiltonian (16). The hyperbolic point is often
referred to as a \lq\lq saddle'' and corresponds to $\tilde{\psi}_+$ or
$\tilde{\psi}_-$ in (19) for even or odd $j$ respectively. The saddle point generates the resonance
separatrix. Using the asymptotic relations (17) and (18), we find that the
resonance Hamiltonian (16) takes the following asymptotic value in the saddle:

\begin{eqnarray}
&&\tilde{H}_{saddle}\simeq \frac{E_s-E_r^{(j)}}{\pi}\omega_f-2h\nonumber\\
&&\quad\quad\quad \simeq  \frac{\omega_f}{\pi}32\exp\left(-\frac{\pi
(2j-1)}{\omega_f}\right)-2h.
\end{eqnarray}

\noindent The second asymptotic equality in (23) takes into account
the relation (22).

As explained in Sec.\ 2 above, $\Delta E^{(-)}(\omega_f)$ possesses a local
maximum at $\omega_f$ for which the resonance separatrix is tangent to the
lower GSS curve (Fig.\ 1(a)). For the relevant frequency range
$\omega_f\rightarrow 0$, the separatrix split (which represents the maximum
deviation of the energy along the GSS curve from $E_s$) approaches the
following value \cite{Zaslavsky:1991} in the asymptotic limit $h\rightarrow 0$

\begin{equation}
\delta\simeq  2\pi h, \quad\quad \omega_f\ll 1.
\end{equation}

\noindent As shown below, the variation of energy along the relevant resonance
trajectories is much larger. Therefore, in the leading-order approximation, the
GSS curve may simply be replaced by the separatrix of the unperturbed system
i.e.\ by the horizontal line $E=E_s$ or, equivalently, $I=I(E_s)$. Then the
tangency occurs at $\tilde{\psi}$, shifted from the saddle by $\pi$, so that
the condition of tangency is written as

\begin{equation}
\tilde{H}_{saddle}=\tilde{H}(I=I(E_s),\tilde{\psi}=\tilde{\psi}_{saddle}+\pi)\equiv
2h.
\end{equation}

Substituting here $\tilde{H}_{saddle}$ (23), we finally obtain the
following transcendental equation for $\omega_{\max}^{(j)}$:

\begin{equation}
x\exp(x)=\frac{8(2j-1)}{h},\quad\quad x\equiv
\frac{(2j-1)\pi}{\omega_{\max}^{(j)}}.
\end{equation}

\noindent Fig.\ 3(b) demonstrates the excellent agreement between Eq.\ (26) and
simulations of the Hamiltonian system over a wide range of $h$.

\begin{figure}[b]
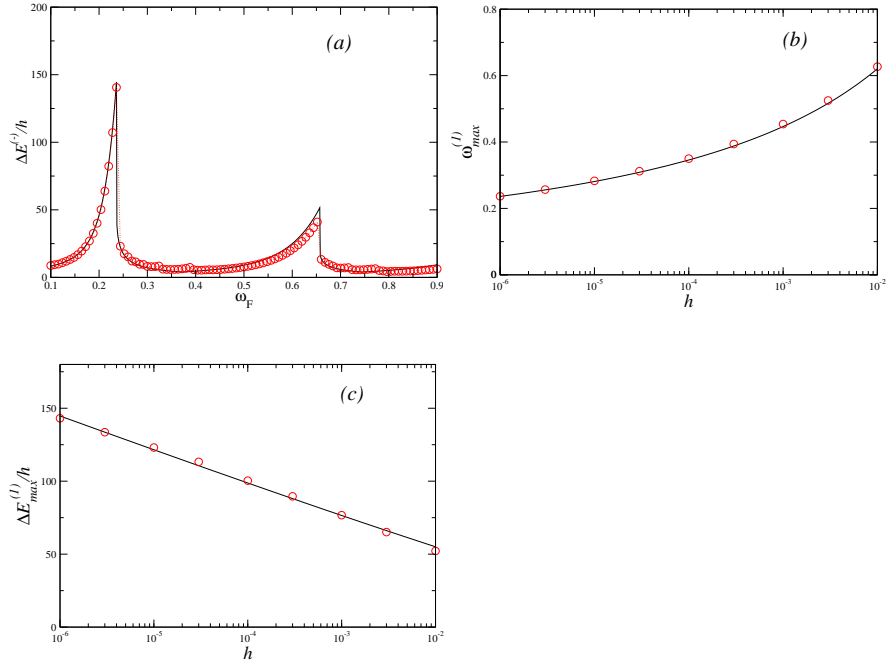

%\sidecaption
\includegraphics[width = 5.8 cm]{soskin_Fig3a.eps}
%\hskip 0.9cm
\includegraphics[width = 5.8 cm]{soskin_Fig3b.eps}
\vskip 0.7cm
\includegraphics[width = 5.8 cm]{soskin_Fig3c.eps}
\caption {An archetypal example of a type I system: the ac-driven pendulum
(14). Comparison of theory (solid lines) and simulations (circles) for: (a) the
deviation $\Delta E^{(-)}(\omega_f)$ of the lower boundary of the chaotic layer
from the separatrix, normalized by the perturbation amplitude $h$, as a
function of the perturbation frequency $\omega_f$, for $h=10^{-6}$; the theory
is from Eqs.\ (26), (31), (32), (38), (39) and (41) (note the discontinuous drop
by the factor
{\rm e} from the maximum to the right wing).
(b) The frequency of the
1st maximum in $\Delta E^{(-)}(\omega_f)$ as a function of $h$; the theory is
from Eq.\ (26). (c) The 1st maximum in $\Delta E^{(-)}(\omega_f)/h$ as a
function of $h$; the theory is from Eqs.\ (34) and (26).} \label{abb:fig3}
\end{figure}

In the asymptotic limit $h\rightarrow 0$, the lowest-order explicit solution of
Eq.\ (26) is

\begin{equation}
\omega_{\max}^{(j)}\simeq
\frac{(2j-1)\pi}{\ln\left(\frac{8(2j-1)}{h}\right)}, \quad\quad
j=1,2,...\ll \ln\left(\frac{1}{h}\right).
\end{equation}

As follows from Eq.\ (26), the value of $E_s-E_r^{(j)}$ (22) for
$\omega_f=\omega_{\max}^{(j)}$ is

\begin{equation}
E_s-E_r^{(j)}(\omega_f=\omega_{\max}^{(j)})=\frac{4\pi
h}{\omega_{\max}^{(j)}}.
\end{equation}

Its leading-order expression is:

\begin{equation}
E_s-E_r^{(j)}(\omega_f=\omega_{\max}^{(j)})\simeq \frac{4
h}{2j-1}\ln\left(\frac{8(2j-1)}{h}\right),
\quad\quad
 h\rightarrow 0.
\end{equation}

If $\omega_f\leq\omega_{\max}^{(j)}$ then, in the chaotic layer, the largest
deviation of energy from the separatrix value corresponds to the minimum energy
$E_{\min}^{(j)}$ on the nonlinear resonance separatrix (Fig.\ 1(a,b)), which
occurs at $\tilde{\psi}$ shifted by $\pi$ from the saddle. The condition of
equality of $\tilde{H}$ at the saddle and at the minimum of the resonance
separatrix is written as

\begin{equation}
\tilde{H}_{saddle}=\tilde{H}(I(E_{\min}^{(j)}),\tilde{\psi}_{saddle}+\pi).
\end{equation}

Let us seek its asymptotic solution in the form

\begin{eqnarray}
&& E_s-E_{\min}^{(j)}\equiv\Delta E_l^{(j)}=(1+y)
(E_s-E_{r}^{(j)})
%\nonumber\\
%&&\quad\quad\quad\quad\quad\quad\quad\quad\quad
\simeq (1+y)32\exp\left(-\frac{\pi (2j-1)}{\omega_f}\right),
\nonumber\\
&& y\stackrel{>}{\sim}1.
\end{eqnarray}

Substituting (31) and (23) into Eq.\ (30), we obtain for $y$ the following
transcendental equation:

\begin{eqnarray}
&& (1+y)\ln(1+y)-y=\frac{h}{8(2j-1)}x_f\exp(x_f),
\\
&&  x_f\equiv \frac{\pi (2j-1)}{\omega_f}, \quad\quad
\omega_f\leq\omega_{\max}^{(j)},\quad\quad y>0, \nonumber
\end{eqnarray}

\noindent where $\omega_{\max}^{(j)}$ is given by Eq.\ (26).

Eqs.\ (31) and (32) describe the left wing of the $j$-th peak of $\Delta
E^{(-)}(\omega_f)$. Fig.\ 3(a) demonstrates the good agreement between our
analytic theory and simulations for the Hamiltonian system.

It follows from Eq.\ (26) that Eq.\ (32) for $\omega_f=\omega_{\max}^{(j)}$
reduces to the relation $\ln(1+y)=1$, i.e.\
\begin{equation}
1+y(\omega_{\max}^{(j)}) = {\rm e}.
\end{equation}

It follows from Eqs.\ (33), (31) and (28) that the maximum for a given peak is:

\begin{equation}
\Delta E^{(j)}_{\max}\equiv
E_s-E^{(j)}_{\min}(\omega_{\max}^{(j)})=\frac{4\pi{\rm
e}h}{\omega_{\max}^{(j)}}.
\end{equation}

Fig.\ 3(c) shows the excellent agreement of this expression with our
simulations of the Hamiltonian system over a wide range of $h$.

The leading-order expression for $\Delta E^{(j)}_{\max}$ is:

\begin{equation}
\Delta E^{(j)}_{\max}\simeq\frac{4{\rm
e}h}{2j-1}\ln(8(2j-1)/h),\quad\quad h\rightarrow 0,
\end{equation}

\noindent which confirms the rough estimate (12).

As $\omega_f$ decreases, it follows from Eq.\ (32) that $y$ increases
exponentially sharply. In order to understand how $\Delta E_l^{(j)}$ decreases
upon decreasing $\omega_f$, it is convenient to rewrite Eq.\ (31) re-expressing
the exponent by means of Eq.\ (32):

\begin{equation}
\Delta E_l^{(j)}(\omega_f)=\frac{4\pi
h}{\omega_f(\ln(1+y)-y/(1+y))}.
\end{equation}

\noindent It follows from Eqs.\ (32) and (36) that $\Delta E_l^{(j)}$ decreases
{\it power-law-like} when $\omega_f$ is decreased. In
particular, $\Delta E_l^{(j)}\propto 1/(\omega_{\max}^{(j)}-\omega_f)$ at the
far part of the wing.

As for the right wing of the peak, i.e.\ for $\omega_f>\omega_{\max}^{(j)}$,
over the chaotic layer, the largest deviation of energy from the separatrix
value corresponds to the minimum of the resonance trajectory tangent to the GSS
curve (Fig.\ 1(c)). The value of $\tilde{\psi}$ at the minimum coincides with
$\tilde{\psi}_{saddle}$. In the leading-order approximation, the GSS curve may
be replaced by the horizontal line $I=I(E_s)$, so that the tangency occurs at
$\tilde{\psi}=\tilde{\psi}_{saddle}+\pi$. Then the energy at the minimum
$E_{\min}^{(j)}$ can be found from the equation

\begin{equation}
\tilde{H}(I(E_s),\tilde{\psi}_{saddle}+\pi)=
\tilde{H}(I(E_{\min}^{(j)}),\tilde{\psi}_{saddle})
\end{equation}

Let us seek its asymptotic solution in the form

\begin{eqnarray}
&& E_s-E_{\min}^{(j)}\equiv\Delta E_r^{(j)}= z
(E_s-E_{r}^{(j)})
%\nonumber\\
%&&\quad\quad\quad\quad\quad\quad
\simeq z32\exp\left(
-\frac{\pi(2j-1)}{\omega_f}\right)
\nonumber\\
&& 0<z< 1, \quad\quad z\sim 1.
\end{eqnarray}

Substituting (38) into (37), we obtain for $z$ the following transcendental
equation:

\begin{eqnarray}
&& z(1+\ln(1/z))=\frac{h}{8(2j-1)}x_f\exp(x_f)\\
&& x_f\equiv\frac{\pi(2j-1)}{\omega_f}, \quad\quad \omega_f>
\omega_{\max}^{(j)},\quad\quad 0<z<1, \nonumber
\end{eqnarray}

\noindent where $\omega_{\max}^{(j)}$ is given by Eq.\ (26). Eqs.\ (38) and
(39) describe the right wing of the $j$-th peak of $\Delta E^{(-)}(\omega_f)$.
Fig.\ 3(a) demonstrates the good agreement between our analytic theory and
simulations.

It follows from Eq.\ (26) that the solution of Eq.\ (39) for
$\omega_f\rightarrow\omega_{\max}^{(j)}$ is $z\rightarrow 1$, so the right wing
starts from the value given by Eq.\ (28) (or, approximately, by Eq.\ (29)).
Expressing the exponent in (38) from (39), we obtain the following equation

\begin{equation}
\Delta E_r^{(j)}(\omega_f)=\frac{4\pi h}{\omega_f(1+\ln(1/z))}.
\end{equation}

\noindent It follows from Eqs.\ (39) and (40) that $\Delta E_r^{(j)}$ decreases
{\it power-law-like} for increasing $\omega_f$. In
particular, $\Delta E_r^{(j)}\propto 1/(\omega_f-\omega_{\max}^{(j)})$ in the
far part of the wing. Further analysis of the asymptotic shape of the peak is
presented in Sec.\ 3.5 below.

Beyond the peaks, the function $\Delta E^{(-)}(\omega_f)$ is logarithmically
small in comparison with the maxima of the peaks. The functions $\Delta
E^{(j)}_l(\omega_f)$ and $\Delta E^{(j)}_r(\omega_f)$ in the ranges beyond the
peaks are also logarithmically small. Hence, nearly any function of
$\Delta E^{(j)}_r(\omega_f)$ and $\Delta E^{(j+1)}_l(\omega_f)$ which
is close to $\Delta E^{(j)}_r(\omega_f)$ in the vicinity of
$\omega_{\max}^{(j)}$ and to $\Delta E^{(j+1)}_l(\omega_f)$ in the vicinity of
$\omega_{\max}^{(j+1)}$
while being sufficiently small beyond the peaks
may be considered as an approximation of the function
$\Delta E^{(-)}(\omega_f)$ to logarithmic accuracy with respect to the maxima
of the peaks, $\Delta E_{\max}^{(j)}$ and $\Delta E_{\max}^{(j+1)}$, in the
whole range $[\omega_{\max}^{(j)},\omega_{\max}^{(j+1)}]$. One of the easiest
options is the following:

\begin{eqnarray}
&& \Delta E^{(-)}(\omega_f)=\Delta E^{(1)}_l(\omega_f)  \;\quad\quad\quad\quad\quad\quad\quad\quad\quad\quad {\rm for}\quad\omega_f<
\omega_{\max}^{(1)},
\nonumber\\
&&\Delta E^{(-)}(\omega_f)=\max\{\Delta E^{(j)}_r(\omega_f),\Delta
E^{(j+1)}_l(\omega_f) \}
%\nonumber\\&&
\quad\quad
{\rm for}\quad\omega_{\max}^{(j)}<\omega_f<
\omega_{\max}^{(j+1)}, \nonumber\\
&&j=1,2,...\ll \frac{\pi}{2\omega_{\max}^{(1)}}.
\end{eqnarray}

\noindent We used this function in Fig.\ 3(a), and the analogous
one
will also be used in the other cases.

In fact, the theory may be generalized in such a way that Eq.\ (41) would
approximate $\Delta E^{(-)}(\omega_f)$ well in the ranges far beyond the peaks
with logarithmic accuracy, even with respect to $\Delta E^{(-)}(\omega_f)$
itself rather than to $\Delta E_{\max}^{(j)}$ only  (cf.\ the next section).
However, we do not do this in the present case, being interested primarily in
the leading-order description of the peaks.

Finally, we demonstrate in Fig.\ 4 that the lowest-order theory describes the
boundary of the layers quite well, even in the Poincar\'{e} section rather than
only in energy/action.

\begin{figure}[t]
\vskip 0.4cm
\includegraphics[width = 5.8 cm]{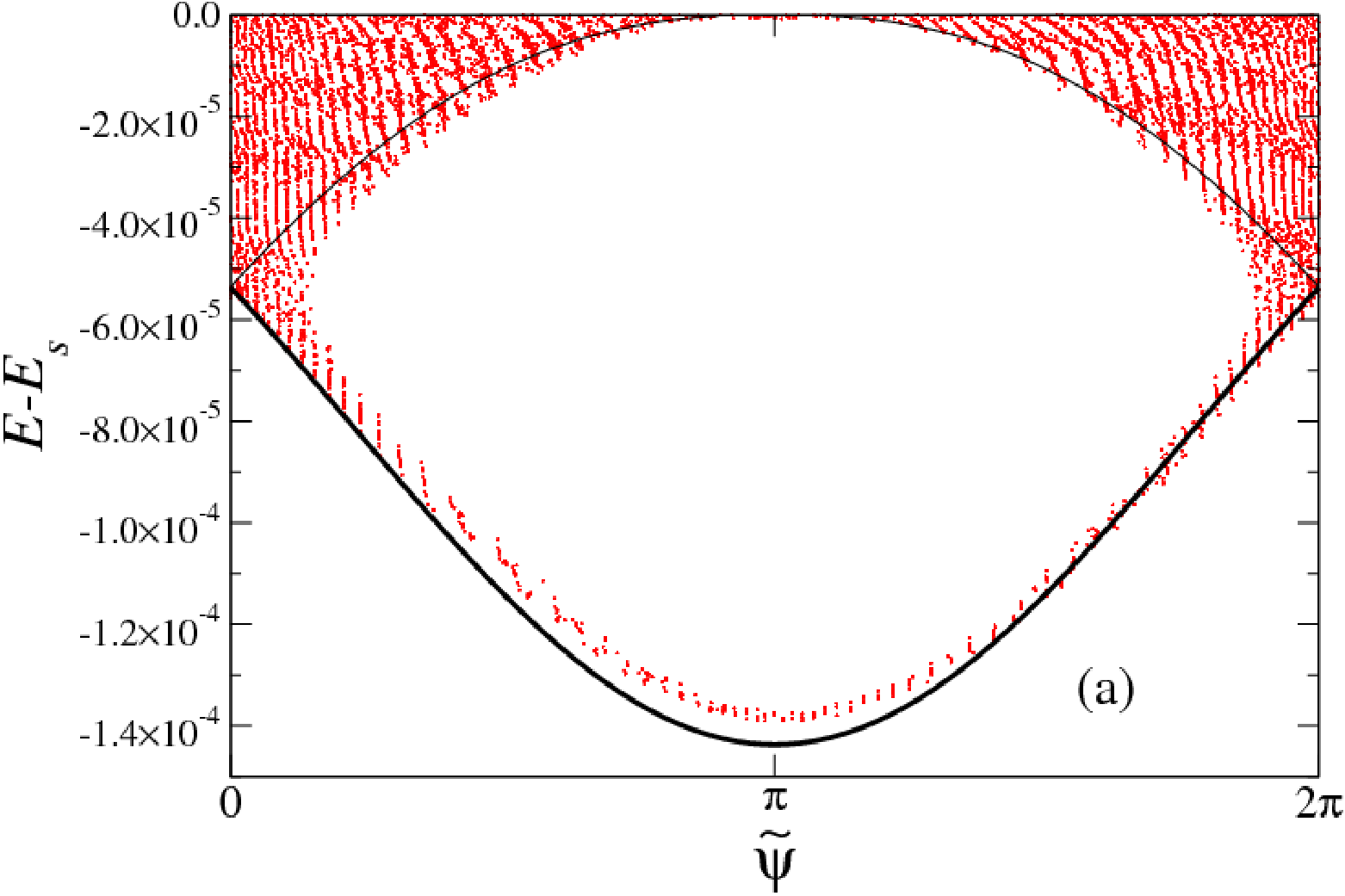}
%\hskip 0.9cm
\includegraphics[width = 5.7 cm]{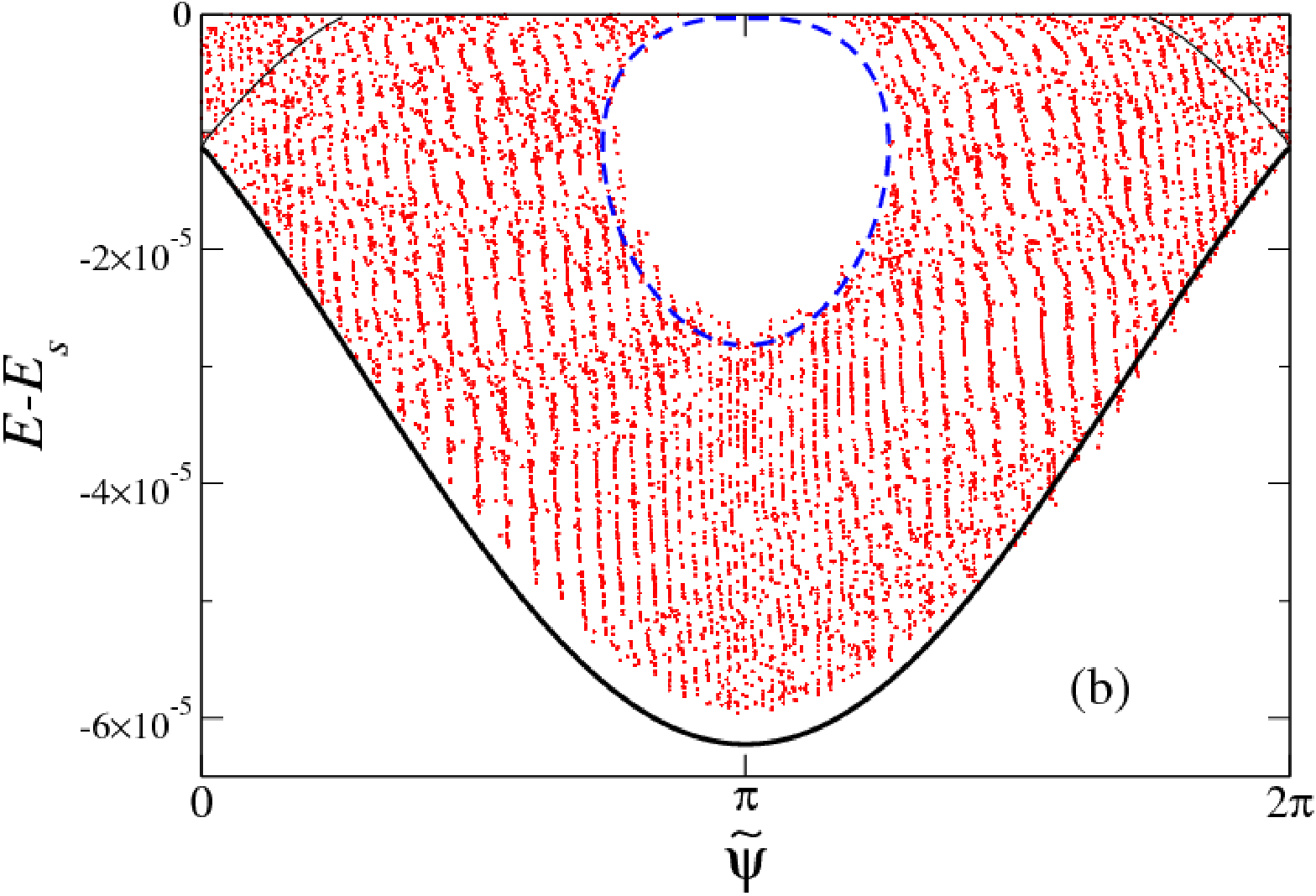}
\vskip 0.7cm
\includegraphics[width = 5.8 cm]{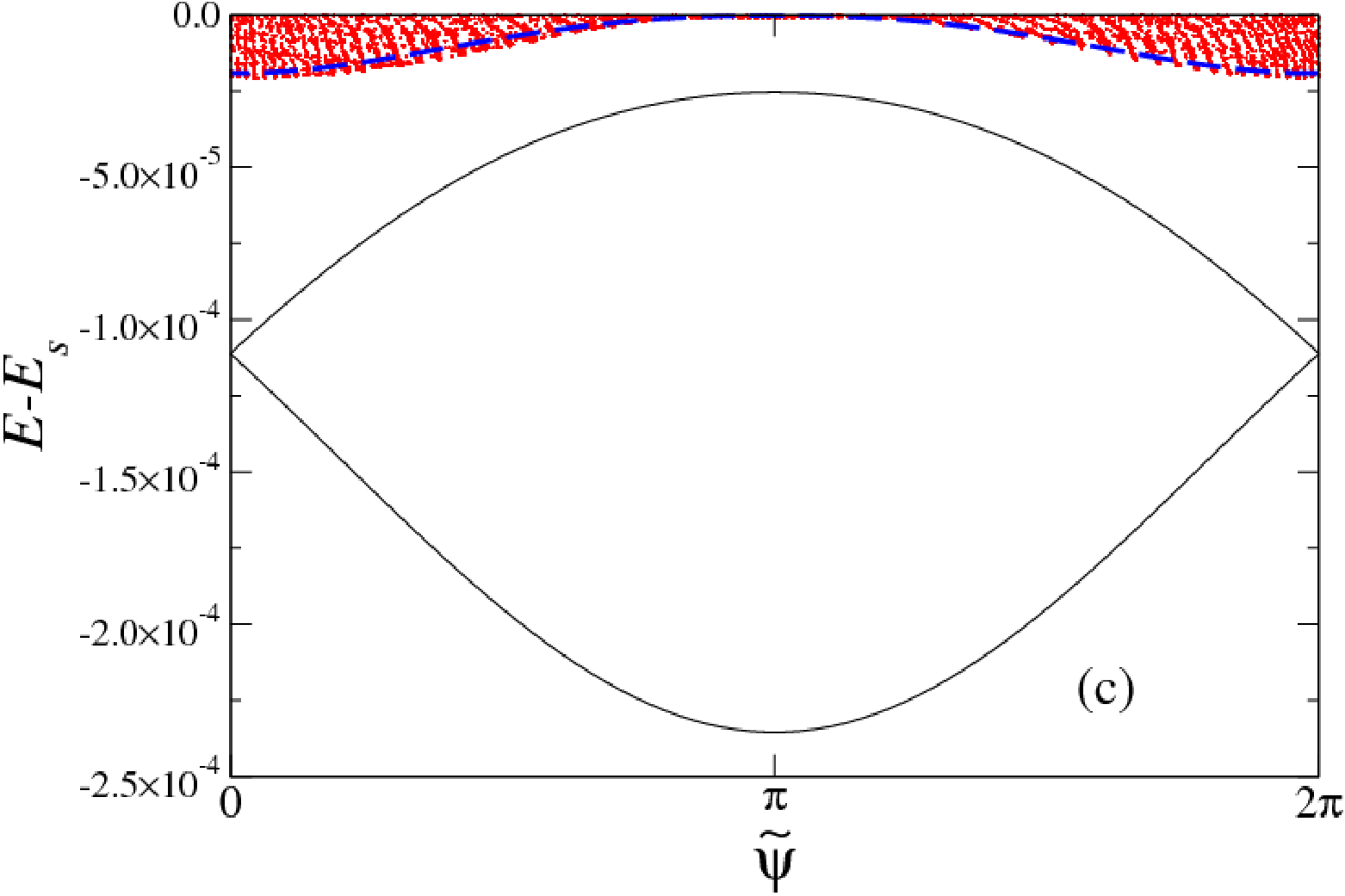}
%\sidecaption
%\includegraphics*[scale=.15]{soskin_Fig4a.eps}
%\includegraphics*[scale=.15]{soskin_Fig4b.eps}
%\includegraphics*[scale=.15]{soskin_Fig4c.eps}
\caption {Some characteristic Poincar\'{e} sections in the $2\pi$-interval of
the energy-angle plane for the system (14) with $h=10^{-6}$ and $\omega_f$
equal to: (a) 0.236 (maximum), (b) 0.21 (left wing), (c) 0.25 (right wing).
Results of the numerical integration of the equations of motion for the
original Hamiltonian (14) are shown by (red) dots. The NR separatrix calculated
in the leading-order approximation (i.e.\ by the integration of the resonant
equations of motion (16) in which $\omega(E)$, $I(E)$ and $q_1(E)$ are
approximated by the explicit formul{\ae} (15), (17) and (18) respectively) is
drawn by the (black) solid line. The NR trajectory (calculated in the
leading-order approximation) tangent to the line $E=E_s$ is drawn by the (blue)
dashed line. The outer boundary (marked by a thicker line) is approximated by:
the lower part of the NR separatrix in cases (a) and (b), and by the tangent NR
trajectory in case (c) The boundary of the island of stability in the cases (a)
and (b) is approximated by the tangent NR trajectory (which coincides in the
case (a) with the NR separatrix).} \label{abc:fig4}
\end{figure}

\subsection{Asymptotic theory for systems of type II.}

We consider two characteristic examples of type II systems, corresponding to
the classification given in Sec.\ 3.1. As an example of a system where the
separatrix of the unperturbed system possesses a single saddle, we consider an
ac-driven Duffing oscillator \cite{abdullaev,gelfreich,treschev,soskin2000}. As
an example of the system where the separatrix possesses more than one saddle,
while the perturbation takes equal values at the saddles, we consider a
pendulum with an oscillating suspension point
\cite{abdullaev,gelfreich,treschev,shevchenko:1998,shevchenko}. The treatment
of these cases is similar in many respects to that presented in Sec.\ 3.2
above. So, we present it in less detail, emphasizing the differences.

\subsubsection{AC-driven Duffing oscillator.}

Consider the following archetypal Hamiltonian
\cite{abdullaev,gelfreich,treschev,soskin2000}:

\begin{eqnarray}
&& H=H_0+hV,
\\
&&  H_0=\frac{p^2}{2}-\frac{q^2}{2}+\frac{q^4}{4}, \quad\quad
V=-q\cos(\omega_ft), \quad\quad h\ll 1. \nonumber
\end{eqnarray}

The asymptotic dependence of $\omega(E)$ on $E$ for $E$ below the
separatrix energy $E_s=0$ is the following
\cite{abdullaev,physica1985}

\begin{eqnarray}
&& \omega(E)\simeq \frac{2\pi}{\ln(16/(E_s-E))},
\\
&&  E_s=0, \quad\quad 0<E_s-E\ll 1. \nonumber
\end{eqnarray}

Correspondingly, the resonance values of energies (determined by the
condition analogous to (21)) are

\begin{equation}
E_s-E_r^{(j)}=16\exp\left(-\frac{2\pi j}{\omega_f}\right),
\quad\quad j=1,2,3,...
\end{equation}

The asymptotic dependence of $I(E)$ is

\begin{equation}
I(E)\simeq I(E_s)-\frac{E_s-E}{2\pi}\left ( \ln \left
(\frac{16}{E_s-E} \right )+1 \right ).
\end{equation}

The nonlinear resonance dynamics is described by the resonance Hamiltonian
$\tilde{H}$ which is identical in form to Eq.\ (16). Obviously, the actual
dependences $\omega(E)$ and $I(E)$ are given by Eq.\ (43) and (45)
respectively. The most important difference is in $q_j(E)$: instead of a
non-zero value (see (18)), it approaches 0 as $E\rightarrow E_s$. Namely, it is
$\propto \omega(E)$ \cite{abdullaev,physica1985}:

\begin{equation}
q_j(E)\simeq \frac{1}{\sqrt{2}}\omega(E),\quad\quad j=1,2,...\ll
\frac{\pi}{\omega(E)},
\end{equation}

\noindent i.e.\ $q_j$ is much smaller than in systems of type I (cf.\ (18)).
Due to this, the resonance is \lq\lq weaker\rq\rq. At the same time, the
separatrix split $\delta$ is also smaller, namely $\sim h\omega_f$ (cf.\
\cite{pre2008}) rather than $\sim h$ as for the systems of type I. That is why
the separatrix chaotic layer is still dominated by resonance dynamics while the
matching of the separatrix map and nonlinear resonance dynamics is still valid
in the asymptotic limit $h\rightarrow 0$ \cite{pre2008}.

Similarly to the previous section, we find the value of $\tilde{H}$ in the
saddle in the leading-order approximation\footnote{The only essential
difference is that $q_n$ at the saddle is described by Eq.\ (46) rather than by
Eq.\ (18).}:

\begin{equation}
\tilde{H}_{saddle}\simeq \omega_f\left(
\frac{E_s-E_r^{(j)}}{2\pi}-\frac{h}{\sqrt{2}} \right),
\end{equation}

\noindent where $E_s-E_r^{(j)}$ is given in (44).

As before, the maximum width of the layer corresponds to $\omega_f$, for which
the resonance separatrix is tangent to the GSS curve (Fig.\ 1(a)). It can be
shown \cite{pre2008} that the angle of tangency asymptotically approaches
$\tilde{\psi}_{saddle}+\pi= \pi$ while the energy still lies in the resonance
range. Here $\omega(E)\approx \omega_{r}^{(-)}\approx \omega_f/j$. Using the
expressions for $\tilde{H}(E,\tilde{\psi})$ (cf.\ (16)), $I(E)$ (45), $q_j(E)$
(46), and taking into account that in the tangency $E <\delta\sim h\omega_f\ll
h$, to leading-order the value of $\tilde{H}$ at the tangency reads
\begin{equation}
\tilde{H}_{tangency}\simeq \omega_f \frac{h}{\sqrt{2}}.
\end{equation}

Allowing for Eqs.\ (47) and (48), the condition for the maximum,
$\tilde{H}_{saddle}=\tilde{H}_{tangency}$, reduces to

\begin{equation}
E_s-E_r^{(j)}(\omega_{\max}^{(j)})\simeq 2\pi \sqrt{2}h.
\end{equation}

Thus these values $E_s-E_r^{(j)}$ are logarithmically smaller than the
corresponding values (28) for systems of type I.

The values of $\omega_f$ corresponding to the maxima of the peaks in
$\Delta E^{(-)}(\omega_f)$ are readily obtained from (49) and
(44):

\begin{equation}
\omega_{\max}^{(j)}\simeq \frac{2\pi j}{\ln(4\sqrt{2}/(\pi
h))},\quad\quad j=1,2,...\ll\ln(1/h).
\end{equation}

The derivation to leading order of the shape of the peaks for the chaotic layer
of the separatrix map, i.e.\ within the nonlinear resonance (NR) approximation,
is similar to that for type I. So, we present only the results, marking them
with the subscript \lq\lq $NR$\rq\rq.

The left wing of the $j$th peak of $\Delta E^{(-)}_{NR}(\omega_f)$
is described by the function

\begin{eqnarray}
&&\Delta E^{(j)}_{l,NR}(\omega_f)=16(1+y)\exp\left(-\frac{2\pi
j}{\omega_f}\right)
%\\&&\quad\quad\quad\quad\quad\quad
\equiv\frac{2\pi\sqrt{2}
h}{\ln(1+y)-y/(1+y)},
\\
%\quad\quad
&&\omega_f\leq\omega_{\max}^{(j)},\nonumber
\end{eqnarray}

\noindent where $y$ is the positive solution of the transcendental
equation

\begin{equation}
(1+y)\ln(1+y)-y=\frac{\pi h}{4\sqrt{2}}\exp\left(\frac{2\pi
j}{\omega_f}\right),\quad\quad y>0.
\end{equation}

In common with the type I case, $1+y(\omega_{\max}^{(j)})={\rm e}$, so that

\begin{equation}
\Delta E^{(j)}_{\max,NR}={\rm e}
(E_s-E_r^{(j)}(\omega_{\max}^{(j)}))\simeq 2\pi{\rm e}\sqrt{2} h.
\end{equation}

\noindent Eq.\ (53) confirms the rough estimate (13). The right wing of the
peak is described by the function

\begin{eqnarray}
&&\Delta E^{(j)}_{r,NR}(\omega_f)=16z\exp\left(-\frac{2\pi
j}{\omega_f}\right)
%\nonumber\\ &&\quad\quad
\equiv\frac{2\pi\sqrt{2} h}{1+\ln(1/z)},\\
% \quad\quad
&&\omega_f>\omega_{\max}^{(j)},
\nonumber
\end{eqnarray}

\noindent where $z<1$ is the solution of the transcendental equation

\begin{equation}
z(1+\ln(1/z))=\frac{\pi h}{4\sqrt{2}}\exp\left(\frac{2\pi
j}{\omega_f}\right),\quad\quad 0<z<1.
\end{equation}

\noindent As in the type I case,
$z(\omega_f\rightarrow\omega_{\max}^{(j)})\rightarrow 1$.

It follows from Eqs.\ (49) and (53) that the typical variation of energy within
the nonlinear resonance dynamics (that approximates the separatrix map
dynamics) is $\propto h$. For the Hamiltonian system, the variation of energy
in between the discrete instants corresponding to the separatrix map
\cite{Zaslavsky:1991,zaslavsky:1998,zaslavsky:2005,abdullaev,pre2008,vered} is
also $\propto h$. Therefore, unlike the type I case, one needs to take it into
account even at the leading-order approximation. Let us consider the right well
of the Duffing potential (the results for the left well are identical), and
denote by $t_k$ the instant at which the energy $E$ at a given $k$-th step of
the separatrix map is taken: it corresponds to the beginning of the $k$-th
pulse of velocity \cite{Zaslavsky:1991,pre2008} i.e.\ the corresponding $q$ is
close to a left turning point $q_{ltp}$ in the trajectory $[q(\tau)]$. Let us
also take into account that the relevant frequencies are small so that the
adiabatic approximation may be used. Thus, the change of energy from $t_k$ up
to a given instant $t$ during the following pulse of velocity ($t-t_k\sim 1$)
may be calculated as

\begin{eqnarray}
&&\Delta E =\int_{t_k}^{t}{\rm
d}\tau\dot{q}h\cos(\omega_f\tau)\simeq
h\cos(\omega_ft_k)\int_{t_k}^{t}{\rm
d}\tau\dot{q}\nonumber\\
&&\quad\quad=h\cos(\omega_ft_k)(q(t)-q_{ltp})
\end{eqnarray}

For the motion near the separatrix, the velocity pulse corresponds
approximately to $\psi=0$ (see the definition of $\psi$ (16)). Thus,
the corresponding slow angle is $\tilde{\psi}\equiv
j\psi-\omega_ft_k\simeq  -\omega_ft_k$.

For the left wing of the peak of $\Delta E^{(-)}(\omega_f)$ (including also the
maximum of the peak), the boundary of the chaotic layer of the separatrix map
is formed by the lower part of the NR separatrix. The minimum energy along this
separatrix occurs at $\tilde{\psi}=\pi$. Taking this into account, and also
noting that $\tilde{\psi}\simeq  -\omega_ft_k$, we conclude that
$\cos(\omega_ft_k)\simeq  -1$. So, $\Delta E\leq 0$, i.e.\ it does lower the
minimum energy of the layer of the Hamiltonian system. The maximum reduction
occurs at the right turning point $q_{rtp}$:

\begin{equation}
\max(|\Delta E|)\simeq  h(q_{rtp}-q_{ltp})=\sqrt{2}h.
\end{equation}

We conclude that the left wing of the $j$-th peak is described as follows:

\begin{equation}
\Delta E_l^{(j)}(\omega_f)\simeq  \Delta
E_{l,NR}^{(j)}(\omega_f)+\sqrt{2}h,\quad\quad\omega_f\leq\omega_{\max}^{(j)},
\end{equation}

\noindent where $\Delta E_{l,NR}^{(j)}(\omega_f)$ is given by Eqs.\ (51)-(52).
In particular, the maximum of the peak is:

\begin{equation}
\Delta E^{(j)}_{\max}\simeq  (2\pi {\rm e}+1)\sqrt{2}h\approx 25.6h.
\end{equation}

For the right wing of the peak, the minimum energy of the layer of the
separatrix map occurs when $\tilde{\psi}$ coincides with
$\tilde{\psi}_{saddle}$ (Fig.\ 1(c)) i.e.\ is equal to 0. As a result,
$\cos(\omega_ft_k)\simeq 1$ and, hence, $\Delta E\geq 0$. So, this variation
cannot lower the minimum energy of the layer for the main part of the wing,
i.e.\ for $\omega_f\leq\omega_{bend}^{(j)}$ where $\omega_{bend}^{(j)}$ is
defined by the condition $\Delta E_{r,NR}^{(j)}= \max(|\Delta E|)\equiv
\sqrt{2}h$. For $\omega_f>\omega_{bend}^{(j)}$, the minimal energy in the layer
occurs at $\tilde{\psi}=\pi$, and it is determined exclusively by the variation
of energy during the velocity pulse (the NR contribution is close to zero at
such $\tilde{\psi}$). Thus, we conclude that there is a bending of the wing at
$\omega_f=\omega_{bend}^{(j)}$:

\begin{eqnarray}
&&\Delta E_r^{(j)}(\omega_f)= \Delta E_{r,NR}^{(j)}(\omega_f),
\quad\quad\omega_{\max}^{(j)}<\omega_f\leq\omega_{bend}^{(j)},\nonumber\\
&&\Delta E_r^{(j)}(\omega_f)= \sqrt{2}h,
\quad\quad\quad\quad\quad\omega_f\geq\omega_{bend}^{(j)},\nonumber\\
&&\omega_{bend}^{(j)}=\frac{2\pi j}{\ln(8\sqrt{2}/h)+1-2\pi},
\end{eqnarray}

\noindent where $\Delta E_{r,NR}^{(j)}(\omega_f)$ is given by Eqs.\ (54) and
(55).

Analogously to the previous case, $\Delta E^{(-)}(\omega_f)$ may be
approximated over the whole frequency range by Eq.\ (41) with $\Delta
E_l^{(j)}$ and $\Delta E_r^{(j)}$ given by Eqs.\ (58) and (60) respectively.
Moreover, unlike the previous case, the theory also describes accurately the
range far beyond the peaks: $\Delta E^{(-)}$ is dominated in this range by the
velocity pulse contribution $\Delta E$, which is accurately taken into account
both by Eqs.\ (58) and (60).

\begin{figure}[b]
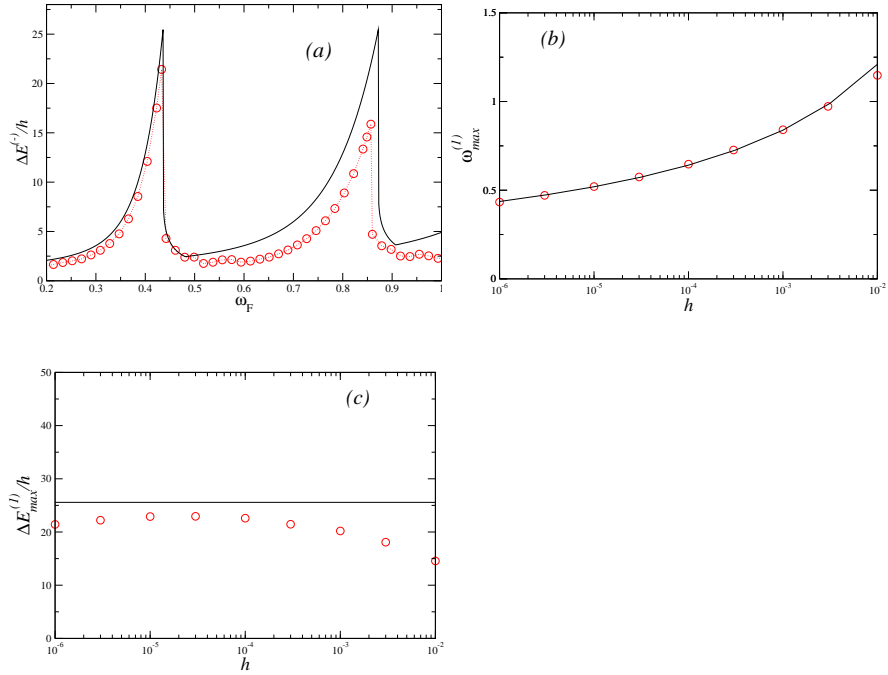

\includegraphics[width = 5.8 cm]{soskin_Fig5a.eps}
%\hskip 0.9cm
\includegraphics[width = 5.8 cm]{soskin_Fig5b.eps}
\vskip 0.7cm
\includegraphics[width = 5.8 cm]{soskin_Fig5c.eps}
%\sidecaption
%\includegraphics*[scale=.156]{soskin_Fig5a.eps}
%\includegraphics*[scale=.156]{soskin_Fig5b.eps}
%\includegraphics*[scale=.156]{soskin_Fig5c.eps}
\caption{An archetypal example of a type II system: the ac driven Duffing
oscillator (42). Comparison of theory (solid lines) and simulations (circles):
(a) the deviation $\Delta E^{(-)}(\omega_f)$ of the lower boundary of the
chaotic layer from the separatrix, normalized by the perturbation amplitude
$h$, as a function of the perturbation frequency $\omega_f$, for $h=10^{-6}$;
the theory is from Eqs.\ (41), (50), (51), (52), (54), (55), (58) and (60)
(note the discontinuous drop from the maximum to the right wing); (b)
the frequency of the 1st maximum in $\Delta E^{(-)}(\omega_f)$ as a function of
$h$; the theory is from Eq.\ (50); (c) the 1st maximum in $\Delta
E^{(-)}(\omega_f)/h$ as a function of $h$; the theory is from Eq.\ (59).}
\label{abd:fig5}
\end{figure}

Fig.\ 5 shows very reasonable agreement between the theory and simulations,
especially for the 1st peak\footnote{The disagreement between theory and
simulations for the magnitude of the 2nd peak is about three times larger than
that for the 1st peak, so that the height of the 2nd peak is about 30$\%$
smaller than that calculated from the asymptotic theory. This occurs because,
for the energies relevant to the 2nd peak, the deviation from the separatrix is
much higher than that for the 1st peak. Due to the latter, the Fourier
coefficient $q_2(E)$ for the relevant $E$ is significantly smaller than that
obtained from the asymptotic formula (42). In addition, the velocity pulse
contribution $\Delta E$ also significantly decreases while the separatrix split
increases as $\omega_f$ becomes $\sim 1$.
%The generalization of the theory for $\omega_f\sim 1$ will be presented elsewhere.
}.

\subsubsection{Pendulum with an oscillating suspension point}

Consider the archetypal Hamiltonian
\cite{abdullaev,gelfreich,treschev,shevchenko:1998,shevchenko}

\begin{eqnarray}
&& H=H_0+hV,
\nonumber\\
&&  H_0=\frac{p^2}{2}+\cos(q), \quad\quad
V=-\cos(q)\cos(\omega_ft), \quad\quad h\ll 1.
\end{eqnarray}

Though the treatment is similar to that used in the previous case, there are
also characteristic differences. One of them is the following: although the
resonance Hamiltonian is similar to the Hamiltonian (16), instead of the
Fourier component of the coordinate, $q_n$, there should be the Fourier
component of $\cos(q)$, $V_n$, which can be shown to be:

\begin{eqnarray}
&& V_{2j}\simeq  (-1)^{j+1}\frac{4}{\pi}\omega(E), \quad\quad
E_s-E\ll 1, \\
% \quad\quad
&& V_{2j-1}=0,
\nonumber\\
&& j=1,2,...\ll\frac{2\pi}{\omega(E)}, \quad\quad
V_n\equiv\frac{1}{2\pi}\int_0^{2\pi}{\rm
d}\psi\cos(q)\cos(n\psi).\nonumber
\end{eqnarray}

The description of the chaotic layer of the separatrix map at the lowest order,
i.e.\ within the NR approximation, is similar to that for the ac-driven Duffing
oscillator. So we present only the results, marking them with the subscript
\lq\lq $NR$\rq\rq.

The frequency of the maximum of a given $j$-th peak is:

\begin{equation}
\omega_{\max}^{(j)}\simeq \frac{2\pi j}{\ln(4/h)},\quad\quad
j=1,2,...\ll\ln(4/h).
\end{equation}

\noindent This expression agrees well with simulations for the Hamiltonian
system (Fig.\ 6(b)). To logarithmic accuracy, Eq.\ (63) coincides with the
formula following from Eq.\ (8) of \cite{shevchenko:1998} (reproduced in
\cite{shevchenko} as Eq.\ (21)) taken in the asymptotic limit $h\rightarrow 0$
(or, equivalently, $\omega_{\max}^{(j)}\rightarrow 0$). However, the numerical
factor in the argument of the logarithm in the asymptotic formula following
from the result of \cite{shevchenko:1998,shevchenko} is half our value: this is
because the nonlinear resonance is approximated in
\cite{shevchenko:1998,shevchenko} by the conventional pendulum model which is
not valid near the separatrix (cf.\ our Sec.\ 3.1 above).

The left wing of the $j$th peak of $\Delta E^{(-)}_{NR}(\omega_f)$
is described by the function

\begin{eqnarray}
&&\Delta E^{(j)}_{l,NR}(\omega_f)=32(1+y)\exp\left(-\frac{2\pi
j}{\omega_f}\right)
%\\&&\quad\quad\quad\quad\quad\quad
\equiv\frac{8
h}{\ln(1+y)-y/(1+y)},\\
%\quad\quad
&&\omega_f\leq\omega_{\max}^{(j)},\nonumber
\end{eqnarray}

\noindent where $y$ is the positive solution of the transcendental
equation

\begin{equation}
(1+y)\ln(1+y)-y=\frac{h}{4}\exp\left(\frac{2\pi
j}{\omega_f}\right),\quad\quad y>0.
\end{equation}

\noindent Similarly to the previous cases, $1+y(\omega_{\max}^{(j)})={\rm e}$.
Hence,

\begin{equation}
\Delta E^{(j)}_{\max,NR}={\rm e}
(E_s-E_r^{(j)}(\omega_{\max}^{(j)}))=8{\rm e} h.
\end{equation}

\noindent Eq.\ (66) confirms the rough estimate (13). The right wing of the
peak is described by the function

\begin{eqnarray}
&&\Delta E^{(j)}_{r,NR}(\omega_f)=32z\exp\left(-\frac{2\pi
j}{\omega_f}\right) \equiv\frac{8
h}{1+\ln(1/z)}, \\
&&\omega_f>\omega_{\max}^{(j)},\nonumber
\end{eqnarray}

\noindent where $z<1$ is the solution of the transcendental
equation

\begin{equation}
z(1+\ln(1/z))=\frac{h}{4}\exp\left(\frac{2\pi
j}{\omega_f}\right),\quad\quad 0<z<1.
\end{equation}

\noindent Similarly to the previous cases,
$z(\omega_f\rightarrow\omega_{\max}^{(j)})\rightarrow 1$.

Now consider the variation of energy during a velocity pulse. Though the final
result looks quite similar to the case with a single saddle, its derivation has
some characteristic differences, and we present it in detail. Unlike the case
with a single saddle, the pulse may start close to either the left or the right
turning point, and the sign of the velocity in such pulses is opposite
\cite{Zaslavsky:1991,pre2008}. The angle $\psi$ in the pulse is close to
$-\pi/2$ or $\pi/2$ respectively. So, let us calculate the change of energy
from the beginning of the pulse, $t_k$, until a given instant $t$ within the
pulse:

\begin{eqnarray}
&&\Delta E =-\int_{t_k}^{t}{\rm d}\tau\dot{q}h\partial V/\partial
q=h\int_{t_k}^{t}{\rm
d}\tau\dot{q}(-\sin(q)\cos(\omega_f\tau))\nonumber\\
&&\simeq  h\cos(\omega_ft_k)\int_{t_k}^{t}{\rm
d}\tau\dot{q}(-\sin(q))\simeq  h\cos(\omega_ft_k)(\cos(q(t))-1) .
\end{eqnarray}

\noindent Here, the third equality assumes adiabaticity while the last equality
takes into account that the turning points are close to the maxima of the
potential i.e.\ close to a multiple of $2\pi$ (where the cosine is equal to 1).

The quantity $\Delta E$ (69) takes its maximal absolute value at $q=\pi$. So,
we shall further consider

\begin{equation}
\Delta E_{\max} =-2h\cos(\omega_ft_k)\equiv
-2h\cos(2j\psi_k-\tilde{\psi}_k)=(-1)^{j+1} 2h\cos(\tilde{\psi}_k).
\end{equation}

\noindent The last equality takes into account that, as mentioned above, the
relevant $\psi_k$ is either $-\pi/2$ or $\pi/2$. For the left wing, the value
of $\tilde{\psi}$ at which the chaotic layer of the separatrix map possesses a
minimal energy corresponds to the minimum of the resonance separatrix. It is
equal to $\pi$ or $0$ if the Fourier coefficient $V_{2j}$ is positive or
negative, i.e.\ for odd or even $j$, respectively: see Eq.\ (63). Thus $\Delta
E_{\max}=-2h$ for any $j$ and, therefore, it does lower the minimal energy of
the boundary. We conclude that

\begin{equation}
\Delta E_l^{(j)}(\omega_f)\simeq  \Delta
E_{l,NR}^{(j)}(\omega_f)+2h,\quad\quad\omega_f\leq\omega_{\max}^{(j)},
\end{equation}

\noindent where $\Delta E_{l,NR}^{(j)}(\omega_f)$ is given by Eqs.\ (64)-(65).
In particular, the maximum of the peak is:

\begin{equation}
\Delta E^{(j)}_{\max}\simeq  (4{\rm e}+1)2h\approx 23.7h.
\end{equation}

\begin{figure}[b]
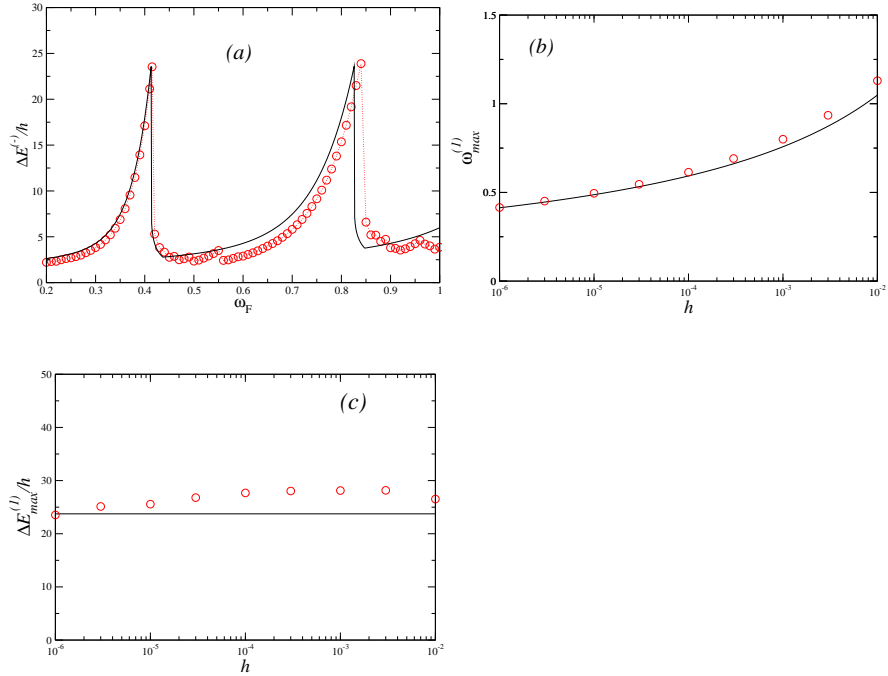

\includegraphics[width = 5.8 cm]{soskin_Fig6a.eps}
%\hskip 0.9cm
\includegraphics[width = 5.8 cm]{soskin_Fig6b.eps}
\vskip 0.7cm
\includegraphics[width = 5.8 cm]{soskin_Fig6c.eps}
%\sidecaption
%\includegraphics*[scale=.156]{soskin_Fig6a.eps}
%\includegraphics*[scale=.156]{soskin_Fig6b.eps}
%\includegraphics*[scale=.156]{soskin_Fig6c.eps}
\caption{An archetypal example of a type II system: the pendulum with an
oscillating suspension point (61). Comparison of theory (solid lines) and
simulations (circles): (a) The deviation $\Delta E^{(-)}(\omega_f)$ of the
lower boundary of the chaotic layer from the separatrix, normalized by the
perturbation amplitude $h$, as a function of the perturbation frequency
$\omega_f$, for $h=10^{-6}$; the theory is by Eqs.\ (41), (63), (64), (65),
(67), (68), (71) and (73)
(note the dicontinuous drop from the maximum to the right wing).
(b) The frequency of the 1st maximum in $\Delta
E^{(-)}(\omega_f)$ as a function of $h$; the theory is from Eq.\ (63). (c) The
1st maximum in $\Delta E^{(-)}(\omega_f)/h$ as a function of $h$; the theory is
from Eq.\ (72).} \label{abe:fig6}
\end{figure}

The expression (72) confirms the rough estimate (13) and agrees well with
simulations (Fig.\ 6(c)). At the same time, it differs from the formula which
can be obtained from Eq.\ (10) of \cite{shevchenko:1998} (using also Eqs.\ (1),
(3), (8), (9) of \cite{shevchenko:1998}) in the asymptotic limit $h\rightarrow
0$: the latter gives for $\Delta E_{\max}^{(j)}$ the asymptotic value $32h$.
Though this result \cite{shevchenko:1998} (referred to also in
\cite{shevchenko}) provides for the correct functional dependence on $h$, it is
quantitatively incorrect because (i) it is based on the pendulum approximation
of the nonlinear resonance while this approximation is invalid in the vicinity
of the separatrix (see the discussion of this issue in Sec.\ 3.1 above), and
(ii) it does not take into account the variation of energy during the velocity
pulse.

The right wing, by analogy to the case of the Duffing oscillator, possesses a
bend at $\omega_f=\omega_{bend}^{(j)}$ where $\Delta E_{r,NR}^{(j)}= |\Delta
E_{\max}|\equiv 2h$, corresponding to the shift of the relevant
$\tilde{\psi}$ for $\pi$. We conclude that:

\begin{eqnarray}
&&\Delta E_r^{(j)}(\omega_f)= \Delta E_{r,NR}^{(j)}(\omega_f),
\quad\quad\omega_{\max}^{(j)}<\omega_f\leq\omega_{bend}^{(j)},\nonumber\\
&&\Delta E_r^{(j)}(\omega_f)= 2h,
\quad\quad\quad\quad\quad\quad\omega_f\geq\omega_{bend}^{(j)},\nonumber\\
&&\omega_{bend}^{(j)}=\frac{2\pi j}{\ln(16/h)-3},
\end{eqnarray}

\noindent where $\Delta E_{r,NR}^{(j)}(\omega_f)$ is given by Eqs.\ (66) and
(67).

Similarly to the previous case, both the peaks and the frequency ranges far
beyond the peaks are well approximated by Eq.\ (41), with $\Delta E_l^{(j)}$
and $\Delta E_r^{(j)}$ given by Eqs.\ (71) and (73) respectively (Fig.\ 6(a)).

\subsection{Estimate of the next-order corrections}

We have calculated explicitly only the leading
%(lowest-order)
term $\Delta E$ in the
asymptotic expansion of the chaotic layer width. Explicit calculation of the
next-order term $\Delta E^{(next)}$ is possible, but it is rather complicated and cumbersome: cf.\
the closely related case with two separatrices \cite{pre2008} (see also Sec.\ 4
below). In the present section, where the perturbation amplitude $h$ in the
numerical examples is 4 orders of magnitude smaller than that in
\cite{pre2008}, there is no particular need to calculate the next-order
correction quantitatively. Let us estimate it, however, in order to
demonstrate that its ratio to the leading term does vanish in the
asymptotic limit $h\rightarrow 0$.

We shall consider separately the contribution $\Delta E^{(next)}_{w}$ stemming from the various
corrections {\it within} the resonance approximation (16) and the contribution
$\Delta E^{(next)}_{t}$ stemming from the corrections {\it to} the resonance approximation.

The former contribution may be estimated in a similar way to the case
considered in \cite{pre2008}: it stems, in particular, from the deviation of the GSS curve from
the separatrix (this deviation reaches $\delta$ at certain angles: see Eq.\
(7)) and from the difference between the exact resonance condition (20) and the
approximate one (21). It can be shown that the absolute value of the
ratio between $\Delta E^{(next)}_w$ and the leading term is logarithmically small (cf.\
\cite{pre2008}):

\begin{equation}
%R_r
\frac{|\Delta E^{(next)}_{w}|}{\Delta E}
\sim\frac{1}{\ln(1/h)}.
\end{equation}

Let us turn to the analysis of the contribution $\Delta E^{(next)}_{t}$, i.e.\
the contribution
stemming from the corrections to the resonance Hamiltonian (16). It is
convenient to consider separately the cases of the left and right wings of the
peak.

As described in Secs. 3.2 and 3.3 above, the left wing corresponds in the
leading-order approximation to formation of the boundary of the layer by the
{\it separatrix} of the resonance Hamiltonian (16). The resonance approximation
(16) neglects time-periodic terms while the frequencies of oscillation of these
terms greatly exceed the frequency of eigenoscillation of the resonance
Hamiltonian (16) around its relevant elliptic point i.e. the elliptic point
inside the area limited by the resonance separatrix. As is well known
\cite{gelfreich,lichtenberg_lieberman,treschev,zaslavsky:1998,zaslavsky:2005,Zaslavsky:1991},
fast-oscillating terms acting on a system with a separatrix give rise to the
onset of an {\it exponentially narrow} chaotic layer in place of the
separatrix. In the present context, this means that the correction to the
maximal action $\tilde{I}$ stemming from fast-oscillating corrections to the
resonance Hamiltonian, i.e. $\Delta E^{(next)}_{t}$, is {\it exponentially
small}, thus being negligible in comparison with the correction $\Delta
E^{(next)}_{w}$ (see (74)).

The right wing, described in Secs. 3.2 and 3.3 above, corresponds in
leading-order approximation to the formation of the boundary of the layer by
the resonance trajectory {\it tangent} to the GSS curve. For the part of the
right wing exponentially close in frequency to the frequency of the maximum,
the tangent trajectory is close to the resonance separatrix, so that the
correction stemming from fast-oscillating terms is exponentially small,
similarly to the case of the left wing. As the frequency further deviates from
the frequency of the maximum, the tangent trajectory further deviates from the
resonance separatrix and the correction $\Delta E^{(next)}_{t}$ differs from
the exponentially small correction estimated above. It may be estimated in the
following way.

It follows from the second-order approximation of the averaging method
\cite{bogmit} that the fast-oscillating terms lead, in the second-order
approximation, to the onset of additional terms
$h^2T_{\tilde{I}}(\tilde{I},\tilde{\psi})$  and
$h^2T_{\tilde{\psi}}(\tilde{I},\tilde{\psi})$ in the dynamic equations for slow
variables $\tilde{I}$ and $\tilde{\psi}$ respectively, where
$T_{\tilde{I}}(\tilde{I},\tilde{\psi})$ and
$T_{\tilde{\psi}}(\tilde{I},\tilde{\psi})$ are of the order of the
power-law-like function of $1/\ln(1/h)$ in the relevant range of $\tilde{I}$.
The corresponding correction to the width of the chaotic layer in energy may be
expressed as

\begin{equation}
\Delta E^{(next)}_{t}=\int_{t_{\min}}^{t_{\max}}{\rm
d}t\;h^2T_{\tilde{I}}\omega(\tilde{I}),
\end{equation}

\noindent where $t_{\min}$ and $t_{\max}$ are instants corresponding to the
minimum and maximum deviation of the tangent trajectory from the separatrix of
the unperturbed system (cf. Figs. 1(c) and 4(c)). The interval
$t_{\max}-t_{\min}$ may be estimated as follows:

\begin{equation}
t_{\max}-t_{\min}\sim\frac{\pi}{|<\dot{\tilde{\psi}}>|},
\end{equation}

\noindent where $<\dot{\tilde{\psi}}>$ is the value of $\dot{\tilde{\psi}}$
averaged over the tangent trajectory. It follows from (16) that

\begin{equation}
|<\dot{\tilde{\psi}}>|\sim\omega_f-\omega(E_s-\delta)
\sim\frac{\omega(E_s-\delta)}{\ln(1/h)}\sim
\frac{\omega_0}{\ln^2(1/h)}.
\end{equation}

Taking together Eqs. (75)-(77) and allowing for the fact that $T_{\tilde{I}}$
is of the order of a power-law-like function of $1/\ln(1/h)$, we conclude that

\begin{equation}
\Delta E^{(next)}_{t}\sim h^2P(\ln(1/h)),
\end{equation}

\noindent where $P(x)$ is some power-law-like function.

The value $\Delta E^{(next)}_{t}$ is still asymptotically smaller than the
absolute value of the correction within the resonance approximation, $|\Delta
E^{(next)}_{w}|$, which is of the order of $h$ or $h/\ln(1/h)$ for systems of
type I or type II respectively.

Thus, we conclude that, both for the left and right wings of the peak, (i) the
correction $\Delta E^{(next)}_{t}$ is determined by the correction within the
resonance approximation $\Delta E^{(next)}_{w}$, and (ii) in the asymptotic
limit $h\rightarrow 0$, the overall next-order correction is negligible in
comparison with the leading term:

\begin{equation}
\frac{|\Delta E^{(next)}|}{\Delta E}\equiv
\frac{|\Delta E^{(next)}_w+\Delta E^{(next)}_t|}{\Delta E}\approx
\frac{|\Delta E^{(next)}_w|}{\Delta E}
\sim\frac{1}{\ln(1/h)}
\stackrel{h\rightarrow
0}{\longrightarrow}0.
\end{equation}

\noindent This estimate well agrees with results in Figs.\ 3-6.

\subsection{Discussion}

In this section, we briefly discuss the following issues: (i) the {\it scaled}
asymptotic shape of the peaks; (ii) peaks in the range of {\it moderate}
frequencies; (iii) {\it jumps} in the amplitude dependence of the layer width;
and (iv) chaotic {\it transport}; (v) smaller peaks at {\it rational}
frequencies; (vi) other separatrix maps; (vii)
an application to the onset of {\it global chaos}.

\begin{enumerate}

\item Let us analyse the scaled asymptotic shape of the peaks. We consider
    first systems of type I. The peaks are then described in the
    leading-order approximation exclusively within separatrix map dynamics
    (approximated, in turn, by the NR dynamics). It follows from Eqs.\
    (32), (34), (36), (39) and (40) that most of the peak fir given $j$ can
    be written in the universal scaled form:

\begin{equation}
\Delta E^{(j)}(\omega_f)=\Delta
E^{(j)}_{\max}S\left(\frac{\pi(2j-1)}{(\omega_{\max}^{(j)})^2}(\omega_f-\omega_{\max}^{(j)})
\right),
\end{equation}

\noindent where the universal function $S(\alpha)$ is strongly
asymmetric:

\begin{eqnarray}
&& S(\alpha)=\left\{^{S_l(\alpha)\quad {\rm for}\quad\alpha\leq 0,
}_{S_r(\alpha)\quad {\rm for}\quad\alpha > 0,}\right.
\\
&& S_l(\alpha)=\frac{1}{{\rm e}(\ln(1+y)-y/(1+y))},
%\nonumber\\&&
\quad\quad(1+y)\ln(1+y)-y=\exp(-\alpha),\nonumber\\
&& S_r(\alpha)=\frac{1}{{\rm e}(1+\ln(1/z))},
%\nonumber\\&&
\quad\quad z(1+\ln(1/z))=\exp(-\alpha).\nonumber
\end{eqnarray}

\noindent It is not difficult to show that

\begin{eqnarray}
&& S_l(\alpha=0)=1,\quad\quad\quad\quad\quad S_r(\alpha\rightarrow
+0)={\rm e}^{-1},
\\
&&\frac{{\rm d}S_l(\alpha= 0)}{{\rm d}\alpha}=1-{\rm
e}^{-1},\quad\frac{{\rm d}S_r(\alpha \rightarrow +0)}{{\rm
d}\alpha}\rightarrow -\infty,\nonumber
\\
&& \quad\quad\quad\quad\quad S(\alpha\rightarrow \pm\infty)\propto
\frac{1}{|\alpha|}.\nonumber
\end{eqnarray}

\noindent Thus, the function $S(\alpha)$ is discontinuous at the maximum.
To the left of the maximum, it approaches the far part of the wing (which
decreases in a power-law-like way) relatively {\it slowly} while, to the
right of the maximum, the function first drops {\it jump-wise} by a factor
${\rm e}$ and then {\it sharply} approaches the far part of the wing (which
again decreases in a power-law-like way).

It follows from Eqs.\ (80), (81), (82) and (27) that the peaks are
logarithmically narrow, i.e.\ the ratio of the half-width of the peak,
$\Delta \omega^{(j)}$, to $\omega^{(j)}_{\max}$ is logarithmically small:

\begin{equation}
\frac{\Delta
\omega^{(j)}}{\omega^{(j)}_{\max}}\sim\frac{1}{\ln\left(8(2j-1)/h\right)}.
\end{equation}

We emphasize that the shape (81) is not restricted to the example
(14): it is valid for any system of type I.

For systems of type II, contributions from the NR and from the variation of
energy during the pulse of velocity, in relation to their $h$ dependence,
are formally of the same order but, numerically, the latter contribution is
usually much smaller than the former. Thus, typically, the function (81)
approximates well the properly scaled shape of the major part of the peak
for systems of type II too.

\item The quantitative theory presented in the paper relates only to the
    peaks of {\it small} order $n$ i.e.\ in the range of logarithmically
    small frequencies. At the same time, the magnitude of the peaks is
    still significant up to frequencies of order of one. This occurs
    because, for motion close to the separatrix, the order of magnitude of
    the Fourier coefficients remains the same up to logarithmically large
    numbers $n$. The shape of the peaks remains the same but their
    magnitude typically decreases (though in some cases, e.g. in case of
    the wave-like perturbation
    \cite{lichtenberg_lieberman,zaslavsky:1998,zaslavsky:2005,Zaslavsky:1991}  it may even
    increase in some range of frequencies). The quantitative description of this
    decrease, together with analyses of more sophisticated cases, requires a
    generalization of our theory.

\item Apart from the frequency dependence of the layer width, our theory is
    also relevant to amplitude dependence: it describes the jumps \cite
    {soskin2000} in the dependence of the width on $h$ and the transition
    between the jumps and the linear dependence. The values of $h$ at which
    the jumps occur, $h_{jump}^{(j)}$, are determined by the same condition
    that determines $\omega_{\max}^{(j)}$ in the frequency dependence of
    the width. The formul{\ae} relevant to the left wings of the peaks in
    the frequency dependence describe the ranges $h>h_{jump}^{(j)}$ while
    the formul{\ae} relevant to the right wings describe the ranges
    $h<h_{jump}^{(j)}$.

\item Apart from the description of the boundaries, the approach allows us
    to describe {\it chaotic transport} within the layer. In particular, it
    allows us to describe quantitatively the effect of the stickiness of
    the chaotic trajectory to boundaries between the chaotic and regular
    areas of the phase space \cite{zaslavsky:1998,zaslavsky:2005}.
    Moreover, the presence of additional (resonance) saddles should give
    rise to an additional slowing down of the transport, despite a widening
    of the area of the phase space involved in the chaotic transport.

\item Our approach can be generalized in order to describe smaller peaks at
    non-integer rational frequencies i.e.\ $\omega_f\approx n/m
    \omega_r^{(\pm)}$ where $n$ and $m$ are integer numbers.

\item Apart from Hamiltonian systems of the one and a half degrees of
    freedom and corresponding Zaslavsky separatrix maps, our approach may
    be useful in the treatment of other chaotic systems and separatrix maps
    (see \cite{treschev} for the most recent major review on various types
    of separatrix maps and related continuous chaotic systems).

\item Finally we note that, apart from systems with a separatrix, our work
    may be relevant to {\it nonlinear resonances} in any system. If the
    system is perturbed by a weak time-periodic perturbation, then
    nonlinear resonances arise and their dynamics is described by the model
    of the auxiliary time-periodically perturbed pendulum
    \cite{Chirikov:79,lichtenberg_lieberman,Zaslavsky:1991,zaslavsky:1998,zaslavsky:2005,abdullaev,gelfreich}.
    If the original perturbation has a single harmonic, then the effective
    perturbation of the auxiliary pendulum is necessarily a high-frequency
    one, and chaotic layers associated with the resonances are
    exponentially narrow
    \cite{Chirikov:79,lichtenberg_lieberman,Zaslavsky:1991,zaslavsky:1998,zaslavsky:2005,abdullaev,gelfreich}
    while our results are irrelevant. But, if either the amplitude or the
    angle of the original perturbation is slowly modulated, or if there is
    an additional harmonic of a slightly shifted frequency, then the
    effective perturbation of the auxiliary pendulum is a low-frequency one
    \cite{pre2008} and the layers become much wider\footnote{This should
    not be confused with the widening occuring with the separatrix chaotic
    layer in the {\it original} pendulum if an originally single-harmonic
    perturbation of a high frequency is completed by one more harmonic of a
    slightly shifted frequency: see \cite{vecheslavov} and references
    therein.} while our theoretical approach becomes relevant. It may allow
    to find optimal parameters of the perturbation for the facilitation of
    the onset of global chaos associated with the overlap in energy between
    different-order nonlinear resonances \cite{Chirikov:79}: the overlap
    may be expected to occur at a much smaller amplitude of perturbation in
    comparison with that one required for the overlap in case of a
    single-harmonic perturbation.

\end{enumerate}

\section{Double-separatrix chaos}

There are many problems in physics where an unperturbed Hamiltonian model
possesses two or more separatrices. A weak perturbation of the system typically
destroys the separatrices, replacing them by thin chaotic layers. As the
magnitude of the perturbation grows, the layers become wider and, at some
critical value, they merge with each other: this may be described as the onset
of {\it global chaos} between the separatrices. Such a connection of regions of
different separatrices is important for transport in the system.

In the present section, following the paper \cite{pre2008}, we consider the
characteristic problem of the onset of global chaos between two close
separatrices of a 1D Hamiltonian system perturbed by a time-periodic
perturbation. As a characteristic example of a Hamiltonian system with two or
more separatrices, we use a spatially periodic potential system with two
different-height barriers per period (Fig.\ 7(a)):

\begin{equation}
H_0(p,q)=\frac{p^2}{2}+U(q), \quad\quad U(q)=\frac{(\Phi-\sin(q))^2}{2},
%\nonumber\\&&
\quad\quad
\Phi={\rm const}<1.
\end{equation}

This model may relate e.g.\ to a pendulum spinning about its vertical axis
\cite{andronov} or to a classical 2D electron gas in a magnetic field spatially
periodic in one of the in-plane dimensions \cite{oleg98,oleg99}. Interest in
the latter system arose in the 1990s due to technological advances allowing to
manufacture magnetic superlattices of high-quality \cite{Oleg12,Oleg10}, and
thus leading to a variety of interesting behaviours of the charge carriers in
semiconductors \cite{oleg98,oleg99,Oleg12,Oleg10,Shmidt:93,shepelyansky}.

Figs.\ 7(b) and 7(c) show respectively the separatrices of the Hamiltonian
system (1) in the $p-q$ plane and the dependence of the frequency $\omega$ of
its oscillation, often called its {\it eigenfrequency}, on its energy $E\equiv
H_0(p,q)$. The separatrices correspond to energies equal to the value of the
potential barrier tops $E_b^{(1)}\equiv (1-\Phi)^2/2$ and $E_b^{(2)}\equiv
(1+\Phi)^2/2$ (Fig.\ 7(a)). The function $\omega(E)$ possesses a local maximum
$\omega_m\equiv\omega(E_m)$. Moreover, $\omega(E)$ is close to $\omega_m$ for
most of the range $[E_b^{(1)},E_b^{(2)}]$ while sharply decreasing to zero as
$E$ approaches either $E_b^{(1)}$ or $E_b^{(2)}$.

\begin{figure}
[b]
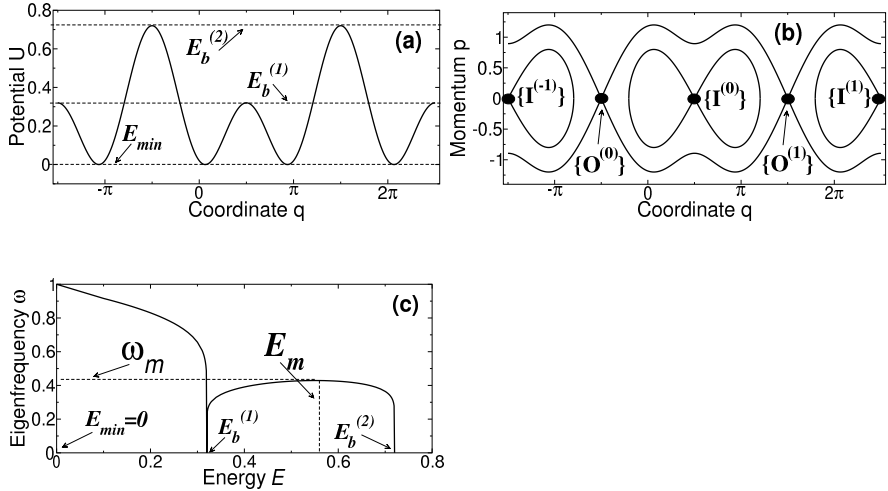

\includegraphics[width = 5.8 cm]{soskin_Fig7a.eps}
%\hskip 0.9cm
\includegraphics[width = 5.8 cm]{soskin_Fig7b.eps}
\vskip 0.7cm
\includegraphics[width = 5.8 cm]{soskin_Fig7c.eps}
\caption {The potential $U(q)$, the separatrices in the phase space, and the
eigenfrequency $\omega(E)$ for the unperturbed system (84) with $\Phi=0.2$, in
(a), (b) and (c) respectively.}
\end{figure}

We now consider the addition of a time-periodic perturbation: as an example, we
use an AC drive, which corresponds to a dipole \cite{Zaslavsky:1991,Landau:76}
perturbation of the Hamiltonian:
\begin{eqnarray}
&&
\dot{q} = \partial H/\partial p, \quad\quad \dot{p} = -\partial
H/\partial q,
%\dot{q} = \partial H/\partial p, \quad\quad \dot{p} = -\partial H/\partial q,
\\
&&
H(p,q)=H_0(p,q)- h q\cos (\omega_f t).
\nonumber
\end{eqnarray}

The {\it conventional} scenario for the onset of global chaos between the
separatrices of the system (84)-(85) is illustrated by Fig.\ 8. The figure
presents the evolution of the stroboscopic Poincar\'{e} section as $h$ grows
while $\omega_f$ is fixed at an arbitrarily chosen value {\it away} from
$\omega_m$ and its harmonics. At small $h$, there are two thin chaotic layers
around the inner and outer separatrices of the unperturbed system. Unbounded
chaotic transport takes place only in the outer chaotic layer i.e.\ in a {\it
narrow} energy range. As $h$ grows, so also do the layers. At some critical
value $h_{gc} \equiv h_{gc} (\omega_f)$, the layers merge. This may be
considered as the onset of global chaos: the whole range of energies between
the barrier levels is involved, with unbounded chaotic transport. The states
$\{I^{(l)} \}\equiv \{p=0,q=\pi/2 +2\pi l \}$ and $\{O^{(l)}
\}\equiv\{p=0,q=-\pi/2 +2\pi l\}$ (where $l$ is any integer) in the
Poincar\'{e} section are associated respectively with the inner and outer
saddles of the unperturbed system, and necessarily belong to the inner and
outer chaotic layers, respectively. Thus, the necessary and sufficient
condition for global chaos onset may be formulated as the possibility for the
system placed initially in the state $\{I^{(0)} \} $ to pass beyond the
neighbourhood of the \lq\lq outer'' states, $\{O^{(0)} \} $ or $\{O^{(1)} \} $,
i.e.\ for the coordinate $q$ to become $<-\pi/2$ or $>3\pi/2$ at sufficiently large
times $t\gg 2\pi/\omega_f$.

\begin{figure}
%[tb]
[t]
\sidecaption[t]
\includegraphics[width = 7 cm]{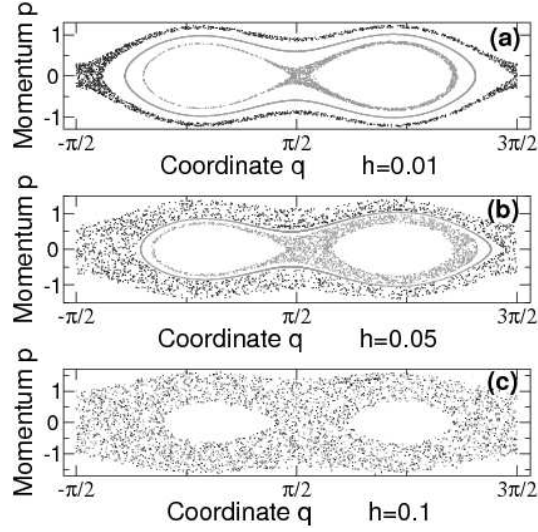}
\caption{The evolution of the stroboscopic (at $t=n2\pi/\omega_f$ with
$n=0,1,2,...$) Poincar\'{e} section of the system (84)-(85) with $\Phi=0.2$ as
$h$ grows while $\omega_f=0.3$. The number of points in each trajectory is
2000. In (a) and (b), three characteristic trajectories are shown: the inner
trajectory starts from the state $\{I^{(0)} \} \equiv \{p=0,q=\pi/2 \}$ and is
chaotic but bounded in space; the outer trajectory starts from $\{O^{(0)} \}
\equiv\{p=0,q=-\pi/2 \}$ and is chaotic and unbounded in coordinate; the third
trajectory is an example of a regular trajectory separating the two chaotic
ones. In (c), the chaotic trajectories mix.}
\end{figure}

A diagram in the $h-\omega_f$ plane,  based on the above criterion, is shown in
Fig.\ 9. The lower boundary of the shaded area represents the function $h_{gc}
(\omega_f)$. It has deep {\it spikes} i.e.\ cusp-like local minima. The most
pronounced spikes are situated at frequencies $\omega_f=\omega_s^{(j)}$ that
are slightly less than the odd multiples of $\omega_m$,

\begin{equation}
\omega_s^{(j)} \approx\omega_m(2j-1), \quad\quad j=1,2,...
\end{equation}

\noindent The deepest minimum occurs at $\omega_s^{(1)}\approx\omega_m$: the
value of $h_{gc}$ at the minimum, $h_s^{(1)}\equiv h_{gc} (\omega_ s^{(1)})$,
is approximately 40 times smaller than the value in the neighbouring pronounced
local maximum of $h_{gc} (\omega_f)$ at $\omega_f\approx 1$. As $n$ increases,
the $n$th minimum becomes shallower. The function $h_{gc} (\omega_f)$ is very
sensitive to $\omega_f$ in the vicinity of the minima: for example, a reduction
of $\omega_f$ from $\omega_s^{(1)}\approx 0.4$ of only 1\% causes an increase
in $h_{gc}$ of $\approx 30\%$.

\begin{figure}
%[tb]
[t]
\sidecaption[t]
\includegraphics[width = 7. cm]{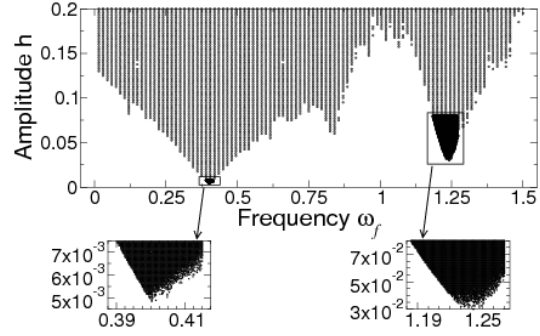}
\caption{Diagram indicating the range of perturbation parameters (shaded) for
which global chaos exists. The integration time for each point of the grid is
$12000\pi$. }
\end{figure}

The origin of the spikes is related to the involvement of the resonance
dynamics in separatrix chaos, similar to that considered in Sec.\ 3. In
particular, the minima of the spikes correspond to the situation when the
resonances almost touch, or slightly overlap with, the separatrices of the
unperturbed system while overlapping each other. This is illustrated by the
evolution of the Poincar\'{e} section as $h$ grows while $\omega_f\approx
\omega_s^{(1)}$ (Fig.\ 10) and by its comparison with the corresponding
evolution of resonance separatrices calculated in the resonance approximation
(Fig.\ 11).

\begin{figure}
%[tb]
[t]
\sidecaption[t]
\includegraphics[width = 7. cm]{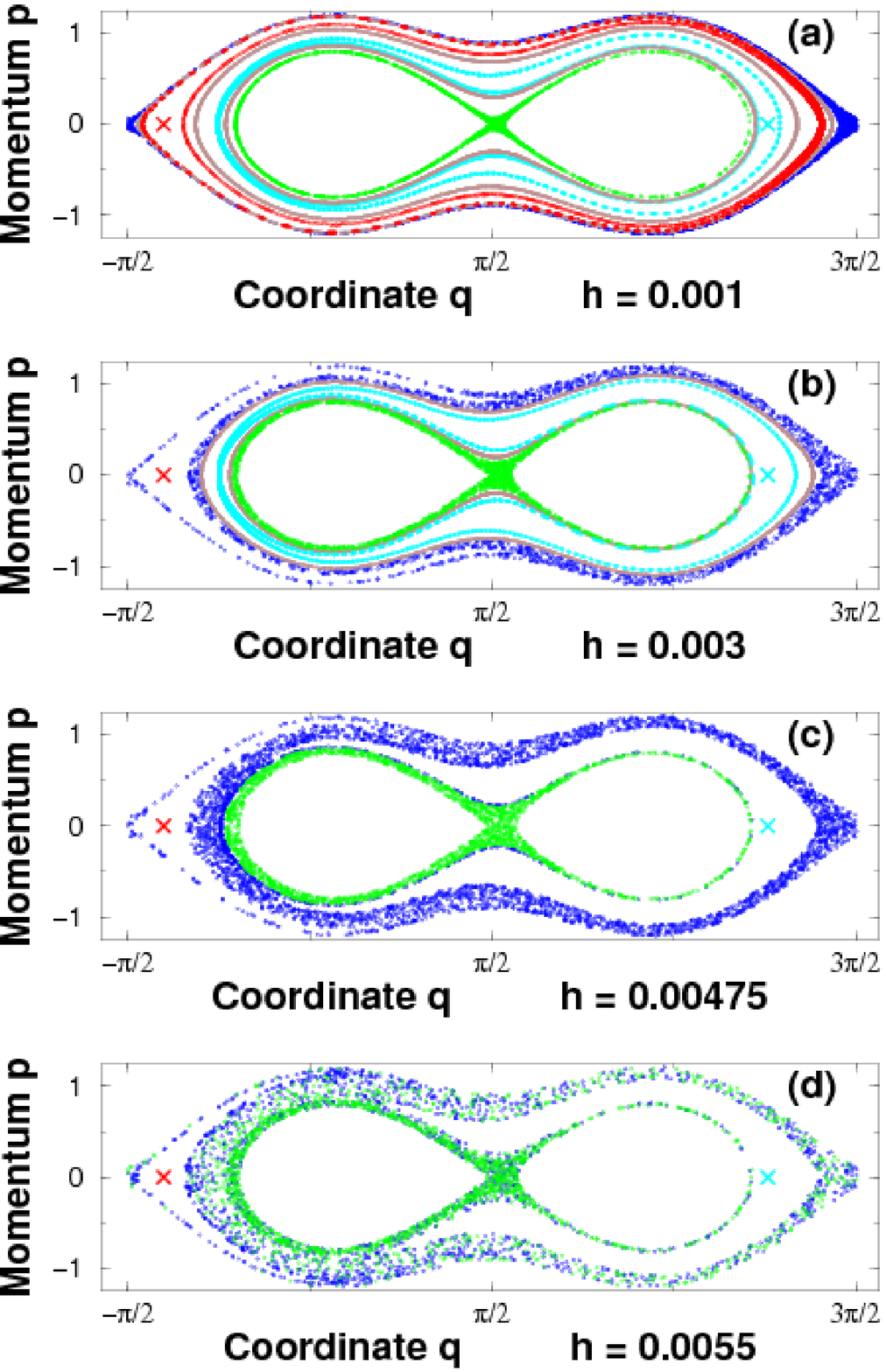}
\caption{(Color version may be found in the online version of \cite{pre2008} as
Fig.\ 5). The evolution of the stroboscopic Poincar\'{e} section of the system
(84)-(85) with $\Phi=0.2$, as the amplitude $h$ of the perturbation grows,
while the frequency remains fixed at $\omega_f=0.401$. The number of points in
each trajectory is 2000. The chaotic trajectories starting from the states
$\{I^{(0)} \} $ and $\{O^{(0)} \} $ are drawn in green and blue respectively.
The stable stationary points of Eq.\ (98) for $n=1$ (i.e.\ for the 1st-order
nonlinear resonances) are indicated by the red and cyan crosses. The chaotic
layers associated with the resonances are indicated in red and cyan
respectively, unless they merge with those associated with the green/blue
chaotic trajectories. Examples of regular trajectories embracing the state
$\{I^{(0)} \} $ while separating various chaotic trajectories are shown in
brown.}
\end{figure}

\begin{figure}
%[tb]
[b]
\sidecaption[b]
\includegraphics[width = 4. cm]{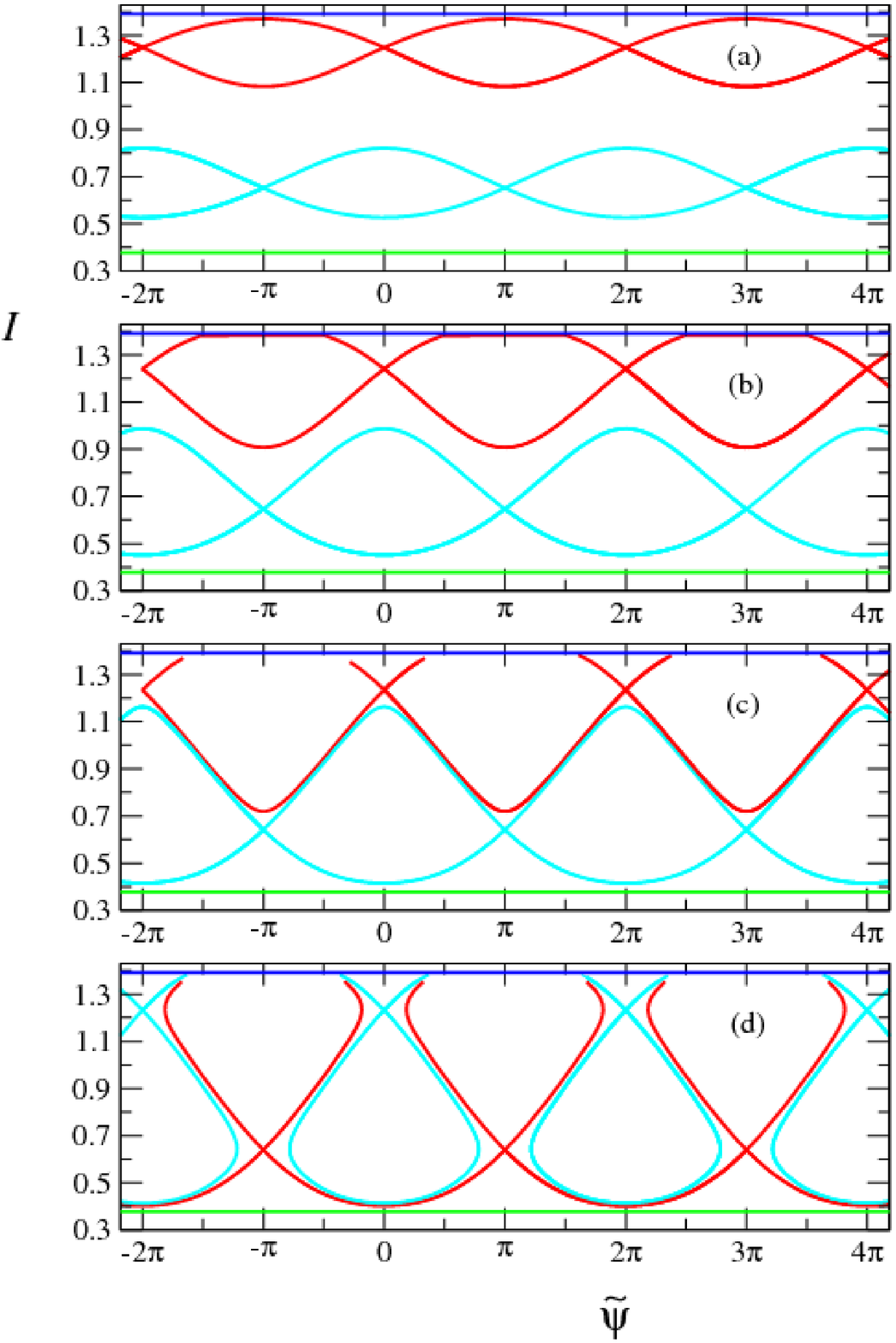}
\caption{(Color version may be found in the online version of \cite{pre2008} as
Fig.\ 6). The evolution of the separatrices of the 1st-order resonances within
the resonance approximation (described by (16) with $n=1$) in the plane of
action $I$ and slow angle $\tilde{\psi}$, for the same parameters as in Fig.\
10 (boxes (a), (b), (c), (d) correspond to those in Fig. 10). Horizontal levels mark the values of $I$ corresponding to the barriers.}
\end{figure}

Sec.\ 4.1 below presents the self-consistent asymptotic theory of the minima of
the spikes, based on an accurate analysis of the overlap of resonances with
each other and on the matching between the separatrix map and the resonance
Hamiltoinian (details of the matching are developed in Appendix). Sec.\ 4.2
presents the theory of the wings of the spikes Generalizations and applications
are discussed in Sec.\ 4.3.

\subsection{Asymptotic Theory For The Minima Of The Spikes}

The eigenfrequency $\omega(E)$ stays close to its local maximum $\omega_m$ for
most of the relevant range $[E_b^{(1)},E_b^{(2)}]$ (Fig.\ 7(c)). As shown
below, $\omega(E)$ approaches a {\it rectangular} form in the asymptotic limit
$\Phi \rightarrow 0$. Hence, if the perturbation frequency $\omega_f $ is close
to $\omega_m$ or its odd multiples, $|\omega_f - (2j-1) \omega_m | \ll \omega_m
$, then the energy widths of nonlinear resonances become comparable to the
width of the whole range between the barriers (i.e.\ $ E_b^{(2)}-E_b^{(1)}
\approx 2\Phi $) at a rather small perturbation magnitude $ h \ll \Phi $. Note
that $\Phi $ determines the characteristic magnitude of the perturbation
required for the conventional overlap of the separatrix chaotic layers, when
$\omega_f$ is not close to any odd multiple of $\omega_m $ (Fig.\ 8 (c)). Thus,
if $ \omega_f\approx \omega_s^{(j)} $, the nonlinear resonances should play a
crucial role in the onset of global chaos (cf.\ Fig.\ 10).

We note that it is not entirely obvious {\it a priori} whether it is indeed
possible to calculate $h_s^{(j)}\equiv h_{gc}(\omega_s^{(j)})$ within the
resonance approximation: in fact, it is essential for the separatrices of the
nonlinear resonances to nearly touch the barrier levels, but the resonance
approximation is invalid in the close vicinity of the barriers;
furthermore, numerical calculations of resonances show that, if
$\omega_f\approx\omega_s^{(j)}$, the perturbation amplitude $h$ at which the
resonance separatrix touches a given energy level in the close vicinity of the
barriers is very sensitive to $\omega_f$, apparently making the calculation of
$h_s^{(j)}$ within the resonance approximation even more difficult.

Nevertheless, we show below in a self-consistent manner that, in the asymptotic
limit $\Phi\rightarrow 0$, the relevant boundaries of the chaotic layers lie in
the range of energies $E$ where $\omega(E)\approx\omega_m$. Therefore, the
resonant approximation is valid and it allows us to obtain {\it explicit}
asymptotic expressions both for $\omega_s^{(j)} $ and $h_s^{(j)}$, and for the
wings of the spikes in the vicinities of $\omega_s^{(j)}$.

The {\it asymptotic} limit $\Phi\rightarrow 0$ is the most interesting one from
a theoretical point of view because it leads to the strongest facilitation of
the onset of global chaos, and it is most accurately described by the
self-contained theory. Most of the theory presented below assumes this limit
and concentrates therefore on the results to the {\it lowest} (i.e.\ leading)
order in the small parameter.

On the applications side, the range of {\it moderately small} $\Phi$  is more
interesting, since the chaos facilitation is still pronounced (and still
described by the asymptotic theory) while the area of chaos between the
separatrices is not too small (comparable with the area inside the inner
separatrix): cf.\ Figs.\ 7, 8 and 10. To increase the accuracy of the
theoretical description in this range, we estimate the next-order corrections
and develop an efficient numerical procedure allowing for further corrections.

\subsubsection{Resonant Hamiltonian and related quantities}

Let $\omega_f$ be close to the $n$th odd\footnote {Even harmonics are absent in
the eigenoscillation due to the symmetry of the potential.} harmonic of
$\omega_m$, $n \equiv (2j-1)$. Over most of the range $[E_b^{(1)},E_b^{(2)}]$,
except in the close vicinities of $E_b^{(1)}$ and $E_b^{(2)}$, the $n$th
harmonic of the eigenoscillation is nearly resonant with the perturbation. Due
to this, the (slow) dynamics of the action $I\equiv I(E) =(2\pi)^{-1}\oint dqp$
and the angle $\psi$
\cite{Chirikov:79,lichtenberg_lieberman,Zaslavsky:1991,zaslavsky:1998,zaslavsky:2005,PR,Landau:76}
can be described by means of a resonance Hamiltonian similar in form to (16).
The lower integration limit in the expression for $\tilde {H}$ may be chosen
arbitrarily, and it will be convenient for us to use presently $I(E_m)$
(instead of $I(E_s)$ in (16)) where $E_m$ is the energy of the local maximum of
$\omega(E)$ (Fig.\ 7(c)). To avoid confusion, we write the resonance
Hamiltonian explicitly below after making this change:

\begin{eqnarray}
&& \tilde{H}(I,\tilde{\psi})=\int_{I(E_m)}^{I}{\rm d}\tilde{I}\;
(n\omega-\omega_f)\;-\; nhq_n\cos(\tilde{\psi})
\\
&& \quad\; \equiv\; n(E-E_m)-\omega_f(I-I(E_m))\;-\;
nhq_n\cos(\tilde{\psi})\;, \nonumber
\\
&& I \equiv I(E) = \int_{E_{\rm min}}^E
\frac{{\rm d}\tilde{E}}{\omega(\tilde{E})}, \quad\quad  E \equiv H_0(p,q),
\nonumber
\\
&& \tilde{\psi}=n\psi-\omega_ft, \quad\quad \nonumber
\\
&& \psi= \pi+{\rm sign}(p)\omega(E)\int^q_{q_{\rm min}(E)}\frac{{\rm
d}\tilde{q}}{\sqrt{2(E-U(\tilde{q}))}}+2\pi l, \nonumber
\\
&& q_n\equiv q_n(E)= \frac{2}{\pi}\int_0^{\pi/2} \!\!\!\!\! {\rm
d}\psi \; q(E,\psi)\cos(n\psi) ,
%\quad i\equiv \sqrt{-1},
\nonumber
\\
&& |n\omega-\omega_f|\ll\omega,\quad\quad n\equiv 2j-1, \quad\quad
j=1,2,3,\ldots \nonumber
\end{eqnarray}

Let us derive explicit expressions for various quantities in (87). In the
unperturbed case ($h=0$), the equations of motion (85) with $H_0$ (84) can be
integrated \cite{oleg99} (see also Eq.\ (144) below), so that we can find
$\omega(E)$:

\begin{equation}
\omega(E)=\frac{\pi(2E)^{1/4}}{2K
\left[k \right]},\quad\quad
k=\frac{1}{2}\sqrt{\frac{(\sqrt{2E}+1)^2-\Phi^2}{\sqrt{2E}}} \, ,
\end{equation}

\noindent
where

%\begin{equation}
$$
K[k]=\int_0^{\frac{\pi}{2}}\frac{{\rm d}\phi}{\sqrt{1-k^2\sin^2(\phi)}},
$$
%\end{equation}

\noindent is the complete elliptic integral of first order
\cite{Abramovitz_Stegun}. Using its asymptotic expression,

\[
K[k\rightarrow 1] \simeq \frac{1}{2} \ln\left( { 16 \over 1-k^2 }
\right),
\]

\noindent we derive $\omega(E)$ in the asymptotic limit $\Phi
\rightarrow 0$:

\begin{eqnarray}
&& \omega(E) \simeq \frac{\pi}{\ln
\left(
\frac{64}{(\Phi-\Delta E)(\Phi+\Delta E)}
\right)},
\\
&&
\Delta E\equiv E-\frac{1}{2},\quad\quad |\Delta E|<\Phi,
\quad\quad
\Phi \rightarrow 0.
\nonumber
\end{eqnarray}

As mentioned above,the function $\omega(E)$ (89) remains close to its maximum

\begin{equation}
\omega_m\equiv \max_{[E_b^{(1)},E_b^{(2)}]}\{\omega(E)\}
       \simeq \frac{\pi}{2\ln (8/\Phi)} \,
\end{equation}

\noindent for most of the interbarrier range of energies
$[1/2-\Phi,1/2+\Phi]$ (note that $E_{b}^{(1,2)}\approx
1/2\mp \Phi$ to first order in $\Phi$.); on the other hand, in the close vicinity of the
barriers, where either $|\ln(1/(1-\Delta E/\Phi))|$ or $|\ln(1/(1+\Delta
E/\Phi))|$ become comparable with, or larger than, $\ln(8/\Phi)$, $\omega(E)$
decreases rapidly to zero as $|\Delta E|\rightarrow \Phi$. The range where this
takes place is $\sim\Phi^2$, and its ratio to the whole interbarrier range,
$2\Phi$, is $\sim\Phi$ i.e.\ it goes to zero in the asymptotic limit
$\Phi\rightarrow 0$: in other words, $\omega(E)$ approaches a {\it rectangular}
form. As it will be clear from the following, {\it it is this almost
rectangular form of $\omega(E)$ which determines many of the characteristic
features of the global chaos onset in systems with two or more separatrices}.

One more quantity which strongly affects $(\omega_s, h_s)$ is the Fourier
harmonic $q_n\equiv q_n(E)$. The system stays most of the time very close to
one of the barriers. Consider the motion within one of the periods of the
potential $U(q)$, between neighboring upper barriers
$[q_{ub}^{(1)},q_{ub}^{(2)}]$ where $q_{ub}^{(2)}\equiv q_{ub}^{(1)}+2\pi$. If
the energy $E\equiv 1/2+\Delta E$ lies in the relevant range
$[E_b^{(1)},E_b^{(2)}]$, then the system will stay close to the lower barrier
$q_{lb}\equiv q_{ub}^{(1)}+\pi$ for a time\footnote {We omit corrections
$\sim(\ln(1/\Phi))^{-1}$ here and in Eq.\ (92) since they vanish in the
asymptotic limit $\Phi\rightarrow 0$.}

\begin{equation}
T_l\approx 2\ln \left( \frac{1}{\Phi+\Delta E} \right)
\end{equation}

\noindent during each period of eigenoscillation, while it will stay close to
one of the upper barriers $q_{ub}^{(1,2)}\equiv q_{lb}\pm \pi$ for most of the
remainder of the eigenoscillation,

\begin{equation}
T_u\approx 2\ln \left( \frac{1}{\Phi-\Delta E} \right)\quad .
\end{equation}

\noindent Hence, the function $q(E,\psi)-q_{lb}$ may be
approximated by the following piecewise even periodic function:

\begin{eqnarray}
&& q(E,\psi)-q_{lb}= \left\{_{0\quad {\rm at} \quad \psi\in
\left.\right]\frac{\pi}{2}\frac{T_u}{T_l+T_u},\pi-\frac{\pi}{2}\frac{T_u}{T_l+T_u}
\left[\right.,}^{\pi\quad {\rm at} \quad \psi\in
\left[0,\frac{\pi}{2}\frac{T_u}{T_l+T_u} \right]\cup
\left[\pi-\frac{\pi}{2}\frac{T_u}{T_l+T_u},\pi \right], } \right.
\\
&& q(E,-\psi)-q_{lb}=q(E,\psi)-q_{lb}, \quad\quad q(E,\psi\pm 2\pi i)=q(E,\psi),
\quad\quad i=1,2,3,... \nonumber
\end{eqnarray}

\noindent
Substituting the above approximation for $q(E,\psi)$ into the
definition of $q_n$ (87), one can obtain:

\begin{eqnarray}
&& q_{2j-1}\equiv q_{2j-1}(E)=\frac{2}{2j-1}\sin \left(
\frac{(2j-1)\pi/2}{1+\frac{\ln \left( \frac{1}{\Phi+\Delta E}
\right) }{\ln \left( \frac{1}{\Phi-\Delta E} \right)}} \right) \;,\\
&& \Phi\rightarrow 0, \quad\quad
q_{2j}=0,\quad\quad
j=1,2,3,...
\nonumber
\end{eqnarray}

At barrier energies, $q_{2j-1}$ takes the values

\begin{equation}
q_{2j-1}(E_b^{(1)})=0, \quad\quad q_{2j-1}(E_b^{(2)}) = -(-1)^{j}{ 2 \over (2j-1) } \, .
\end{equation}

As $E$ varies in between its values at the barriers, $q_{2j-1}$ varies monotonically if
$j=1$ and non-monotonically otherwise (cf.\ Fig.\ 16). But in any case, the
significant variations occur mostly in the close vicinity of the barrier
energies $E_b^{(1)}$ and $E_b^{(2)}$ while, for most of the range
$[E_b^{(1)},E_b^{(2)}]$, the argument of the sine in Eq.\ (94) is close to $\,
\pi/4$ and $\, q_{2j-1}$ is then almost constant:

\begin{eqnarray}
&& q_{2j-1}\approx
(-1)^{\left[\frac{2j-1}{4}\right]}\frac{\sqrt{2}}{2j-1}, \quad \
j= 1, 2, 3, \, \ldots,
\\
&& \left| \ln \left( \frac{1+\Delta E/\Phi}{1-\Delta E/\Phi}
\right) \right| \ll 2\ln \left( \frac{1}{\Phi} \right), \nonumber
\end{eqnarray}

\noindent where $[\ldots]$ means the integer part.

In the asymptotic limit $\Phi\rightarrow 0$, the range of $\Delta E$ for which
the approximate equality (96) for $q_{2j-1}$ is valid approaches the whole
range $]-\Phi,\Phi[$.

We emphasize that $|q_n|$ determines the \lq\lq strength'' of the nonlinear
resonances: therefore, apart from the nearly rectangular form of $\omega(E)$,
the non-smallness of $|q_n|$ is an important additional factor strongly
facilitating the onset of global chaos.

We shall need also an asymptotic expression for the action $I$. Substituting
$\omega(E)$ (89) into the definition of $I(E)$ (87) and carrying out the
integration, we obtain

\begin{equation}
I(E)=I(1/2)+ \frac {\Delta E \ln \left( \frac{64{\rm
e}^2}{\Phi^2-(\Delta E)^2} \right)  + \Phi \ln \left(
\frac{\Phi-\Delta E}{\Phi+\Delta E} \right)}{\pi}
  \; , \quad\quad
\Phi\rightarrow 0.
\end{equation}

\subsubsection{Reconnection of resonance separatrices}

We now turn to analysis of the {\it phase space} of the resonance Hamiltonian
(87). The evolution of the Poincar\'{e} section (Fig.\ 10) suggests that we
need to find a {\it separatrix} of (87) that undergoes the following evolution
as $h$ grows: for sufficiently small $h$, the separatrix does not overlap
chaotic layers associated with the barriers while, for $h>h_{gc}(\omega_f)$, it
does overlap them. The relevance of such a condition will be further justified.

Consider $\omega_f\approx n\omega_m$ with a given odd $n$. For the sake of
convenience, let us write down the equations of motion (87) explicitly:

\begin{equation}
\dot{I}=-\frac{\partial \tilde{H}}{\partial \tilde{\psi}}\equiv
-nhq_n\sin(\tilde{\psi}),
\quad\quad
\dot{\tilde{\psi}}=\frac{\partial \tilde{H}}{\partial I}\equiv
n\omega - \omega_f-nh\frac{{\rm d}q_n}{{\rm
d}I}\cos(\tilde{\psi}).
\end{equation}

\noindent Any separatrix necessarily includes one or more unstable stationary
points. The system of dynamic equations (98) may have several stationary points
per $2\pi$ interval of $\tilde{\psi}$. Let us first exclude those points which
are irrelevant to a separatrix undergoing the evolution described above.

Given that $q_n(E_b^{(1)})=0$, there are two unstable stationary points with
$I$ corresponding to $E=E_b^{(1)}$ and $\tilde{\psi}=\pm\pi/2$. They are
irrelevant because, even for an infinitely small $h$, each of them necessarily
lies inside the corresponding barrier chaotic layer.

If $E\neq E_b^{(1)}$, then $q_n\neq 0$, so $\dot{I}=0$ only if
$\tilde{\psi}$ is equal either to 0 or to $\pi$. Substituting these
values into the second equation of (98) and putting
$\dot{\tilde{\psi}}=0$, we obtain the equations for the
corresponding actions:

\begin{equation}
 X_{\mp}(I) \equiv  n\omega-\omega_f\mp nh{\rm d}q_n/{\rm d}I =0,
\end{equation}

\noindent where the signs \lq\lq -" and \lq\lq +" correspond to
$\tilde{\psi}=0$ and $\tilde{\psi}=\pi$ respectively. A typical example of the
graphical solution of equations (99) for $n=1$ is shown in Fig.\ 12. Two of the
roots corresponding to $\tilde{\psi}=\pi$ are very close to the barrier values
of $I$ (recall that the relevant values of $h$ are small). These roots arise
due to the divergence of ${\rm d} q/{\rm d} I$ as $I$ approaches any of the
barrier values. The lower/upper root corresponds to a stable/unstable point,
respectively. However, for any $n$, both these points and the separatrix
generated by the unstable point necessarily lie in the ranges covered by the
barrier chaotic layers. Therefore, they are also irrelevant\footnote{For
sufficiently small $\Phi$ and $h$, the separatrix generated by the unstable
point forms the boundary of the upper chaotic layer, but this affects only the
higher-order terms in the expressions for the spikes minima (see below).}. For
$n>1$, the number of roots of (99) in the vicinity of the barriers may be
larger (due to oscillations of the modulus and sign of ${\rm d}q_n/{\rm d}I$ in
the vicinity of the barriers) but they all are irrelevant for the same reason,
at least to leading-order terms in the expressions for the spikes' minima.

\begin{figure}[t]
\sidecaption[t]
\includegraphics[width = 7. cm]{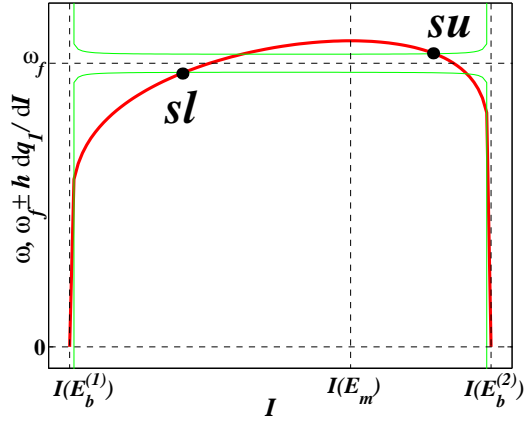}
\caption{(Color version may be found in the online version of \cite{pre2008} as
Fig.\ 7).  A schematic example illustrating the graphical solutions of Eqs.\
(99) for $n=1$, as intersections of the curve $\omega(I)$ (thick solid red
line) with the curves $\omega_f\pm h{\rm d}q_n(I)/{\rm d}I$ (thin solid green
lines). The solutions corresponding to the lower and upper relevant saddles
(defined by Eq.\ (100)) are marked by dots and by the labels $sl$ and $su$
respectively (we do not mark other solutions because they are irrelevant).}
\end{figure}

Consider the stationary points corresponding to the remaining four roots of
equations (99). Just these points are conventionally associated with {\it
nonlinear resonances}
\cite{Chirikov:79,lichtenberg_lieberman,Zaslavsky:1991,zaslavsky:1998,zaslavsky:2005,PR}.
It follows from the analysis of equations (98) linearized near the stationary
points (cf.\
\cite{Chirikov:79,lichtenberg_lieberman,Zaslavsky:1991,zaslavsky:1998,zaslavsky:2005,PR}),
two of them are stable (elliptic) points\footnote{In the Poincar\'{e} sections
shown in Fig.\ 10, the points which correspond to such stable points of
equations (98) are indicated by the crosses.}, while two others are unstable
(hyperbolic) points, often called {\it saddles}. These saddles are of central
interest in the context of our work. They belong to the {\it separatrices}
dividing the $I-\tilde{\psi}$ plane for regions with topologically different
trajectories.

We shall identify the relevant saddles as those with the {\it lower}
action/energy (using the subscript \lq\lq$sl$'') and {\it upper} action/energy
(using the subscript \lq\lq$su$''). The positions of the saddles in the
$I-\tilde{\psi}$ plane are defined by the following equations (cf.\ Figs.\ 11
and 12):

\begin{eqnarray}
&& g\equiv {\rm
sgn}(q_n(I_{su,sl}))={\rm
sgn}\left(
(-1)^{\left[\frac{n}{4}\right]} \right),
\\
&& \tilde{\psi}_{sl}=\pi(1+g)/2, \quad\quad
\tilde{\psi}_{su}=\pi(1-g)/2,\nonumber
\\
&& X_g(I_{sl})=X_{-g}(I_{su})=0, \quad\quad\frac {{\rm d}X_g(I_{sl})}{{\rm d}I_{sl}}>0, \quad\quad \frac
{{\rm d}X_{-g}(I_{su})}{{\rm d}I_{su}}<0, \nonumber
\end{eqnarray}

\noindent where $[...]$ means an integer part, $X_{\pm}(I)$ are defined in Eq.\ (99) while $I_{sl}$ and
$I_{su}$ are closer to $I(E_m)$ than any other solution of (100) (if any) from
below and from above, respectively.

Given that the values of $h$ relevant to the minima of the spikes
asymptotically approach 0 in the asymptotic limit $\Phi\rightarrow 0$, one may
neglect the last term in the definition of $X_{\mp}$ in Eq.\ (99) in the
lowest-order approximation\footnote {As will become clear in what follows, the
remaining terms are much larger in the asymptotic limit than the neglected
term: cf.\ the standard theory of the nonlinear resonance \cite
{Chirikov:79,lichtenberg_lieberman,zaslavsky:1998,zaslavsky:2005,Zaslavsky:1991}.},
so that the equations $X_{\mp}=0$ reduce to the simplified resonance condition

\begin{equation}
n\omega(I_{su,sl})= \omega_f.
\end{equation}

\noindent Substituting here Eq.\ (89) for $\omega$, we obtain explicit
expressions for the energies in the saddles:

\begin{eqnarray}
&&
E_{su,sl}\approx \frac{1}{2}\pm \Delta E^{(1)},
\\
&& \Delta E^{(1)}\equiv
\sqrt{\Phi^2-64\exp\left(-\frac{n\pi}{\omega_f}\right)}, \quad\quad
\omega_f\leq n\omega_m. \nonumber
\end{eqnarray}

\noindent The corresponding actions $I_{su,sl}$ are expressed via $E_{su,sl}$
by means of Eq.\ (97).

For $\omega_f \approx n \omega_m $, the values of $E_{su,sl}$ (102) lie in the
range where the expression (96) for $q_n$ holds true. This will be confirmed by
the results of calculations based on this assumption.

Using (100) for the angles and (102) for the energies, and the asymptotic
expressions (89), (96) and (97) for $\omega(E)$, $q_n(E)$ and $I(E)$
respectively, and allowing for the resonance condition (101), we obtain
explicit expressions for the values of the Hamiltonian (87) at the saddles:

\begin{equation}
\tilde{H}_{sl}=-\tilde{H}_{su}=\frac{\omega_f}{\pi}
\left[
2\Delta E^{(1)}-\Phi\ln
\left(\frac{\Phi+\Delta E^{(1)}}{\Phi-\Delta E^{(1)}}
\right)
\right]
+h\sqrt{2}.
\end{equation}

As the analysis of simulations suggests and as it is self-consistently shown
further, one of the main conditions which should be satisfied in the spikes is
the overlap in phase space between the separatrices of the nonlinear
resonances, which is known as {\it separatrix reconnection}
\cite{PR,Howard:84,Howard:95,Diego,James,Albert}. Given that the Hamiltonian
$\tilde{H}$ is constant along any trajectory of the system (87), the values of
$\tilde{H}$ in the lower and upper saddles of the {\it reconnected}
separatrices are equal to each other:

\begin{equation}
\tilde{H}_{sl}=\tilde{H}_{su} \,.
\end{equation}

\noindent This may be considered as the necessary and sufficient\footnote{ Eq.\
(104) is the {\it sufficient} (rather than just necessary) condition for
separatrix reconnection since there is no any other separatrix which would lie
in between the separatrices generated by the saddles \lq\lq {\it sl}'' and
\lq\lq {\it su}''.} condition for the reconnection. Taking into account that
$\tilde{H}_{sl}=-\tilde{H}_{su}$ (see (103)), it follows from (104) that

\begin{equation}
\tilde{H}_{sl}=\tilde{H}_{su}=0.
\end{equation}

Explicitly, the relations in (105) reduce to

\begin{eqnarray}
&& h\equiv h(\omega_f)=\frac{\omega_f}{\sqrt{2}\pi} \left[\Phi\ln
\left(\frac{\Phi+\Delta E^{(1)}}{\Phi-\Delta E^{(1)}} \right)
-2\Delta E^{(1)} \right],
\\
&& \Delta E^{(1)}\equiv
\sqrt{\Phi^2-64\exp(-\frac{n\pi}{\omega_f})},
\quad\quad
0<\omega_m-\omega_f/n\ll \omega_m\equiv \frac{\pi}{2\ln(8/\Phi)},
\nonumber
\\
&&
n=1,3,5,...
\nonumber
\end{eqnarray}

\noindent The function $h(\omega_f)$ (106) decreases monotonically to zero as
$\omega_f$ grows from $0$ to $n\omega_m$, where the line abruptly stops. Fig.\
15 shows the portions of the lines (106) relevant to the left wings of the 1st
and 2nd spikes (for $\Phi=0.2$).

\subsubsection {Barrier chaotic layers}

The next step is to find the minimum value of $h$ for which the resonance
separatrix overlaps the chaotic layer related to a potential barrier. With this
aim, we study how the relevant outer boundary of the chaotic layer behaves as
$h$ and $\omega_f$ vary. Assume that the relevant $\omega_f$ is close to
$n\omega_m$ while the relevant $h$ is sufficiently large for $\omega(E)$ to be
close to $\omega_m$ at all points of the outer boundary of the layer (the
results will confirm these assumptions). Then the motion along the regular
trajectory infinitesimally close to the layer boundary may be described within
the resonance approximation (87). Hence the boundary may also be described as a
trajectory of the resonant Hamiltonian (87). This is explicitly proved in the
Appendix, using a separatrix map analysis allowing for the validity of the
relation $\omega(E)\approx\omega_m$ for all $E$ relevant to the boundary of the
chaotic layer. The main results are presented below. For the sake of clarity,
we present them for each layer separately, although they are similar in
practice.

\paragraph{4.1.3.1 Lower Layer}

Let $\omega_f$ be close to any of the spikes' minima.

One of the key roles in the formation of the upper boundary of the layer is
played by the angle-dependent quantity $\delta_l|\sin( \tilde{\psi})|$ which we
call the {\it generalized separatrix split} (GSS) for the lower layer, alluding
to the conventional {\it separatrix split} \cite{zaslavsky:1998} for the lower
layer $\delta_l \equiv |\epsilon^{(low)}( \omega_f)|h$ with $\epsilon^{(low)}$
given by Eq.\ (172)\footnote{The quantity $\delta_l$ may also be interpreted as
the magnitude of the corresponding Melnikov integral
\cite{Chirikov:79,lichtenberg_lieberman,Zaslavsky:1991,zaslavsky:1998,zaslavsky:2005},
sometimes called as the Poincar\'{e}-Melnikov integral \cite{treschev}.} (cf.\
also (4)). Accordingly, we use the term \lq\lq lower GSS curve'' for the
following curve in the $I-\tilde{\psi}$ plane:

\begin{equation}
I=I_{\rm GSS}^{(l)}(\tilde{\psi})\equiv
I(E_b^{(1)}+\delta_l|\sin(\tilde{\psi})|).
\end{equation}

\subparagraph{4.1.3.1.1 Relatively Small Values Of $h$}

If $h < h_{cr}^{(l)} (\omega_f)$, where the critical value $ \,
h_{cr}^{(l)}(\omega_f) \, $ is determined by Eq.\ (125) (its origin will be
explained further), then there are differences in the boundary formation for
the frequency ranges of {\it odd} and {\it even} spikes. We describe these
ranges separately.

\vskip 0.2cm

\hskip 2.9cm {\it 1. Odd spikes}

\vskip 0.2cm

In this case, the boundary is formed by the trajectory of the Hamiltonian (87)
{\it tangent} to the GSS curve (see Fig.\ 22(a); cf.\ also Figs.\ 1(c), 13(a),
14(b), 14(c)). There are two tangencies in the angular range $]-\pi,\pi[$: they
occur at the angles $\pm\tilde{\psi}_t^{(l)}$ where $\tilde{\psi}_t^{(l)}$ is
determined by Eq.\ (182).

In the ranges of $h$ and $\omega_f$ relevant to the spike minimum, the
asymptotic expressions for $\delta_l$ and $\tilde{\psi}_t^{(l)}$ are:

\begin{eqnarray}
&&
\delta_l = \sqrt{2}\pi h,
\\
&& \tilde{\psi}_t^{(l)}= (-1)^{\left[\frac{n}{4}\right]}
        \sqrt{\frac{n\pi}{8\ln\left(1/\Phi\right)}} +\pi\frac{1-(-1)^{\left[\frac{n}{4}\right]}}{2}.
\end{eqnarray}
Hence, the asymptotic value for the deviation of the tangency
energy $E_t^{(l)}$ from the lower barrier reduces to:
\begin{equation}
E_t^{(l)}-E_b^{(1)} \equiv \delta_l\sin(\tilde{\psi}_t^{(l)} ) =
\frac{\pi^{3/2}} {2}\frac{h}{\sqrt{\ln\left(1/\Phi\right)/n}}.
\end{equation}

The minimum energy on the boundary, $E_{\min}^{(l)}$, corresponds to
$\tilde{\psi}=$ $0$ or $\pi$ for even or odd values of $[n/4]$ respectively.
Thus, it can be found from the equality
\begin{equation}\label{27}
\tilde{H}\left(I(E_{\min}^{(l)}),
\tilde{\psi}=\pi(1-(-1)^{\left[\frac{n}{4}\right]})/2\right)=
\tilde{H}\left(I_t^{(l)}\equiv
I(E_t^{(l)}),\tilde{\psi}_t^{(l)}\right).
\end{equation}

At $\Phi\rightarrow 0$, Eq.\ (111) yields the following expression for the
minimal deviation of energy on the boundary from the barrier:
\begin{equation}
\delta_{\min}^{(l)}\equiv
E_{\min}^{(l)}-E_b^{(1)}=(E_t^{(l)}-E_b^{(1)}) /\sqrt{{\rm
e}}=\frac{\pi^{3/2}} {2\sqrt{{\rm
e}}}\frac{h}{\sqrt{\ln\left(1/\Phi\right)/n}}.
\end{equation}

In the context of the onset of global chaos, the most important property of the
boundary is that the {\it maximum} deviation of its energy from the barrier,
$\delta_{\max}^{(l)}$, should greatly exceed both $\delta_{\min}^{(l)}$ and
$\delta_{l}$. As $h\rightarrow h_{cr}^{(l)}$, the maximum of the boundary
approaches the saddle \lq\lq{\it sl}''.

\vskip 0.2cm

\hskip 2.9cm {\it 2. Even spikes}

\vskip 0.2cm

In this case, the Hamiltonian (87) possesses saddles \lq\lq {\it s}'' in the
close vicinity to the lower barrier (see Fig.\ 22(b)). Their angles differ by
$\pi$ from those of \lq\lq {\it sl}'':
\begin{equation}
\tilde{\psi}_s=\pi\frac{1-(-1)^{\left[\frac{n}{4}\right]}}{2}+2\pi
m,
\quad\quad m=0,\pm 1, \pm 2, \ldots,
\end{equation}

\noindent while the deviation of their energies from the barrier
still lies in the relevant (resonant) range and reads, in the
lowest-order approximation,

\begin{equation}
\delta_s=\frac{\pi}{2\sqrt{2}}\frac{h}{\ln(\ln(1/\Phi))}.
\end{equation}

The lower whiskers of the separatrix generated by these saddles intersect the
GSS curve while the upper whiskers in the asymptotic limit do not intersect it
(Fig.\ 22(b)). Thus, it is the upper whiskers of the separatrix which form  the
boundary of the chaotic layer in the asymptotic limit. The energy on the
boundary takes the minimal  value right on the saddle \lq\lq {\it s}'', so that

\begin{equation}
\delta_{\min}^{(l)}=\delta_s=\frac{\pi}{2\sqrt{2}}\frac{h}{\ln(\ln(1/\Phi))}.
\end{equation}

Similar to the case of the odd spikes, the {\it maximal} deviation of the
energy from the barrier (measured along the boundary) greatly exceeds both
$\delta_{\min}^{(l)}$ and $\delta_{l}$. As $h\rightarrow h_{cr}^{(l)}$, the
maximum of the boundary approaches the saddle \lq\lq{\it sl}''.

\subparagraph{4.1.3.1.2 Relatively Large Values Of $h$}

If $h>h_{cr}^{(l)}(\omega_f)$, the previously described trajectory (either the
tangent one or the separatrix, for the odd or even spike ranges respectively)
is encompassed by the separatrix of the lower nonlinear resonance and typically
forms the boundary of a major stability island inside the lower layer
(reproduced periodically in $\tilde{\psi}$ with the period $2\pi$). The upper
{\it outer} boundary of the layer is formed by the upper part of the {\it
resonance separatrix}. This may be interpreted as the absorption of the lower
resonance by the lower chaotic layer.

\begin{figure}[t]
%\sidecaption[t]
\includegraphics[width = 5.5 cm]{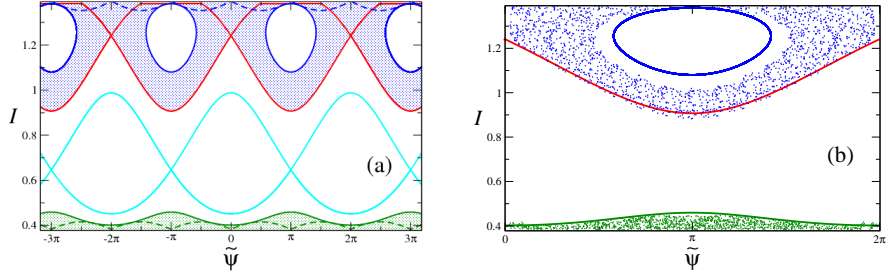}
\hskip 0.6 cm
\includegraphics[width = 5.5 cm]{soskin_Fig13b.eps}
%\vskip 0.75 cm
\caption{(Color version may be found in the online version of \cite{pre2008} as
Fig.\ 8). (a) Chaotic layers (shaded in green and blue, for the upper and lower
layers respectively) in the plane of action $I$ and slow angle $\tilde{\psi}$,
as described by our theory. Parameters are the same as in Figs.\ 10(b) and
11(b). The lower and upper boundaries of the figure box coincide with
$I(E_b^{(1)})$ and $I(E_b^{(2)})$ respectively. The resonance separatrices are
drawn by the cyan and red solid lines (for the lower and upper resonances
respectively). Dashed green and blue lines mark the curves $I=I_{\rm
GSS}^{(l)}(\tilde{\psi})\equiv I(E=E_b^{(1)}+\delta_l|\sin(\tilde{\psi})|)$ and
$I=I_{\rm GSS}^{(u)}(\tilde{\psi})\equiv
I(E=E_b^{(2)}-\delta_u|\sin(\tilde{\psi})|)$ respectively, where $\delta_l$ and
$\delta_u$ are the values of the separatrix split related to the lower and
upper barrier respectively. The upper boundary of the lower layer is formed by
the trajectory of the resonant Hamiltonian system (87) tangent to the curve
$I=I_{\rm GSS}^{(l)}(\tilde{\psi})$. The lower boundary of the upper layer is
formed by the lower part of the upper (red) resonance separatrix. The periodic
closed loops (solid blue lines) are the trajectories of the system (87) tangent
to the curve $I_{\rm GSS}^{(u)}(\tilde{\psi})$: they form the boundaries of the
major stability islands inside the upper chaotic layer. (b) Comparison of the
chaotic layers obtained from computer simulations (dots) with the theoretically
calculated boundaries (solid lines) shown in the box (a).}
\end{figure}

\paragraph{4.1.3.2 Upper Layer}

Let $\omega_f$ be close to any of the spikes' minima.

One of the key roles in the formation of the lower boundary of the layer is
played by the angle-dependent quantity $\delta_u|\sin( \tilde{\psi})|$ which we
call the {\it generalized separatrix split} (GSS) for the upper layer;
$\delta_u$ is the separatrix split for the upper layer: $\delta_u =
|\epsilon^{(up)}( \omega_f)|h$ with $\epsilon^{(up)}$ given by Eq.\ (204).
Accordingly, we use the term \lq\lq upper GSS curve'' for the following curve
in the $I-\tilde{\psi}$ plane:

\begin{equation}
I=I_{\rm GSS}^{(u)}(\tilde{\psi}) \equiv
I(E_b^{(2)}-\delta_u|\sin(\tilde{\psi})|).
\end{equation}

\subparagraph {4.1.3.2.1 Relatively Small Values Of $h$}

If $h < h_{cr}^{(u)} (\omega_f)$, where the critical value $ \,
h_{cr}^{(u)}(\omega_f) \, $ is determined by Eq.\ (126) (its origin will be
explained further), then there are some differences in the boundary formation
in the frequency ranges of {\it odd} and {\it even} spikes: for odd spikes, the
formation is similar to the one for even spikes in the lower-layer case and
vice versa.

\vskip 0.2cm

\hskip 2.9cm {\it 1. Odd spikes}

\vskip 0.2cm

In the case of odd spikes, the Hamiltonian (87) possesses saddles \lq\lq {\it
$\tilde{s}$}'' in the close vicinity of the upper barrier, analogous to the
saddles \lq\lq {\it s}'' near the lower barrier in the case of even spikes.
Their angles are shifted by $\pi$ from those of \lq\lq {\it s}'':
\begin{equation}
\tilde{\psi}_{\tilde{s}}=\tilde{\psi}_s+\pi=\pi\frac{1+(-1)^{\left[\frac{n}{4}\right]}}{2}+2\pi
m,
\quad\quad m=0,\pm 1, \pm 2, \ldots
\end{equation}

\noindent The deviation of their energies from the upper barrier
coincides, in the lowest-order approximation, with $\delta_{s}$:

\begin{equation}
\delta_{\tilde{s}}
=\delta_s=\frac{\pi}{2\sqrt{2}}\frac{h}{\ln(\ln(1/\Phi))}.
\end{equation}

The upper whiskers of the separatrix generated by these saddles intersect the
upper GSS curve while the lower whiskers in the asymptotic limit do not
intersect it. Thus, it is the lower whiskers of the separatrix which form the
boundary of the chaotic layer in the asymptotic limit. The deviation of the
energy from the upper barrier takes its minimal value (measured along the
boundary) right on the saddle \lq\lq {\it $\tilde{s}$}'',

\begin{equation}
\delta_{\min}^{(u)}=\delta_{\tilde{s}}=\frac{\pi}{2\sqrt{2}}\frac{h}{\ln(\ln(1/\Phi))}.
\end{equation}

The {\it maximal} deviation of the energy from the barrier (along the boundary)
greatly exceeds both $\delta_{\min}^{(u)}$ and $\delta_{u}$. As $h\rightarrow
h_{cr}^{(u)}$, the maximum of the boundary approaches the saddle \lq\lq{\it
su}''.

\vskip 0.2cm

\hskip 2.9cm {\it 2. Even spikes}

\vskip 0.2cm

The boundary is formed by the trajectory of the Hamiltonian (87) {\it tangent}
to the GSS curve. There are two tangencies in the angle range $]-\pi,\pi[$:
they occur at the angles $\pm\tilde{\psi}_t^{(u)}$ where $\tilde{\psi}_t^{(u)}$
is determined by Eq.\ (202).

In the ranges of $h$ and $\omega_f$ relevant to the spike minimum, the
expressions for $\delta_u$ and $\tilde{\psi}_t^{(u)}$ in the asymptotic limit
$\Phi\rightarrow 0$ are similar to the analogous quantities in the lower-layer
case:
\begin{eqnarray}
&&
\delta_u = \sqrt{2}\pi h,
\\
&& \tilde{\psi}_t^{(u)}=-(-1)^{\left[\frac{n}{4}\right]}
        \sqrt{\frac{n\pi}{8\ln\left(\frac{1}{\Phi}\right)}}
        +\pi\frac{1+(-1)^{\left[\frac{n}{4}\right]}}{2} .
\end{eqnarray}
Hence, the asymptotic value for the deviation of the tangency
energy $E_t^{(u)}$ from the upper barrier reduces to:
\begin{equation}
E_b^{(2)}-E_t^{(u)}\delta_u\left|\pi\frac{1+(-1)^{\left[\frac{n}{4}\right]}}{2}-\tilde{\psi}_t^{(u)}\right|=
\frac{\pi^{3/2}} {2}\frac{h}{\sqrt{\ln\left(1/\Phi\right)/n}}.
\end{equation}

The maximal energy on the boundary, $E_{\max}^{(u)}$, corresponds to
$\tilde{\psi}=\pi(1+(-1)^{[n/4]})/2$. Thus, it can be found from the equality
\begin{equation}
\tilde{H}(I=I(E_{\max}^{(u)}),\tilde{\psi}=\pi(1+(-1)^{[n/4]})/2)=
\tilde{H}(I_t^{(u)}\equiv I(E_t^{(u)}),\tilde{\psi}_t^{(u)}).
\end{equation}

At $\Phi\rightarrow 0$, Eq.\ (123) yields the following expression for the
minimal deviation of energy from the barrier (measured along the boundary):
\begin{equation}
\delta_{\min}^{(u)}\equiv
E_b^{(2)}-E_{\max}^{(u)}=(E_b^{(2)}-E_t^{(u)})/\sqrt{{\rm e}}=
\frac{\pi^{3/2}} {2{\rm
e}^{1/2}}\frac{h}{\sqrt{\ln\left(1/\Phi\right)/n}}.
\end{equation}

\subparagraph {4.1.3.2.2 Relatively Large Values Of $h$}

If $h>h_{cr}^{(u)}(\omega_f)$ (cf.\ Fig.\ 13(a)), the previously described
trajectory (either the tangent one or the separatrix, for the even and odd
spikes ranges respectively) is encompassed by the separatrix of the upper
nonlinear resonance and typically forms the boundary of a major stability
island inside the upper layer (reproduced periodically in $\tilde{\psi}$ with
the period $2\pi$). The lower {\it outer} boundary of the layer is formed in
this case by the lower part of the {\it resonance separatrix}. This may be
interpreted as the absorption of the upper resonance by the upper chaotic
layer.

\vskip 0.5cm

The self-consistent description of chaotic layers given above, and in more
detail in the Appendix, is the first main result of this section. It provides a
{\it rigorous basis} for our intuitive assumption that the minimal value of $h$
at which the layers overlap corresponds to the reconnection of the nonlinear
resonances with each other while the reconnected resonances touch one of the
layers and also touch/overlap another layer. It is gratifying that we have
obtained a {\it quantitative} theoretical description of the chaotic layer
boundaries in the {\it phase space}, including even the major stability
islands, that agrees well with the results of simulations (see Fig.\ 13(b)). To
the best of our knowledge it was the first ever \cite {pre2008} quantitative
description of the layer boundaries in the phase space.

\subsubsection{Onset of global chaos: the spikes' minima}

The condition for the merger of the lower resonance and the lower
chaotic layer may be written as

\begin{equation}
\tilde{H}(I=I(E=E_b^{(1)}+\delta^{(l)}_{\min}),\tilde{\psi}=\pi(1-(-1)^{[n/4]})/2)=\tilde{H}_{sl}.
\end{equation}

\noindent The condition for the merger of the upper resonance and the upper
chaotic layer may be written as

\begin{equation}
\tilde{H}(I=I(E=E_b^{(2)}-\delta^{(u)}_{\min}),\tilde{\psi}=\pi(1+(-1)^{[n/4]})/2)=\tilde{H}_{su}.
\end{equation}

For the onset of global chaos related to the spike minimum, either of Eqs.\
(125) and (126) should be combined with the condition of the separatrix
reconnection (104). Let us seek first only the leading terms of $h_s\equiv
h_s(\Phi)$ and $\omega_s\equiv \omega_s(\Phi)$. Then (104) may be replaced by
its lowest-order approximation (105) or, equivalently, (106). Using also the
lowest-order approximation for the barriers ($E_b^{(1,2)}\approx 1/2\mp \Phi$),
we reduce Eqs.\ (125), (126) respectively to

\begin{equation}
\tilde{H}(I=I(E=1/2-\Phi+\delta^{(l)}_{\min}),\tilde{\psi}=\pi(1-(-1)^{[n/4]})/2)=0,
\end{equation}
\begin{equation}
\tilde{H}(I=I(E=1/2+\Phi-\delta^{(u)}_{\min}),\tilde{\psi}=\pi(1+(-1)^{[n/4]})/2)=0,
\end{equation}

\noindent where $\delta^{(l)}_{\min}$ is given by (112) or (115) for the odd or
even spikes respectively, while $\delta^{(u)}_{\min}$ is given by (119) or
(124) for the odd or even spikes respectively.

To the {\it leading} order, the solution $(h_s^{(l)},\omega_s^{(l)})$ of the system of equations
(106),(127) and the solution $(h_s^{(u)},\omega_s^{(u)})$ of the system of
equations (106),(128) turn out {\it identical}. For the
sake of brevity, we derive below just $(h_s^{(l)},\omega_s^{(l)})$, denoting
the latter, in short, as $(h_s,\omega_s)$\footnote{With account taken of the
next-order corrections, the spike minimum $(h_s,\omega_s)$ coincides with
$(h_s^{(l)},\omega_s^{(l)})$ in case of an odd spike, or with
$(h_s^{(u)},\omega_s^{(u)})$ in case of an even spike. This occurs because, in
case of odd spikes, $|q_n(E)|$ increases/decreases as $E$ approaches the
relevant vicinity of the upper/lower barrier, while it is {\it vice versa} in
the case of even spikes. And the larger $|q_n|$ the further the resonance
separatrix extends: in other words, the reconnection of the barrier chaotic
layer with the resonance separatrix requires a smaller value of $h$ at the
barrier where $|q_n|$, in the relevant vicinity of the barrier, is larger.}.

The system of algebraic equations (106) and (127) is still too complicated for
us to find its exact solution. However, we need only the {\it lowest-order}
solution -- and this simplifies the problem. Still, even this simplified
problem is not trivial, both because the function $h_s(\Phi)$ turns out to be
non-analytic and because $\Delta E^{(1)}$ in (106) is very sensitive to
$\omega_f$ in the relevant range. Despite these difficulties, we have found the
solution in a {\it self-consistent} way, as briefly described below.

Assume that the asymptotic dependence $h_s(\Phi) $ is:

\begin{eqnarray}
&&
h_s=a\frac{\Phi}{\ln(4{\rm e}/\Phi)} \, ,
\end{eqnarray}

\noindent where the constant $a$ may be found from the asymptotic solution of
(106), (127), (129).

Substituting the energies $E = 1/2-\Phi+\delta^{(l)}_{\min}$ and $ E =1/2+
\Phi- \delta^{(u)}_{\max}$ in (89) and taking account of (112), (115), (119),
(124) and (129), we find that, both for the odd and even spikes, the inequality

\begin{equation}
\omega_m-\omega(E)\ll \omega_m
\end{equation}

\noindent holds over the whole relevant range of energies, i.e.\ for

\begin{equation}
\Delta E\in [-\Phi+\delta^{(l)}_{\min},\Phi-\delta^{(u)}_{\min}].
\end{equation}

\noindent Thus, the resonant approximation is valid over the whole range (131).
Eq.\ (96) for $q_n(E)$ is valid over the whole relevant range of energies too.

Consider Eq.\ (127) in an explicit form. Namely, we express $\omega_f$ from
(127), using Eqs.\ (87), (96), and (97), using also (112) or (115) for odd or
even spikes and (129):

\begin{equation}
\omega_f =  \frac{n\pi}{2 \ln \left(\frac{4{\rm e}}{\Phi} \right)
} \left\{ 1+\frac{h\sqrt{2}}{n\Phi} + O \left(
\frac{1}{\ln^2(4{\rm e}/\Phi)}\right) \right\}.
\end{equation}

\noindent We emphasize that the value of $\delta_{\min}^{(l)}$ enters
explicitly only the term $O(\ldots)$ while, as is clear from the consideration
below, this term does not affect the leading terms in $(h_s,\omega_s)$. Thus,
$\delta_{\min}^{(l)}$ does not affect the leading term of $\omega_s$ at all,
while it affects the leading term of $h_s$ only {\it implicitly}:
$\delta_{\min}^{(l)}$ lies in the range of energies where $nq_n(E)\approx
\sqrt{2}$. This latter quantity is present in the second term in the curly
brackets in (132) and, as becomes clear from further consideration, $h_s$ is
proportional to it.

Substituting (132) into the expression for $\Delta E^{(1)}$ in (106), using
(129) and keeping only the leading terms, we obtain

\begin{equation}
\Delta E^{(1)}=\Phi\sqrt{1-4{\rm e}^{c-2}}, \quad\quad c\equiv
\frac{2\sqrt{2}}{n}a.
\end{equation}

\noindent Substituting $\Delta E^{(1)}$ from (133) into Eq.\ (106) for
$h(\omega_f)$ and allowing for (129) once again, we arrive at a transcendental
equation for $c$:

\begin{equation}
\ln
\left(
\frac{1+\chi(c)}{1-\chi(c)}
\right)
-2\chi(c)=c,
\quad\quad
\chi(c)\equiv \sqrt{1-4{\rm e}^{c-2}}.
\end{equation}

\noindent An approximate numerical solution of Eq.\ (134) is:

\begin{equation}
c \simeq0.179.
\end{equation}

Thus, the final leading-order asymptotic formul{\ae} for $\omega_f$ and $h$ in
the minima of the spikes are the following:

\begin{eqnarray}
&&
\omega_{s0}\equiv \omega_{s0}^{
\left(\frac{n+1}{2}
\right)}=n\frac{\pi}{2\ln
\left(\frac{4{\rm e}}{\Phi}
\right) },
\quad\quad
h_{s0}\equiv h_{s0}^{\left(\frac{n+1}{2}\right)}=n\frac{c}{2\sqrt{2}}\frac{\Phi}{\ln
\left(\frac{4{\rm e}}{\Phi}
\right) },
\\
&& n=1,3,5,..., \quad\quad \Phi\rightarrow 0, \nonumber
\end{eqnarray}

\noindent where the constant $c\simeq 0.179$ is the solution of Eq.\ (134).

The self-consistent derivation of the explicit asymptotic formul{\ae} for the
minima of $h_{gc}(\omega_f)$ constitutes the second main result of this
section. These formul{\ae} allow one to predict immediately the parameters for
the weakest perturbation that may lead to global chaos.

\subsubsection{Numerical example and next-order corrections}

For $\Phi=0.2$, the numerical simulations give the following values for the
frequencies at the minima of the first two spikes (see Fig.\ 9):

\begin{equation}
\omega_{s}^{(1)}\approx 0.4005\pm 0.0005, \quad\quad
\omega_{s}^{(2)}\approx 1.24\pm 0.005.
\end{equation}

\noindent By the lowest-order formula (136), the values are:

\begin{equation}
\omega_{s0}^{(1)}\approx 0.393, \quad\quad \omega_{s0}^{(2)}\approx
1.18,
\end{equation}

\noindent in rather good agreement with the simulations.

The next-order correction for $\omega_s$ can immediately be found from Eq.\
(132) for $\omega_f$ and from Eq.\ (136) for $h_{s0}$, so that

\begin{equation}
\omega_{s1}\simeq\omega_{s0}(1+\frac{c}{2\ln \left(\frac{4{\rm
e}}{\Phi} \right)})\approx \frac{n\pi \left( 1+\frac{0.09}{\ln
\left(\frac{4{\rm e}}{\Phi} \right) } \right) }{2\ln
\left(\frac{4{\rm e}}{\Phi} \right) } \, ,
\quad\quad
 n=1,3,5,...
\end{equation}

\noindent The formula (139) agrees with the simulations even better than the
lowest-order approximation:

\begin{equation}
\omega_{s1}^{(1)}\approx 0.402, \quad\quad
\omega_{s1}^{(2)}\approx 1.21.
\end{equation}

For $h$ in the spikes minima, the simulations give the following values
(see Fig.\ 9):

\begin{equation}
h_{s}^{(1)}\approx 0.0049, \quad\quad h_{s}^{(2)}\approx 0.03.
\end{equation}

\noindent The values according to the lowest-order formula (52) are:

\begin{equation}
h_{s0}^{(1)}\approx 0.0032, \quad\quad h_{s0}^{(2)}\approx 0.01.
\end{equation}

\noindent The theoretical value $h_{s0}^{(1)}$ gives reasonable agreement with
the simulation value $h_{s}^{(1)}$. The theoretical value $h_{s0}^{(2)}$ gives
the correct order of magnitude for the simulation value $h_{s}^{(2)}$. Thus,
the accuracy of the lowest-order formula (136) for $h_s$ is much lower than
that for $ \omega_s $: this is due to the steepness of $h_{gc}(\omega_f)$ in
the ranges of the spikes (the steepness, in turn, is due to the flatness of the
function $\omega(E)$ near its maximum). Moreover, as the number of the spike
$j$ increases, the accuracy of the lowest-order value $h_{s0}^{(j)}$
significantly decreases. The latter can be explained as follows. For the
next-order correction to $h_{s0}^{(j)}$, the dependence on $\Phi$ reads as:

\begin{equation}
\frac{h_{s1}^{(j)}-h_{s0}^{(j)}}{h_{s0}^{(j)}}\propto \frac{1}{\ln(4{\rm
e}/\Phi)}.
\end{equation}

\noindent At least some of the terms in this correction are positive and
proportional to $h_{s0}^{(j)}$ (e.g.\ due to the difference between the exact
equation (99) and its approximate version (101)), while $h_{s0}^{(j)}$ is
proportional to $n\equiv 2j-1$. Thus, for $\Phi=0.2$, the relative correction
for the 1st spike is $\sim 0.25$ while the correction for the 2nd spike is a
few times larger i.e.\ $\sim 1$. But the latter means that, for $\Phi=0.2$, the
asymptotic theory for the 2nd spike cannot pretend to be a quantitative
description of $h_s^{(2)}$, but only provides the correct order of magnitude.
Besides, if $n>1$ while $\Phi$ exceeds some critical value, then the search of
the minimum involves Eq.\ (150) rather than Eq.\ (104), as explained below in
Sec.\ 4.2 (cf.\ Figs.\ 15(b) and 16). Altogether, this explains why
$h_{s}^{(1)}$ is larger than $h_{s0}^{(1)}$ only by 50$\%$ while $h_{s}^{(2)}$
is larger than $h_{s0}^{(2)}$ by 200$\%$.

To provide a consistent explicit derivation of the correction to $h_{s0}^{(j)}$
is complicated. A reasonable alternative may be a proper {\it numerical}
solution of the algebraic system of Eqs.\ (104)\footnote{For $n>1$, it is also
necessary to check if the solution lies above the line (150). If it does not,
then (104) should be replaced here by (150).} and
(125) for the odd spikes, or (126) for the even spikes$^{16}$. To this end, in
Eqs.\ (104)$^{17}$ and (125) or (126) we use: (i) the exact values of the
saddle energies obtained from the exact relations (100) instead of the
approximate relations (101); (ii) the exact formula (88) for $\omega(E)$
instead of the asymptotic expression (89); (iii) the exact functions $q_n(E)$
instead of the asymptotic formula (86); (iv) the relation (111) and the
calculation of the \lq\lq tangent'' state $(\tilde{\psi}_t^{(l)}, I_t^{(l)})$
by Eqs.\ (172), (183) for the odd spikes, or relation (123) and the calculation
of the \lq\lq tangent'' state $(\tilde{\psi}_t^{(u)}, I_t^{(u)})$ by Eqs.\
(202)-(204) for the even spikes. Note that, to find the exact function
$q_n(E)$, we substitute into the definition of $q_n(E)$ in (87) the
explicit\footnote{In the general case of an arbitrary potential $U(q)$, when
the explicit expression for $q(E,\psi)$ and $\omega(E)$ cannot be obtained,
these functions can be calculated numerically.} solution for $q(E,\psi)$:

\begin{eqnarray}
&& q(E,\psi)=\arcsin \left( \frac{\eta-\sqrt{2E}+\Phi}{1-\eta}
\right) \quad {\rm for} \quad \psi\in\left[0,\frac{\pi}{2}\right],
\nonumber
\\
&& q(E,\psi)=\pi-q(E,\pi-\psi) \quad\quad\quad\quad {\rm for}
\quad \psi\in\left[\frac{\pi}{2},\pi\right], \nonumber
\\
&& q(E,\psi)=q(E,2\pi-\psi) \quad\quad\quad\quad\quad\quad {\rm
for} \quad \psi\in\left[\pi,2\pi\right], \nonumber
\\
&& \eta\equiv \frac{1}{2}(\sqrt{2E}-\Phi+1){\rm sn}^2
\left(\frac{2K}{\pi}\psi \right) \, ,
\end{eqnarray}
where ${\rm sn}(x)$ is the elliptic sine \cite{Abramovitz_Stegun} with the same
modulus $k$ as the full elliptic integral $K$ defined in (88). The numerical
solution described above gives:

%\begin{equation}
\begin{eqnarray}
&&   \left(\omega_{s}^{(1)}\right)_{num}\approx 0.401 \ , \qquad
   \left(h_{s}^{(1)}\right)_{num} \approx 0.005 \ ,
   \nonumber
\\
&&\\
 &&   \left(\omega_{s}^{(2)}\right)_{num}\approx 1.24 \ , \qquad
   \left(h_{s}^{(2)}\right)_{num} \approx 0.052 \ .
\nonumber
\end{eqnarray}
%\end{equation}

The agreement with the simulation results is: (i) excellent for
$\omega_{s}$ for the both spikes and for $h_s$ for the 1st spike,
(ii) reasonable for $h_s$ for the 2nd spike. Thus, if $\Phi$ is {\it
moderately} small, a much more accurate prediction for $h_s$ than
that by the lowest-order formula is provided by the numerical
procedure described above.

\subsection{Theory of the Spikes' Wings}

The goal of this section is to find the mechanisms responsible for the
formation of the spikes' {\it wings} (i.e.\ the function $h_{gc}(\omega_f)$ in
the ranges of $\omega_f$ slightly deviating from $\omega_s^{(j)}$), and to
provide for their theoretical description.

Before developing the theory, we briefly analyze the simulation data (Fig.\ 9),
concentrating on the 1st spike. The left wing of the spike is smooth and nearly
straight. The initial part of the right wing is also nearly
straight\footnote{Provided $h_{gc}(\omega_f)$ is smoothed over small
fluctuations.}, though less steep. But at some small distance from
$\omega_s^{(1)}$ its slope changes jump-wise by a few times: compare the
derivative ${\rm d}h_{gc}/{\rm d}\omega_f\approx 0.1$ at $\omega_f= 0.4 \div 0.41$
(see the left inset in Fig.\ 9) and ${\rm
d}h_{gc}/{\rm d}\omega_f\approx 0.4$ at $\omega_f = 0.45 \div 0.55$ (see the
main part of Fig.\ 9). Thus, even prior to the theoretical analysis, one may
assume that there are a number of different mechanisms responsible for
formation of the wings.

Consider the arbitrary $j$th spike. We have shown in the previous section that,
in the asymptotic limit $\Phi\rightarrow 0$, the minimum of the spike
corresponds to the intersection between the line (104) with (125) or (126) for
odd or even spikes respectively. We recall that: (i) Eq.\ (104) corresponds to
the overlap in phase space between nonlinear resonances of the same order
$n\equiv 2j-1$; (ii) Eq.\ (125) or (126) corresponds to the onset of the
overlap between the resonance separatrix associated respectively with the lower
or upper saddle and the chaotic layer associated with the lower or upper
potential barrier; (iii) for $\omega_f=\omega_s^{(j)}$, the condition (125) or
(126) also guarantees the overlap between the upper or lower resonance
separatrix, respectively, and the chaotic layer associated with the upper or
lower barrier$^{16}$.

\begin{figure}[b]
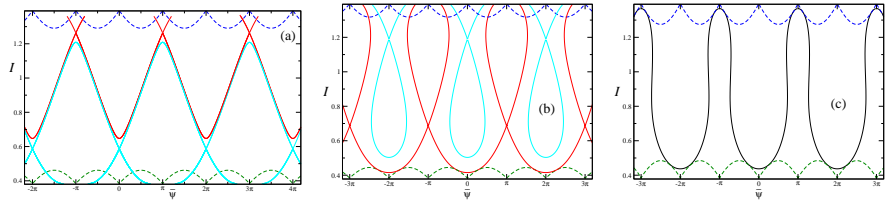

\includegraphics[width = 3.9 cm]{soskin_Fig14a.eps}
\hskip 0.2 cm
\includegraphics[width = 3.6 cm]{soskin_Fig14b.eps}
\hskip 0.2 cm
\includegraphics[width = 3.6 cm]{soskin_Fig14c.eps}
%\vskip 0.75 cm
\caption{(Color version may be found in the online version of \cite{pre2008} as
Fig.\ 9). Illustrations of the mechanisms of the formation of the 1st spike
wings and of the corresponding theoretical lines in Fig.\ 15(a). Boxes (a), (b)
and (c) illustrate the lines of Eqs.\ (104), (125) and (148) respectively: the
corresponding perturbation parameters are ($\omega_f=0.39,h=0.0077$),
($\omega_f=0.41,h=0.00598$) and ($\omega_f=0.43,h=0.01009$) respectively.
Resonance separatrices are drawn in red and cyan. The dashed lines show the
functions $I_{\rm GSS}^{(l)}(\tilde{\psi})$ and $I_{\rm
GSS}^{(u)}(\tilde{\psi})$. The black line in (c) is the trajectory of the
resonant Hamiltonian system (87) which is tangent to both dashed lines.}
\end{figure}

If $\omega_f$ becomes slightly smaller than $\omega_s^{(j)}$, the resonances
shift closer to the barriers while moving apart from each other. Hence, as $h$
increases, the overlap of the resonances with the chaotic layers associated
with the barriers occurs earlier than with each other. Therefore, at
$0<\omega_s^{(j)}- \omega_f\ll \omega_m$, the function $h_{gc}(\omega_f)$
should correspond approximately to the reconnection of resonances of the order
$n\equiv 2j-1$ (Fig.\ 14(a)). Fig.\ 15(a) demonstrates that even the asymptotic
formula (106) for the separatrix reconnection line fits the left wing of the
1st spike quite well, and that the numerically calculated line (104) agrees
with the simulations perfectly.

If $\omega_f$ becomes slightly larger than $\omega_s^{(j)}$ then, on the
contrary, the resonances shift closer to each other and further from the
barriers. Therefore, the mutual overlap of the resonances occurs at smaller $h$
than the overlap between any of them and the chaotic layer associated with the
lower/upper barrier (cf.\ Figs.\ 10(c) and 10(d) as well as 11(c) and 11(d)).
Hence, it is the latter overlap which determines the function
$h_{gc}(\omega_f)$ in the relevant range of $\omega_f$ (Fig.\ 14(b)). Fig.\ 15
shows that $h_{gc}(\omega_f)$ is indeed well-approximated in the close vicinity
to the right from $\omega_s^{(j)}$ by the numerical solution of Eq.\ (125) or
(126), for an odd or even spike respectively and, for the 1st spike and the
given $\Phi$, even by its asymptotic form,

\begin{eqnarray}
&& h\equiv h(\omega_f)=
n\frac
{-\Phi +\frac{\omega_f}{n\pi}
\left[
\Phi
\left\{
2\ln\left(\frac{4{\rm e}}{\Phi}\right)+\ln\left(\frac{\Phi+\Delta E^{(1)}}{\Phi-\Delta E^{(1)}} \right)
\right\}
-2\Delta E^{(1)}
\right]
}
{2\sqrt{2}},
\nonumber\\
&& \Delta E^{(1)}\equiv
\sqrt{\Phi^2-64\exp(-\frac{n\pi}{\omega_f})}, n\equiv 2j-1,\quad
|\omega_f-\omega_s^{(j)}|\ll \omega_m \, .
\end{eqnarray}

\begin{figure}[b]
\includegraphics[width = 5.7 cm]{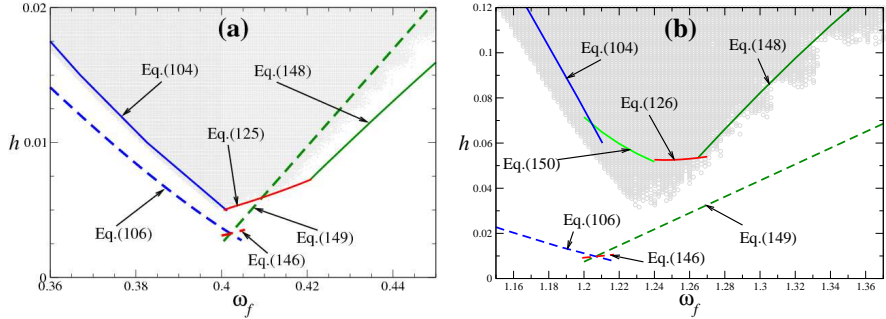}
\hskip 0.2 cm
\includegraphics[width = 5.7 cm]{soskin_Fig15b.eps}
\caption{(Color version may be found in the online version of \cite{pre2008} as
Fig.\ 10). The 1st (a) and 2nd (b) spike in $h_{gc}(\omega_f)$: comparison
between the results of the numerical simulations (the lower boundary of the
shaded area) and the theoretical estimates. The estimates are indicated by the
corresponding equation numbers and are drawn by different types of lines, in
particular the dashed lines represent the explicit asymptote for the solid line
of the same color.}
\end{figure}

The mechanism described above determines $h_{gc}(\omega_f)$ only in the close
vicinity of $\omega_s^{(j)}$. If $\omega_f/n$ becomes too close to $\omega_m$
or exceeds it, then the resonances are not of immediate relevance: they may
even disappear or, if they still exist, their closed loops shrink, so that they
can no longer provide for connection of the chaotic layers in the relevant
range of $h$. At the same time, the closeness of the frequency to $\omega_m$
may still give rise to a large variation of action along the trajectory of the
Hamiltonian system (87). For the odd/even spikes, the boundaries of the chaotic
layers in the asymptotic limit $\Phi\rightarrow 0$ are formed in this case by
the trajectory of (87) which is tangent to the lower/upper GSS curves (for the
lower/upper layer) or by the lower/upper part of the separatrix of (87)
generated by the saddle \lq\lq $\tilde{s}$''/\lq\lq $s$'' (for the upper/lower
layer). The overlap of the layers occurs when these trajectories coincide with
each other, which may be formulated as the equality of $\tilde{H}$ in the
corresponding tangency and saddle:

\begin{eqnarray}
&& \tilde{H}(I_t^{(l)},\tilde{\psi}_t^{(l)})=
\tilde{H}(I_{\tilde{s}},\tilde{\psi}_{\tilde{s}}) \quad
{\rm for}
\quad j=1,3,5,\ldots, \nonumber
\\
&&\tilde{H}(I_s,\tilde{\psi}_s)=
\tilde{H}(I_t^{(u)},\tilde{\psi}_t^{(u)}) \quad {\rm for} \quad
j=2,4,6,\ldots, \nonumber
\\
&& I_{\tilde{s}}\equiv I(E_b^{(2)}-\delta_{\tilde{s}}), \quad\quad
I_{s}\equiv I(E_b^{(1)}+\delta_{s}).
\end{eqnarray}

\noindent Note however that, for {\it moderately} small $\Phi$, the tangencies
may be relevant both to the lower layer and to the upper  one (see the
Appendix). Indeed, such a case occurs for our example with $\Phi=0.2$: see
Fig.\ 14(c). Therefore, the overlap of the layers corresponds to the equality
of $\tilde{H}$ in the tangencies:

\begin{equation}
\tilde{H}(I_t^{(l)},\tilde{\psi}_t^{(l)})=
\tilde{H}(I_t^{(u)},\tilde{\psi}_t^{(u)})
\, .
\end{equation}

\noindent To the lowest order, Eq.\ (147) and Eq.\ (148) read as:

\begin{equation}
h\equiv h(\omega_f)=\frac{\sqrt{2}\Phi\ln\left(\frac{4{\rm
e}}{\Phi}\right)}
{\pi}\left(\omega_f-\frac{n\pi}{2\ln\left(\frac{4{\rm
e}}{\Phi}\right)}
 \right).
\end{equation}

\noindent Both the line (148) and the asymptotic line (149) well agree with the
part of the right wing of the 1st spike situated to the right from the fold at
$\omega_f\approx 0.42$ (Fig.\ 15(a)). The fold corresponds to the change of the
mechanisms of the chaotic layers overlap.

\begin{figure}[b]
\sidecaption[b]
\includegraphics[width = 7. cm]{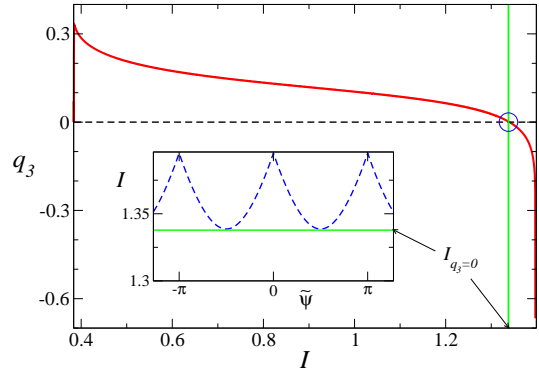}
\caption{(Color version may be found in the online version of \cite{pre2008} as
Fig.\ 11). Amplitude of the 3rd Fourier harmonic as a function of action (solid
red line). The dashed black line shows the zero level. Its intersection with
the solid red line is marked by the circle. The green line indicates the value
of action where $q_3=0$. The inset illustrates the line (150) in Fig.\ 15(b):
the GSS curve touches the horizontal line $I=I_{q_3=0}$.}
\end{figure}

If $\Phi$ is moderately small while $n>1$, the description of the far wings by
the numerical lines (104) and (148) may be still quite good but the asymptotic
lines (106) and (149) cannot pretend to describe the wings quantitatively any
more (Fig.\ 15(b)). As for the minimum of the spike and the wings in its close
vicinity, one more mechanism may become relevant for their formation (Figs.\
15(b) and 16). It may be explained as follows. If $n>1$, then $q_n(E)$ becomes
zero in the close vicinity ($\sim \Phi^2$) of the relevant barrier (the upper
or lower barrier, in the case of even or odd spikes respectively: cf.\ Fig.\
16). It follows from the equations of motion (98) that no trajectory can cross
the line $I=I_{q_n=0}$. In the asymptotic limit $\Phi\rightarrow 0$, provided
$h$ is from the range relevant for the spike minimum, almost the whole GSS curve is further from the
barrier than the line $I=I_{q_n=0}$, and the latter becomes irrelevant. But,
for a moderately small $\Phi$, the line may separate the whole GSS curve from
the rest of the phase space. Then the resonance separatrix cannot connect to
the GSS curve even if there is a state on the latter curve with the same value
of $\tilde{H}$ as on the resonance separatrix. For a given $\omega_f$, the
connection then requires a higher value of $h$: for such a value, the GSS curve
itself crosses the line $I=I_{q_n=0}$. In the relevant range of $h$, the
resonance separatrix passes very close to this line, so that the connection is
well approximated by the condition that the GSS curve {\it touches} this line
(see the inset in Fig.\ 16):

\begin{eqnarray}
&& \delta_u=E_b^{(2)}-E_{q_{2j-1}=0}\quad\quad {\rm for}\quad
j=2,4,6,\ldots,
\nonumber\\
&& \delta_l=E_{q_{2j-1}=0}-E_b^{(1)}\quad\quad {\rm for}\quad
j=3,5,7,\ldots.
\end{eqnarray}

\noindent This mechanism is relevant to the formation of the minimum of the 2nd
spike at $\Phi=0.2$, and in the close vicinity of the spike on the left (Fig.\
15(b)).

Finally, let us find explicitly the {\it universal asymptotic shape} of the
spike in the vicinity of its minimum. First, we note that the lowest-order
expression (146) for the spike between the minimum and the fold can be written
as the {\it half-sum} of the expressions (106) and (149) (which represent the
lowest-order approximations for the spike to the left of the minimum, and to
the right of the fold respectively). Thus, all three lines (106), (146) and
(149) intersect at a single point. This means that, in the asymptotic limit
$\Phi\rightarrow 0$, the fold merges with the minimum: $\omega_f$ and $h$ in
the fold asymptotically approach $\omega_s$ and $h_s$ respectively. Thus,
though the fold is a generic feature of the spikes, it is not of major
significance: the spike is formed basically from two straight lines. The ratio
between their slopes is universal. So, introducing a proper scaling, we reduce
the spike shape to the universal function (Fig.\ 17):

\begin{eqnarray}
&& \tilde{h}(\Delta \tilde {\omega}_f)=\tilde{h}^{(lw)}(\Delta
\tilde {\omega}_f)\equiv 1-\sqrt {1-4{\rm e}^{c-2}}\Delta \tilde
{\omega}_f\approx
\nonumber\\
&& \quad\quad\quad\quad\quad\quad\quad\quad\approx
1-0.593\Delta
\tilde {\omega}_f \quad {\rm for}\quad \Delta \tilde {\omega}_f<0,
\nonumber
\\
&& \tilde {h}(\Delta \tilde {\omega}_f) = \tilde{h}^{(rw)}(\Delta
\tilde {\omega}_f)\equiv 1+\Delta \tilde
 {\omega}_f\quad {\rm for}\quad\Delta \tilde {\omega}_f>0, \nonumber
\\
&&  \\
&&
 \tilde{h}^{(fold)}(\Delta
\tilde {\omega}_f)= \frac{ \tilde{h}^{(lw)}(\Delta \tilde
{\omega}_f)+ \tilde{h}^{(rw)}(\Delta
\tilde {\omega}_f)}{2}\equiv
\nonumber \\
&&\equiv
 1+\frac{1-\sqrt {1-4{\rm
e}^{c-2}}}{2}\Delta \tilde {\omega}_f \approx 1+0.203\Delta \tilde
{\omega}_f, \nonumber\\
&& \nonumber \\
&& \tilde {h}\equiv \frac{h}{h_{s0}},  \quad\quad\Delta \tilde
{\omega}_f\equiv
\frac{\omega_f-\omega_{s1}}{\omega_{s1}-\omega_{s0}}, \quad\quad
\Phi\rightarrow 0,\nonumber
\end{eqnarray}

\noindent where $\omega_{s0}$ and $h_{s0}$ are the lowest-order expressions
(136) respectively for the frequency and amplitude in the spike minimum,
$\omega_{s1}$ is the expression (139) for the frequency in the spike minimum,
including the first-order correction, and $c$ is a constant (135).

\begin{figure}[b]
\sidecaption[b]
\includegraphics[width = 4.5 cm]{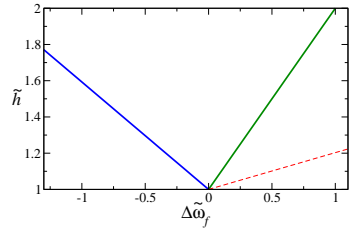}
\caption{(Color version may be found in the online version of \cite{pre2008} as
Fig.\ 12). The universal shape of the spike minimum (151) (solid lines). The
dashed line indicates the universal slope of the spike in between the minimum
and the fold, which have merged in the universal (asymptotic) function (151).}
\end{figure}

In addition to the left and right wings of the universal shape (the solid lines
in Fig.\ 17), we include in (151) the function $\tilde{h}^{(fold)}(\Delta
\tilde {\omega}_f)$ (the dashed line in Fig.\ 17): its purpose is to show, on
one hand, that the fold merges asymptotically with the minimum but, on the
other hand, that the fold is generic and the slope of the spike between the
minimum and the fold has a universal ratio to any of the slopes of the major
wings.

Even for a moderately small $\Phi$, as in our example, the ratios between the
three slopes related to the 1st spike in the simulations are reasonably well
reproduced by those in Eq.\ (151): cf.\ Figs.\ 15(a) and 17. It follows from
(151) that the asymptotic scaled shape is universal i.e.\ independent of
$\Phi$ (but still assuming the asymptotic limit $\Phi\rightarrow 0$), $n$ or any other parameter.

The description of the wings of the spikes near their minima, in particular the
derivation of the spike universal shape, constitutes the third main result of
this section.

\subsection{Generalizations and Applications}
The {\it facilitation of the onset of global chaos} between adjacent
separatrices has a number of possible generalizations and applications. We
discuss an application in Sec.\ 5, but first list some of generalizations
below.

\begin{enumerate}

\item The spikes in $h_{gc}(\omega_f)$ may occur for an {\it arbitrary
    Hamiltonian} system with two or more separatrices. The asymptotic
    theory can be generalized accordingly.

\item The absence of pronounced spikes at {\it even} harmonics $2j\omega_m$
    is explained by the symmetry of the potential (84): the even Fourier
    harmonics of the coordinate, $q_{2j}$, are equal to zero. For
    time-periodic perturbation of the dipole type, as in Eq.\ (85), there
    are no resonances of even order on account of this symmetry
    \cite{Chirikov:79,lichtenberg_lieberman,Zaslavsky:1991,zaslavsky:1998,zaslavsky:2005,PR}.
    If either the potential is {\it non-symmetric}, or the additive
    perturbation of the Hamiltonian is not an {\it odd} function of the
    coordinate, then even-order resonances do exist, resulting in the
    presence of the spikes in $h_{gc}(\omega_f)$ at $\omega_f \approx
    2j\omega_m$.

\item There may also be an additional facilitation of the onset of global
    chaos that could reasonably be described as a \lq\lq secondary''
    facilitation. Let the frequency $\omega_f$ be close to the frequency
    $\omega_s$ of the spike minimum, while the amplitude $h$ be $\sim h_s$
    but still lower than $h_{gc}(\omega_f)$. Then there are two resonance
    separatrices in the $I-\tilde{\psi}$ plane that are not connected by
    chaotic transport (cf.\ Fig.\ 11(b)). This system possesses
    the zero-dispersion property. The trajectories of the resonant
    Hamiltonian (87) which start in between the separatrices oscillate in
    $I$ (as well as in ${\rm d}\tilde{\psi}/{\rm d}t$). The frequency
    $\tilde{\omega}$ of such oscillations along a given trajectory depends
    on the corresponding value of $\tilde{H}$ analogously to the way in
    which $\omega$ depends on $E$ for the original Hamiltonian $H_0$:
    $\tilde{\omega}(\tilde{H})$ is equal to zero for the values of
    $\tilde{H}$ corresponding to the separatrices (being equal in turn to
    $\tilde{H}_{sl}$ and $\tilde{H}_{su}$: see Eq.\ (103)) while possessing
    a nearly rectangular shape in between, provided the quantity
    $|\tilde{H}_{sl}-\tilde{H}_{su}|$ is much smaller than the variation of
    $\tilde{H}$ within each of the resonances,

\begin{equation}
|\tilde{H}_{sl}-\tilde{H}_{su}| \ll \tilde{H}_{var}\sim
|\tilde{H}_{sl}-\tilde{H}_{el}|\sim
|\tilde{H}_{su}-\tilde{H}_{eu}|,
\end{equation}

\noindent where $\tilde{H}_{el}$ and $\tilde{H}_{eu}$ are the values of
$\tilde{H}$ at the elliptic point of the lower and upper resonance
respectively. The maximum of $\tilde{\omega}(\tilde{H})$ in between
$\tilde{H}_{sl}$ and $\tilde{H}_{su}$ is described by the asymptotic
formula:

\begin{equation}
\tilde{\omega}_m\approx\frac{\pi}{\ln\left(
\tilde{H}_{var}/|\tilde{H}_{sl}-\tilde{H}_{su}|\right)}.
\end{equation}

If we additionally perturb the system in such a way that an additional
time-periodic term of frequency $\tilde{\omega}_f\approx \tilde{\omega}_m$
arises in the resonance Hamiltonian, then the chaotic layers associated
with the resonance separatrices may be connected by chaotic transport even
for a rather small amplitude of the additional perturbation, due to a
scenario similar to the one described in this paper.

There may be various types of such additional perturbation \cite{webs}. For
example, one may {\it add} to $H$ (85) one more dipole time-periodic
perturbation of {\it mixed} frequency (i.e.\ $\approx
\omega_m+\tilde{\omega}_m$). Alternatively, one may directly perturb the {\it
angle} of the original perturbation by a {\it low-frequency} perturbation,
i.e.\ the time-periodic term in $H$ (85) is replaced by the term

\begin{equation}
-hq\cos(\omega_ft+A\cos(\tilde{\omega}_ft)),
\quad\quad
\omega_f\approx \omega_m,\quad\quad \tilde{\omega}_f \approx
\tilde{\omega}_m.
\end{equation}

Recently discussed physical problems where a similar situation is
relevant are: chaotic mixing and transport in a meandering jet
flow \cite{prants} and reflection of light rays in a corrugated
waveguide \cite{leonel}.

\item If the time-periodic perturbation is {\it multiplicative} rather than
    additive,  the resonances become {\it parametric} (cf.\
    \cite{Landau:76}). Parametric resonance is more complicated and much
    less studied than nonlinear resonance. Nevertheless, the main mechanism
    for the onset of global chaos remains the same, namely the combination
    of the reconnection between resonances of the same order and of their
    overlap in energy with the chaotic layers associated with the barriers.
    At the same time, the frequencies of the main spikes in
    $h_{gc}(\omega_f)$ may change (though still being related to
    $\omega_m$). We consider below an example when the periodically driven
    parameter is\footnote{In the case of a 2D electron gas in a magnetic
    superlattice, this may correspond e.g.\ to the time-periodic electric
    force applied perpendicular to the direction of the periodic magnetic
    field \cite{oleg98,oleg99}.} $\Phi$ in (84). The Hamiltonian is

\begin{eqnarray}
&& H= p^2/2 +(\Phi-\sin(q))^2/2,
\nonumber\\
&&
\Phi=\Phi_0+h\cos(\omega_f t), \quad\quad \Phi_0={\rm const}<1.
\end{eqnarray}

\noindent The term $(\Phi-\sin(q))^2/2$ in $H$ (155) may be rewritten as
$(\Phi_0-\sin(q))^2/2
+(\Phi_0-\sin(q))h\cos(\omega_ft)+h^2\cos^2(\omega_ft)/2$. The last term in
the latter expression does not affect the equations of motion. Thus, we end
up with an additive perturbation $(\Phi_0-\sin(q))h\cos(\omega_ft)$. In the
asymptotic limit $\Phi_0\rightarrow 0$, the $n$th-order Fourier component
of the function $(\Phi_0-\sin(q))$ can be shown to differ from zero only
for the orders $n=2,6,10,...$ Therefore one may expect the main spikes in
$h_{gc}(\omega_f)$ to be at frequencies twice larger than those for the
dipole perturbation (85):

\begin{equation}
\omega_{sp}^{(j)}\approx 2 \omega_{s}^{(j)}\approx
2(2j-1)\omega_m, \quad\quad j=1,2,3,...
\end{equation}

\noindent This agrees well with the results of simulations (Fig.\ 18).

\begin{figure}[t]
\sidecaption[t]
\includegraphics[width = 5. cm]{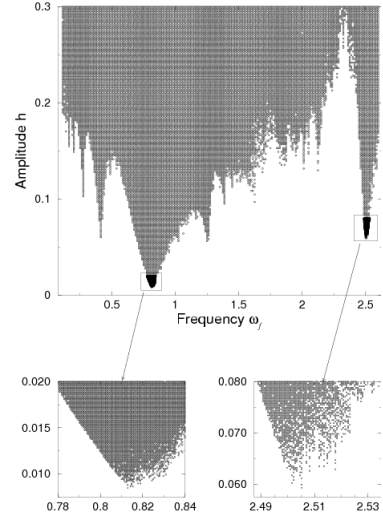}
\caption{Diagram analogous to that in Fig.\ 9, but for the system (155)
(with $\Phi_0=0.2$).}
\end{figure}

Moreover, the asymptotic theory for the dipole perturbation may immediately
be generalized to the present case: it is necessary only to replace the
Fourier component of the coordinate $q$ by the Fourier component of the
function $(\Phi_0-\sin(q))$:

\begin{equation}
%&&
(\Phi_0-\sin(q))_n= \left\{_{0\quad \quad {\rm at} \quad n\neq
2(2j-1),}^{\frac{4}{\pi n}\quad  {\rm at} \quad n= 2(2j-1),} \right.
%\\&&
\quad\quad
j=1,2,3,...,
%\nonumber\\&&
\quad\quad
\Phi_0\rightarrow 0
%\nonumber
\end{equation}

\noindent (cf.\ Eq.\ (96) for $q_n$). We obtain:

\begin{eqnarray}
&&\omega_{sp0}\equiv \omega_{sp0}^{ \left(\frac{n+2}{4}
\right)}=n\frac{\pi}{2\ln \left(\frac{4{\rm e}}{\Phi_0} \right) },
\quad\quad
h_{sp0}\equiv
h_{sp0}^{\left(\frac{n+2}{4}\right)}=n\frac{c\pi}{8}\frac{\Phi_0}{\ln
\left(\frac{4{\rm e}}{\Phi_0} \right) },
\\
&& n=2,6,10,..., \qquad \Phi_0\rightarrow 0, \nonumber
\end{eqnarray}

\noindent where $c$ is given in Eqs.\ (134) and (135).

For $\Phi_0=0.2$, Eq.\ (158) gives, for the 1st spike, values differing
from the simulation data by about $3\%$ in frequency and by about $10\%$ in
amplitude. Thus, the lowest-order formul{\ae} accurately describe the 1st
spike even for a moderately small $\Phi$.

\item One more generalization relates to {\it multi-dimensional}
    Hamiltonian systems with two or more saddles with different energies:
    the perturbation may not necessarily be time-periodic, in this case.
    The detailed analysis has not yet been done.

\end{enumerate}

The paper \cite {pre2008} presents a rather detailed discussion of possible
applications to the telectron gas in a magnetic superlattice, a spinning
pendulum, cold atoms in an optical lattice as well as to problems of
noise-induced escape and the stochastic web formation. We review briefly in the
next section the further development of the latter application.

\section{Enlargement of a low-dimensional stochastic web}

The stochastic web concept dates back to the 1960s when Arnold showed
\cite{arnold} that, in non-degenerate Hamiltonian systems of dimension
exceeding 2, resonance lines necessarily intersect, forming an infinite-sized
web in the Poincar\'{e} section. It provides in turn for a slow chaotic
(sometimes called \lq\lq stochastic'') diffusion for infinite distances in
relevant dynamical variables.

It was discovered towards the end of 1980s
\cite{zaslavsky:1986,A.A.,chernikov:1987_,chernikov:1988} that, in degenerate
or nearly-degenerate systems, a stochastic web may arise even if the dimension
is 3/2. One of the archetypal examples of such a low-dimensional stochastic web
arises in the 1D harmonic oscillator perturbed by a weak traveling wave the
frequency of which coincides with a multiple of the natural frequency of the
oscillator \cite{zaslavsky:1998,chernikov:1987_,Zaslavsky:1991}. Perturbation
plays a dual role: on the one hand, it gives rise to a slow dynamics
characterized by an auxiliary Hamiltonian that possesses an infinite web-like
separatrix; on the other hand, the perturbation destroys this self-generated
separatrix, replacing it by a thin chaotic layer. Such a low-dimensional
stochastic web may be relevant to a variety of physical systems and plays an
important role in corresponding transport phenomena: see
\cite{zaslavsky:1998,chernikov:1987_,Zaslavsky:1991} for reviews on relevant
classical systems. In addition, there are quantum systems in which the dynamics
of transport reduces to that in the classical model described above. The latter
concerns e.g.\ nanometre-scale semiconductor superlattices with an applied
voltage and magnetic field \cite{fromhold,fromhold_nature}.

One might assume that, like the Arnold web, the low-dimensional stochastic web
described above should be infinite, so that it can provide for transport
between the centre of the web and states situated arbitrarily far away in
coordinate and momentum. However the numerical integration of the equations of
motion shows that this is not so: even for a rather non-weak perturbation, the
real web is limited to the region within {\it a few} inner loops of the
infinite web-like resonant separatrix (Fig.\ 19(a)) while chaotic layers
associated with outer loops are distinctly separated from each other and from
the web-like chaotic layer formed by the few inner loops. The reason is
apparently as follows. The single infinite web-like separatrix is possessed by
the resonant Hamiltonian only in the first-order approximation of the averaging
method \cite{bogmit} whereas, with the account taken of the next-order
approximations, the separatrix apparently splits into many separate complex
loops successively embedded into each other. Non-resonant terms of the
perturbation dress the separatrices by exponentially narrow chaotic layers. If
the perturbation is not small, the chaotic layers manage to connect
neighbouring separatrix loops situated close to the centre. However, the width
of the chaotic layer decreases exponentially sharply as the distance from the
centre grows \cite{zaslavsky:1998,chernikov:1987_,Zaslavsky:1991}. As a result,
the merger between chaotic layers associated with neighbouring loops takes
place only within the few loops closest to the centre, provided that the
perturbation is not exponentially strong.

If the resonance between the perturbation and the oscillator is inexact, or if
the oscillator is nonlinear, the splitting between the neighbouring loops is
typically much larger: it appears even in the first-order approximation of the
averaging method \cite{zaslavsky:1998,chernikov:1988,Zaslavsky:1991}. So the
number of loops connected to the centre by chaotic transport is even smaller
\cite{zaslavsky:1998,chernikov:1988,Zaslavsky:1991} than in the case of the
exact resonance.

A natural question arises:
how can the perturbation be modified in order for the transport to be unlimited
or, at least, significantly extended? One of the answers was obtained in the
very beginning of studies of the low-dimensional webs
\cite{zaslavsky:1986,A.A.}: if the perturbation consists of repeated in time short
kicks that are also periodic space, and if the frequency
of the kicks is equal to a multiple of the natural frequency, then a so-called
uniform web covering the whole of phase space is formed. However such a
perturbation is absent in many cases and, even where present, the chaotic
transport is still exponentially slow if the perturbation is weak
\cite{zaslavsky:1998,Zaslavsky:1991}.

It is reasonable then to pose the following question: is it possible to obtain
a web of form similar to the original one \cite{chernikov:1987_} but
substantially extended in phase space? A positive answer was suggested in
\cite{pre2008} and explicitly realized recently \cite{icnf_enlargement} using
the following simple idea. The chaotic layer in the webs is {\it exponentially}
narrow since the frequency of the non-resonant perturbation of the resonant
Hamiltonian is necessarily much higher than the frequency of small
eigenoscillation in the cell of the web-like separatrix
\cite{zaslavsky:1998,zaslavsky:1986,A.A.,chernikov:1987_,chernikov:1988,Zaslavsky:1991}.
So we need to modify the perturbation in such a way that the resonant
Hamiltonian does not change while its perturbation contains, in addition to the
conventional terms, a low-frequency one. One may do this modulating the wave
angle with a low frequency or adding one more wave with the frequency slightly
shifted from the original one. The latter option, together with a
generalization for the uniform web leading to a huge enhancement of the chaotic
transport through it, have not yet been considered in detail while the work
\cite{icnf_enlargement} and the present section concentrate on the former
option since it may have immediate applications to nanometre-scale
semiconductor superlattices in electric and magnetic fields
\cite{fromhold,fromhold_nature}.

\subsection{Slow modulation of the wave angle}

Fig.\ 19 demonstrates the validity of our idea. We integrate the equation
\begin{equation}
\ddot q+q = 0.1 \sin [15 q-4 t - h \sin (0.02 t)],
\end{equation}
\noindent first for $h=0$ (i.e.\ for the conventional case with parameters as
in  \cite{zaslavsky:1998,chernikov:1987_,Zaslavsky:1991}), and secondly for
$h=0.1$. Although the modulation in the latter case is weak (its amplitude is
about 63 times smaller than the $2\pi$ period of the wave angle which is a
characteristic scale in this problem), the resultant increase in the size of
the web in coordinate and momentum is large: by a factor of $\sim$6.

An analytic theory can be developed to account for these results.
It can be generalized for the off-resonant case
\cite{zaslavsky:1998,chernikov:1988,Zaslavsky:1991} too, using the general
method developed in \cite{pre2008,proceedings,icnf_approach,pre_submitted} and
described above in the previous sections.

It is anticipated that the method can also be generalized for uniform webs
\cite{zaslavsky:1998,zaslavsky:1986,Zaslavsky:1991} too, leading to an
exponentially strong enhancement of chaotic transport through them.
\begin{figure}[b]
\includegraphics[width=6.cm]{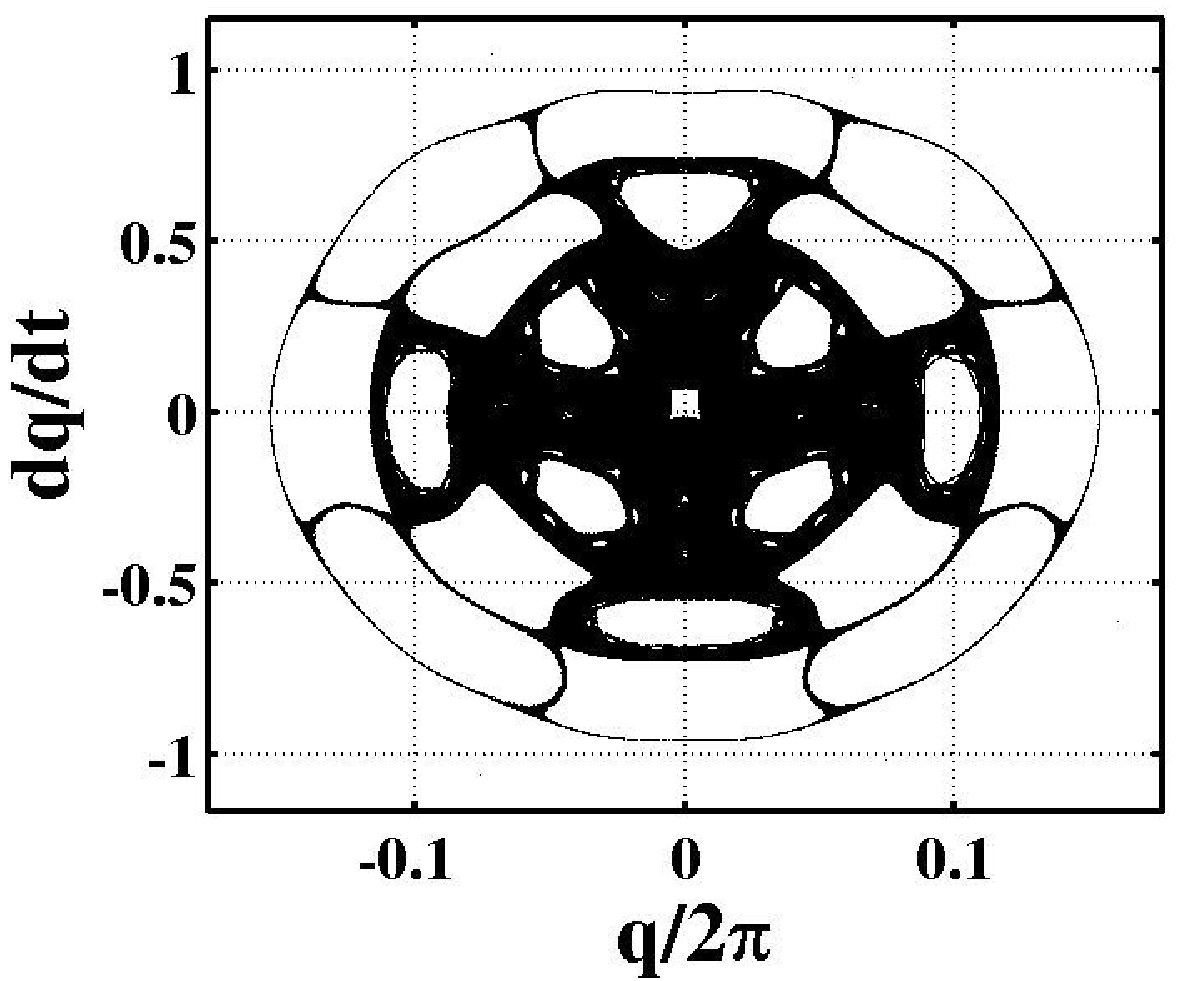}
\includegraphics[width=6.cm]{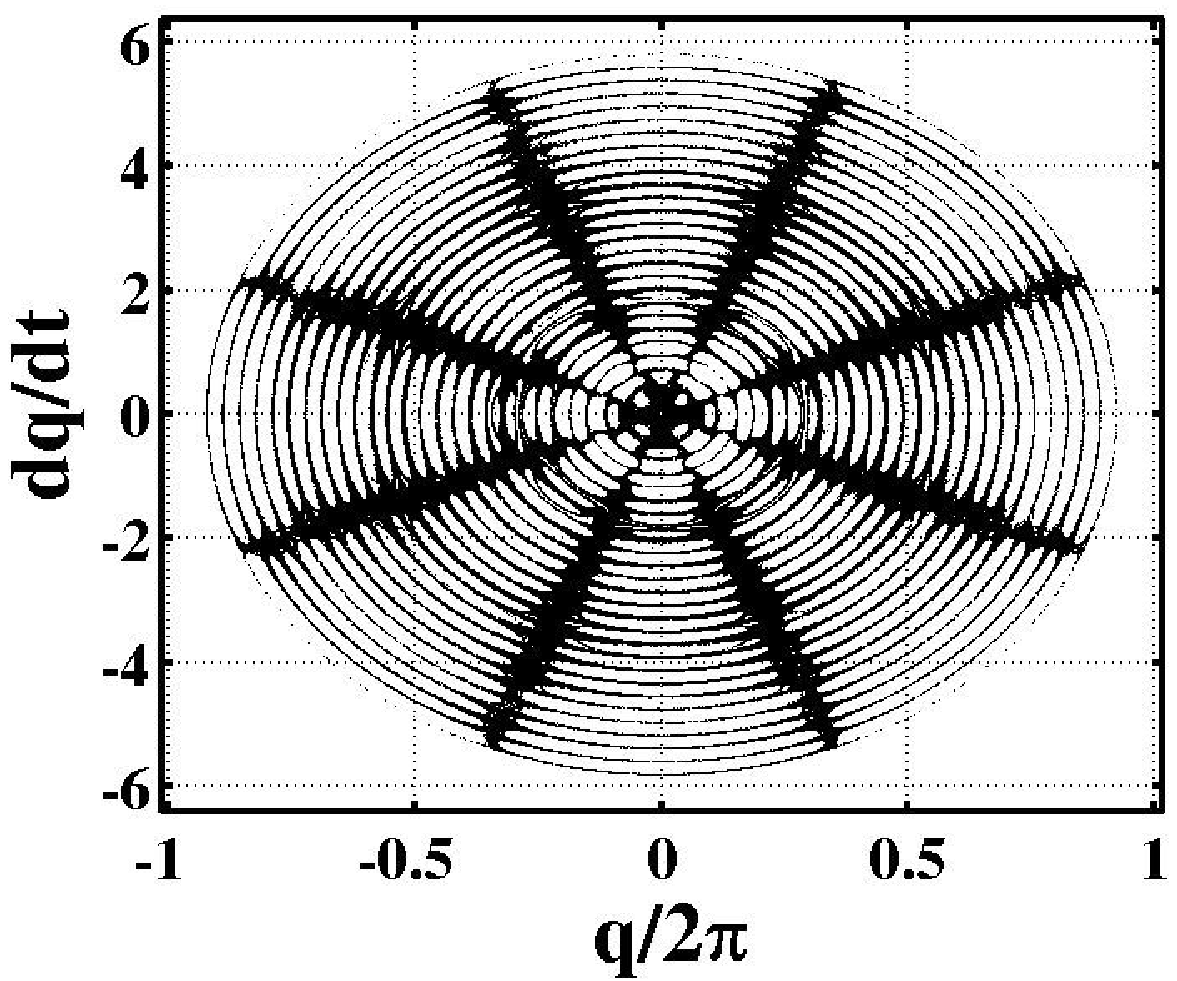}\\
\caption{ The Poincar{\'e} section for a trajectory of the system (159) with
initial state $q=0.1,\;\dot{q}=0$ (at instants $t_n=nT$ where $T\equiv 2
\pi/0.02$ is the period of the modulation and $n=1,2,3,...600000$) for $h=0$
(left panel) and $h=0.1$ (right panel). A sympletic integration scheme of the
fourth order is used, with an integration step $t_{int}=\frac{2
\pi}{40000}\approx 1.57\times 10^{-4}$, so that the inaccuracy at each step is
of the order of $t_{int}^5\approx \times 10^{-19}$. The left panel corresponds
to the example of the conventional case considered in
\cite{zaslavsky:1998,chernikov:1987_,Zaslavsky:1991}. The right panel
demonstrates that the modulation, although weak, greatly enlarges the web
sizes.}
\end{figure}

\subsection{Application to semiconductor superlattices}

The works \cite{fromhold,fromhold_nature} consider quantum electron transport
in 1D semiconductor superlattices (SLs) on the nanometre scale, subject to a
constant electric field along the SL axis and to a constant magnetic field. The
spatial periodicity with a period of the nanometre scale gives rise to the onset of
minibands for electrons. In the tight-binding approximation, the electron
motion in the lowest mini-band is described by the following dispersion
relation for the electron energy $E$ versus momentum $\vec{p}$:

\begin{equation}
E(\vec{p})=\frac{\Delta[1-\cos(p_xd/\hbar)]}{2}+\frac{p_y^2+p_z^2}{2m^{*}},
\end{equation}

\noindent where $x$ is the direction along the SL axis, $\Delta$ is the
miniband width, $d$ is the SL period, $m^{*}$ is the electron effective mass
for the motion in the transverse (i.e.\ $y-z$) direction.

Thus, the quasi-classical motion of electron in an electric field $\vec{F}$ and
a magnetic field $\vec{B}$ is described by the following equation:

\begin{equation}
\frac{{\rm d}\vec{p}}{{\rm d}t}=-e\{\vec{F} + [\nabla_{\vec{p}}
E(\vec{p}) \times \vec{B}]\}.
\end{equation}

\noindent where $e$ is the electron charge

It was shown in \cite{fromhold} that, with a constant electric field along the
SL axis $\vec{F}=(-F_0,0,0)$ and a constant magnetic field with a given angle
$\theta$ to the axis $\vec{B}=(B\cos(\theta),0,B\sin(\theta))$, the dynamics of
the $z$-component of momentum $p_z$ reduces to the equation of motion of an
auxiliary harmonic oscillator in a plane wave. At certain values of the
parameters, the ratio of the wave and oscillator frequencies takes integer
values (like in Eq.\ (159) with $h=0$) which gives rise to the onset of the
stochastic web, leading in turn to a delocalization of the electron in the
$x$-direction and, as a result, to an increase of the dc-conductivity along the
SL axis. The experiment \cite{fromhold_nature} appears to provide evidence in
favor of this exciting hypothethis.

At the same time, the finite size of the web and, yet more so, the
exponentially fast decrease in the transport rate as the distance from the
centre of the web increases, seems to put strong limitations on the use of the
effect. We suggest a simple and efficient way to overcome these limitations.
Indeed, one can show that, if we add to the original (constant) electric field
$F_0$ a small time-periodic (ac) component $F_{\rm ac}\sin(\Omega_{\rm ac}t)$,
then the wave angle in the equation of motion of $p_z$ is modulated by the term
(cf.\ Eq.\ (159)):

\begin{equation}
h\sin(\Omega t)\equiv \frac{F_{\rm ac}}{F_0}\frac{\Omega_0}{\Omega_{\rm
ac}}\sin
\left(\frac{\Omega_{\rm ac}}
{\Omega_0}t\right),
\quad\quad \Omega_0\equiv \frac{eF_0d}{\hbar}.
\end{equation}

This allows us to increase drastically the size of the web and the rate of
chaotic transport through it. For example, for the case shown in Fig.\ 19,
where we have an increase of the web size by a factor of 6$\times$, it is
sufficient to add an ac component of the electric field with the frequency
$0.02\cdot\Omega_{0}$ and an amplitude $F_{\rm ac}=0.1\cdot0.02\cdot F_0$ i.e.\
an amplitude smaller than that of the original constant field $F_0$ by a factor
of 500$\times$.

\subsection{Discussion}

We have presented above just initial results on the subject
\cite{icnf_enlargement}. There are still many unsolved interesting problems --

\begin{enumerate}

\item It can be shown that, in the off-resonance
    case, there may be a facilitation of the onset of global chaos similar
    to that described in Sec.\ 4 above, i.e.\ the critical value of the
    modulation amplitude $h$ required for the onset of global chaos between
    neighbouring separatrix loops possesses deep spikes (minima) as a
    function of the modulation frequency $\Omega_{ac}$. The detailed theory
    of this facilitation has yet to be developed.

\item Our conjecture that, in the resonant case, taking account of the
    next-order approximations of the averaging method could explain the
    split between different separatrix loops, should be proved rigorously.
    If the corresponding theory is developed, it will provide the
    possibility of calculating both the optimal modulation frequency, i.e.\
    that at which the web sizes are maximal, for a given amplitude of
    modulation, and the maximum sizes themselves.

\item It would be interesting to study the case with an additive perturbation (rather than an angular
    modulation) in detail, both numerically and
    theoretically.

\end{enumerate}

\section{Conclusions}

We have reviewed the recently developed method for the theoretical treatment of
separatrix chaos in regimes when it involves resonance dynamics. It has been
applied both to single-separatrix chaotic layers and to the onset of global
chaos between two close separatrices. The method is based on a matching between
the discrete chaotic dynamics of the separatrix map and the continuous regular
dynamics of the resonance Hamiltonian. For single-separatrix chaos, the method
has allowed:

\begin{enumerate}

\item Development of the first asymptotic (i.e.\ for $h\rightarrow
    0$) description of the high peaks in the width of the separatrix
    chaotic layer as a function of the perturbation frequency, thus
    describing its dominant feature and, in particular, its maxima.

\item Classification of all systems into two types, based on the
    asymptotic dependence of the maximum width on the perturbation
    amplitude $h$: the maximum width is proportional to $h\ln(1/h)$ or $h$
    for systems of type I or type II respectively.

\end{enumerate}

\noindent For systems with two or more separatrices, the method has allowed us
to develop an accurate asymptotic theory of the facilitation of the onset of
global chaos between neighbouring separatrices which occurs at frequencies
close to multiples of a local maximum in the eigenoscillation frequency as a
function of the energy: the local maximum necessarily exists
in the range between the separatrices.

Finally, for an oscillator perturbed by a plane wave of frequency equal to or
close to the frequency of a small eigenoscillation, the method has allowed us
to suggest how to enlarge substantially the size of the stochastic web using a
rather weak perturbation, and it promises to provide an accurate theoretical
description of the enlargement.

\begin{acknowledgement}
We are very much indebted to George Zaslavsky: numerous discussions with him,
and his friendly attitude and interest in our research, have greatly stimulated
the work. We also acknowledge financial support from the grant within the
Convention between the Institute of Semiconductor Physics and University of
Pisa for 2008-2009, from a Royal Society International Joint Project grant
2007/R2-IJP, and from the German Physical Society grant SFB-TR12.
Finally, S.M.S. acknowledges the hospitality of Pisa and Lancaster
Universities during his visits to both places and, in turn, R.M. and P.V.E.McC.
acknowledge the hospitality of the Institute of Semiconductor Physics during
their visits there.
\end{acknowledgement}

\section{Appendix}

This appendix follows the appendix of the paper \cite{pre2008}. The chaotic
layers of the system (85) associated with the separatrices of the unperturbed
system (84) are described here by means of the separatrix map. To derive the
map, we follow the method described in \cite{Zaslavsky:1991}, but the analysis
of the map significantly differs from formerly existing ones
\cite{lichtenberg_lieberman,Zaslavsky:1991,zaslavsky:1998,zaslavsky:2005,treschev}
(cf.\ also the recently published paper \cite{shevchenko} where the analysis of
the map has some similarity to ours but still differs significantly). Using our
approach, we are able to calculate the chaotic layer boundaries in the {\it
phase space} (rather than only in energy), throughout the resonance frequency
ranges, and we can quantitatively describe the {\it transport} within the layer
in a manner different from existing ones (cf.\ \cite{treschev,vered1} and
references therein).

\subsection {Lower chaotic layer}.

We now present a detailed consideration of the lower chaotic layer. The upper
layer may be considered in a very similar way.

\subsubsection {Separatrix map}

A typical form of trajectory $\dot{q}(t)$ close to the inner separatrix (that
corresponding to the lower potential barrier) is shown in Fig.\ 20. One can
resolve pulses in $\dot{q}(t)$. Each of them consists of two approximately
antisymmetric spikes\footnote{Spikes correspond to motion over any of the
minima of the potential, first in one and then (after the reflection from one
of the upper barriers) in the opposite direction. If $\Phi$ is small, then the
spikes within the pulse are separated by long intervals since the reflection
point is situated close to the top of the upper barrier, where the motion is
slow.}. The pulses are separated by intervals during which $|\dot{q}|$ is
relatively small. In general, successive intervals differ between each other.
Let us introduce the pair of variables $ \, E \, $ and $ \, \varphi $:

\begin{figure}[t]
\sidecaption[t]
\includegraphics[width = 7.0 cm]{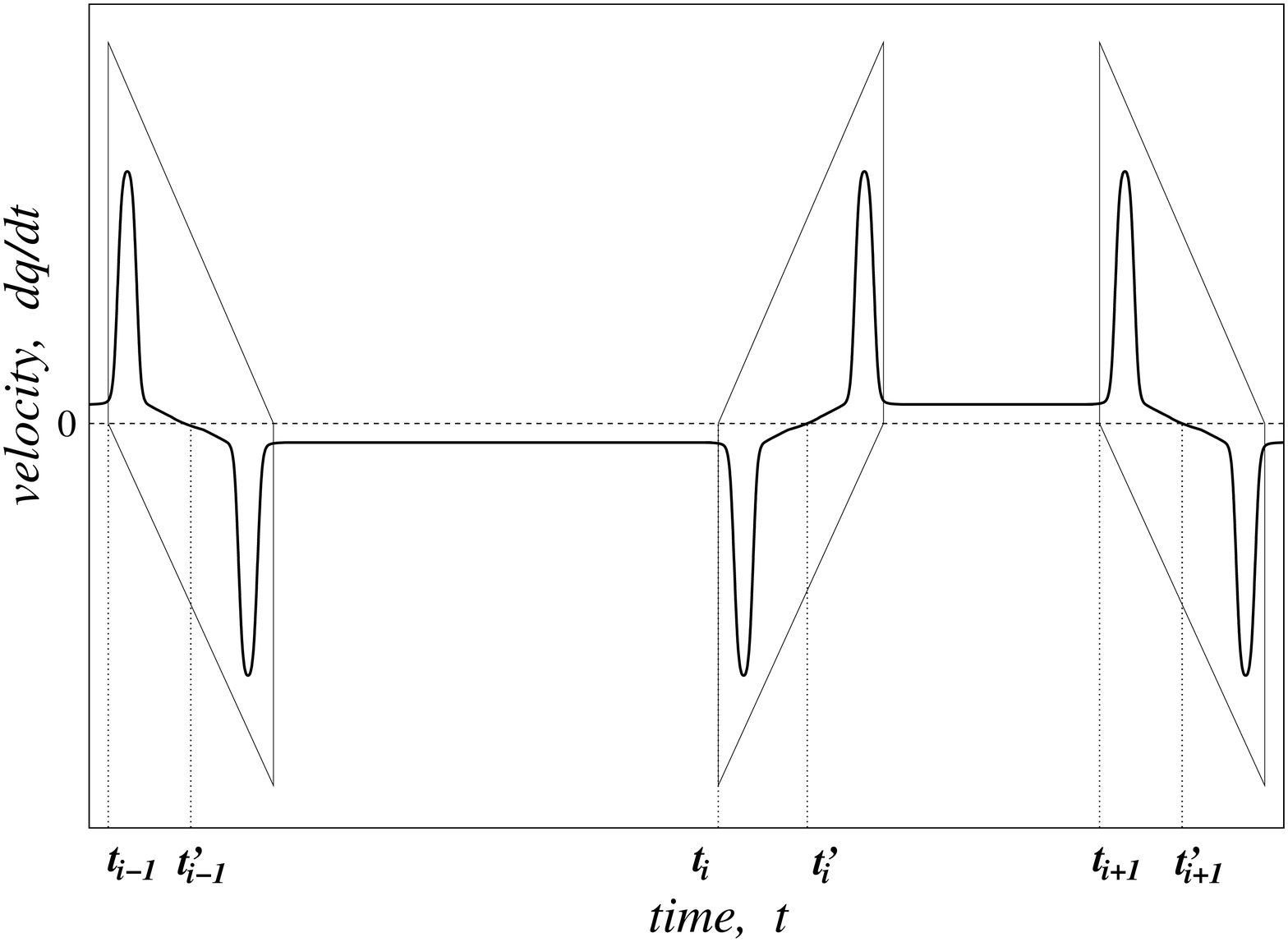}
\caption{Schematic example of the time dependence of the
velocity of the perturbed system (thick solid line) in the case
when the energy of motion varies in the close vicinity of the top of the lower
potential barrier. The dashed line marks
the zero level of the velocity. Pulses of the velocity are schematically
singled out by the
parallelograms (drawn by a thin solid line). The two sequences of time
instants $(...,t_{i-1}, t_i, t_{i+1}, ...)$ and
$(...,t_{i-1}^{\prime}, t_i^{\prime}, t_{i+1}^{\prime}, ...)$
correspond to beginnings and centers of the pulses, respectively.}
\end{figure}

\begin{equation}
  E \equiv H_0, \quad \varphi\equiv \omega_ft+\varphi_a \, ,
\end{equation}

\noindent where the constant $\varphi_a$ may be chosen
arbitrarily.

The energy $E$ changes only during the pulses of $\dot{q}(t)$ and remains
nearly unchanged during the intervals between the pulses, when $|\dot{q}(t)|$
is small \cite{Zaslavsky:1991}. We assign numbers $ \, i \, $ to the pulses and
introduce the sequences of $(E_i,\varphi_i)$ corresponding to the initial
instants $t_i$ of the pulses. In such a way, we obtain the following map (cf.\
\cite{Zaslavsky:1991}):

\begin{eqnarray}
&&
E_{i+1}=E_i+\Delta E_i,
\quad\quad
\varphi_{i+1}=\varphi_i+\frac{\omega_f\pi(3- {\rm sign}(E_{i+1}-E_b^{(1)}))
}{2\omega(E_{i+1})},
\nonumber
\\
&&
\Delta E_i\equiv h\int_{\hbox{\small{{\it i\,}th pulse}}} \!\!\!\!\!
{\rm d}t\;\dot{q}(t)\cos(\omega_ft),
%\nonumber
\end{eqnarray}

\noindent where $ \, \int_{\hbox{\small{{\it i\,}th pulse}}} \, $ means
integration over the $i$th pulse. Before deriving a more explicit expression
for $\Delta E_i$, we make two remarks.

\begin{enumerate}

\item Let us denote with $t_i^{\prime}$ the instant within the $i$th pulse
    when $\dot{q}$ is equal to zero (Fig.\ 20). The function
    $\dot{q}(t-t_i^{\prime})$ is an odd function within the $i$th pulse and
    it is convenient to transform the cosine in the integrand in $\Delta
    E_i$ (164) as

$$
\cos(\omega_ft)\equiv\cos(\omega_f(t-t_i^{\prime})+\omega_ft_i^{\prime})\equiv
\cos(\omega_f(t-t_i^{\prime}))
\cos(\omega_ft_i^{\prime})-
\sin(\omega_f(t-t_i^{\prime}))
\sin(\omega_ft_i^{\prime}),
$$

\noindent and to put $ \varphi_a = \omega_f(t_i^{\prime}-t_i) $, so
that $ \varphi_i\equiv \omega_ft_i^{\prime} $.

\item Each pulse of $\dot{q}$ contains one positive and one negative spike.
    The first spike can be either positive or negative. If $E$ changes
    during the given $n$th pulse so that its value at the end of the pulse
    is {\it smaller} than $E_b^{(1)}$, then the first spikes of the $i$th
    and $(i+1)$st pulses have the {\it same} signs. On the contrary, if $E$
    at the end of the $i$th pulse is {\it larger} than $E_b^{(1)}$, then
    the first spikes of the $i$th and $(i+1)$st pulses have {\it opposite}
    signs. Note that Fig.\ 20 corresponds to the case when the energy
    remains above $E_b^{(1)}$ during the whole interval shown in the
    figure. This obviously affects the sign of $\Delta E_i$, and it may be
    explicitly accounted for in the map if we introduce a new discrete
    variable $\sigma_i=\pm 1$ which characterizes the sign of $\dot{q}$ at
    the beginning of a given $i$th pulse,

\begin{equation}
   \sigma_i\equiv {\rm sign} (\dot{q}(t_i)) \ ,
\end{equation}

\noindent
and changes from pulse to pulse as

\begin{equation}
\sigma_{i+1}=\sigma_i \, {\rm sign}(E_b^{(1)}-E_{i+1}) \ .
\end{equation}

\end{enumerate}

\noindent With account taken of the above remarks, we can rewrite
the map (164) as follows:

\begin{eqnarray}
&&E_{i+1}=E_i+\sigma_ih\epsilon^{(low)}\sin(\varphi_i),
\\
&&\varphi_{i+1}=\varphi_i+\frac{\omega_f\pi(3- {\rm
sign}(E_{i+1}-E_b^{(1)})) }{2\omega(E_{i+1})}, \nonumber
\\
&&\sigma_{i+1}=\sigma_i \, {\rm sign}(E_b^{(1)}-E_{i+1}), \nonumber
\\
&&\quad\quad\;\epsilon^{(low)}\equiv \epsilon^{(low)}(\omega_f)=
-\sigma_i \int_{i{\rm th}\;{\rm pulse}} {\rm
d}t\;\dot{q}(t-t_i^{\prime})\sin(\omega_f(t-t_i^{\prime}))
\nonumber\\
&&\quad\quad\quad\quad\quad\quad
\approx  -2\sigma_i \int_{t_i^{\prime}}^{t_{i+1}}{\rm
d}t\;\dot{q}(t-t_i^{\prime})\sin(\omega_f(t-t_i^{\prime})).
\nonumber
\end{eqnarray}

A map similar to (167) was introduced in \cite{ZF:1968}, and it is often called
the Zaslavsky separatrix map. It was re-derived mathematically rigorously in
\cite{vered}; see also the recent mathematical review \cite{treschev}. The
latter review also describes generalizations of the Zaslavsky map as well as
other types of separatrix map. The analysis presented below relates immediately
to the Zaslavsky map but it is hoped that it will prove possible to generalize
it for other types of separatrix maps too.

The variable $\epsilon^{(low)}$ introduced in (167) will be convenient for the
further calculations since it does not depend on $i$ in the lowest-order
approximation. A quantity like $\delta_l\equiv h |\epsilon^{(low)}|$ is
sometimes called the {\it separatrix split} \cite{zaslavsky:1998} since it is
conventionally assumed that the maximal deviation of energy on the chaotic
trajectory from the separatrix energy is of the order of $\delta_l$
\cite{lichtenberg_lieberman,Zaslavsky:1991,zaslavsky:1998,zaslavsky:2005}. As
in the main text, we shall use this term, but we emphasize that the maximal
deviation may be much larger.

In the adiabatic limit $\omega_f\rightarrow 0$, the excess of the upper
boundary $E_{cl}^{(1)}$ of the lower layer over the lower barrier $E_b^{(1)}$
does not depend on angle and is equal to $2\pi h$  (cf.\ \cite{E&E:1991}). But
$\omega_f$ relevant for the spike of $h_{gc}(\omega_f)$ cannot be considered as
an adiabatic frequency, despite its smallness, because it is close to
$\omega_m$ or to its multiple while all energies at the boundary lie in the
range where the eigenfrequency is also close to $\omega_m$:

\begin{equation}
\omega_f\approx(2j-1)\omega_m\approx
(2j-1)\omega(E_{cl}^{(1)}),\quad\quad j=1,2,3,...
\end{equation}

\noindent The validity of (168) (confirmed by the results) is {\it crucial} for
the description of the layer boundary in the relevant case.

\subsubsection {Separatrix split}

Let us evaluate $\epsilon^{(low)}$ explicitly. Given that the energy is close
to $E_b^{(1)}$, the velocity $\dot{q}(t-t_i^{\prime})$ in $\epsilon^{(low)}$
(167) may be replaced by the corresponding velocity along the separatrix
associated with the lower barrier, $\dot{q}_s^{(low)}(t-t_i^{\prime})$, while
the upper limit of the integral may be replaced by infinity. In the asymptotic
limit $\Phi\rightarrow 0$, the interval between spikes within the pulse becomes
infinitely long$^{23}$ so that only the short ($\sim\omega_0^{-1}$) intervals
corresponding to the spikes contribute to the integral in $\epsilon^{(low)}$
(167). In the scale $\omega_f^{-1}$, they may be considered just as the two
instants:

\begin{equation}
t_{sp}^{(1,2)}-t_i^{\prime}\approx\pm\frac{\pi}{4\omega_m}
,\quad
\Phi\rightarrow
0.
\end{equation}

\noindent In the definition of $\epsilon^{(low)}$ (167), we substitute the
argument of the sine by the corresponding constants for the positive and
negative spikes respectively:

\begin{eqnarray}
&&
\epsilon^{(low)}\approx 2\sin
\left(\frac{\pi\omega_f}{4\omega_m}
\right)
\int_{\rm positive \, spike}{\rm
d}t\;\dot{q}_s^{(low)}(t-t_i^{\prime})
\approx 2\pi\sin
\left(\frac{\pi\omega_f}{4\omega_m}
\right),
\\
&&
\Phi\rightarrow
0.
\nonumber
\end{eqnarray}

\noindent In the derivation of the first equality in (170), we have also taken
into account that the function $\dot{q}_s^{(low)}(x)$ is odd. In the derivation
of the second equality in (170), we have taken into account that the right
turning point of the relevant separatrix is the top of the lower barrier and
the distance between this point and the left turning point of the separatrix
approaches $\pi$ in the limit $\Phi\rightarrow 0$.

For the frequencies relevant to the minima of the spikes of $h_{gc}(\omega_f)$,
i.e.\ for $\omega_f=\omega_s^{(j)}\approx(2j-1)\omega_m$, we obtain:

\begin{eqnarray}
&& \epsilon^{(low)}(\omega_s^{(j)}) \approx 2 \pi \sin
\left((2j-1)\frac{\pi}{4} \right) =\sqrt{2}\pi (-1)^{ \left[
\frac{ 2j-1}{ 4 } \right] },\nonumber
\\
&& j =  1,2,3,...,\quad \Phi\rightarrow 0.
\end{eqnarray}

\noindent For moderately small $\Phi$, it is better to use the more accurate
formula:

\begin{equation}
\epsilon^{(low)}(\omega_f)= 2\int_0^{\infty}{\rm
d}t\;\dot{q}_s^{(low)}(t)\sin(\omega_ft),
\end{equation}

\noindent where the instant $t=0$ corresponds to the turning point of the
separatrix to the left of the lower barrier, i.e.\ $\dot{q}_s^{(low)}(t=0)=0$
while $\dot{q}_s^{(low)}>0$ for all $t>0$. The dependence
$\left|\epsilon^{(low)}(\omega_f)\right|$ by Eq.\ (172) is shown for $\Phi=0.2$
in Fig.\ 21(a). For small frequencies, the asymptotic formula (170) fits well
the formula (172).

\begin{figure}[b]
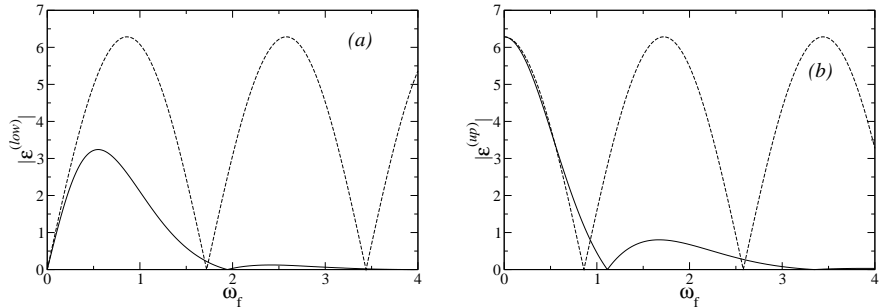

\includegraphics[width = 5.5 cm]{soskin_Fig21a.eps}
\hskip 0.5 cm
\includegraphics[width = 5.5 cm]{soskin_Fig21b.eps}
%\vskip 0.5 cm
\caption{ Theoretical estimates for the normalized separatrix split (for
$\Phi=0.2$) as a function of the perturbation frequency, for the lower and
upper layers in (a) and (b) respectively. The solid lines are calculated from
Eqs.\ (172) and (204) for (a) and (b) respectively, while the dashed lines
represent the asymptotic expressions (170) and (205) respectively. }
\end{figure}

\subsubsection {Dynamics of the map}

Consider the {\it dynamics} of the map (167) when $\omega_f$ is close to the
spikes' minima: $\omega_f\approx n\omega_m$ where $n\equiv 2j-1$ while
$j=1,2,3,\ldots$. Let the energy at the step $i=-1$ be equal to $E_b^{(1)}$.
The trajectory passing through the state with this energy is chaotic since
$(\omega(E))^{-1}$ diverges as $E\rightarrow E_b^{(1)}$ and, therefore, the
angle $\varphi_{-1}$ is not correlated with the angle on the previous step
$\varphi_{-2}$. The quantity $\sigma_{-1}$ is not correlated with $\sigma_{-2}$
either. Thus, $\sin(\varphi_{-1})$ may take any value in the range $[-1,1]$ and
$\sigma_{-1}$ may equally take the values 1 or -1. Therefore, the energy may
change on the next step by an arbitrary value in the interval
$[-h|\epsilon^{(low)}|,h|\epsilon^{(low)}|]$. Thus, $E_0-E_b^{(1)}$ may have a
positive value\footnote{The latter is valid for any $\varphi_{-1}$ except in
the close vicinity of multiples of $\pi$ while the state $E_0,\varphi_0$ (167)
in the latter range of $\varphi_{-1}$ turns out irrelevant to the boundary, as
shown further down.} $\sim h|\epsilon^{(low)}|$. Then, the approximate equality
$n\omega(E_0)\approx \omega_m$ holds, provided that the value of $h$ is from
the relevant range. Allowing for this and recalling that we are interested only
in those realizations of the map such that $E_0>E_b^{(1)}$, the relevant
realization of the map $i=-1\;\rightarrow\; i=0$ may be written as:

\begin{eqnarray}
E_{0} & =&E_b^{(1)}+\sigma_{-1}h\epsilon^{(low)}\sin(\varphi_{-1})=
E_b^{(1)}+h|\epsilon^{(low)}\sin(\varphi_{-1})|, \nonumber
\\
\varphi_{0} & \approx &\varphi_{-1}+n\pi,\nonumber
\\
\sigma_0 &&=-\sigma_{-1}.
\end{eqnarray}

One may expect that further evolution of the map will, for some time,
approximately follow the trajectory of the system (87) with the initial energy
$E_0$ (173) and an arbitrary $\varphi_{-1}$ and initial slow angle
$\tilde{\psi}$ somehow related to $\varphi_0\approx \varphi_{-1}+n\pi$. Let us
prove this explicitly.

Consider two subsequent iterations of the map (167): $2i\rightarrow 2i+1$ and
$2i+1\rightarrow 2i+2$ with an arbitrary $i\ge 0$. While doing this, we shall
assume the validity of (168) (it will be clarified below when this is true)
from which it follows that: (i) $\omega(E_{k+1})\approx \omega(E_k)$, (ii)
$\varphi_{k+1}-\varphi_{k}\approx n\pi\equiv (2j-1)\pi$. It will follow from
the results that the neglected corrections are small in comparison with the
characteristic scales of the variation of $E$ and $\varphi$ (cf.\ the
conventional treatment of the nonlinear resonance dynamics
\cite{Chirikov:79,lichtenberg_lieberman,Zaslavsky:1991,zaslavsky:1998,zaslavsky:2005,PR}).
Furthermore it follows from (167) that, while the energy remains above the
barrier energy, $\sigma_{k}$ oscillates, so that $\sigma_{2i}=\sigma_{0}$ and
$\sigma_{2i+1}=-\sigma_{0}$. Then,

\begin{eqnarray}
E_{2i+1} & =&E_{2i}+\sigma_0h\epsilon^{(low)}\sin(\varphi_{2i}),
\nonumber
\\
\varphi_{2i+1} &
=&\varphi_{2i}+\frac{\omega_f}{\omega(E_{2i+1})}\pi\approx \varphi_{2i}+n\pi
+\pi\frac{\omega_f-n\omega(E_{2i})}{\omega(E_{2i})},
\end{eqnarray}

\begin{eqnarray}
E_{2i+2}&=&E_{2i+1}-\sigma_0h\epsilon^{(low)}\sin(\varphi_{2i+1})=\nonumber
\\
&
=&E_{2i+1}+\sigma_0h\epsilon^{(low)}\sin(\varphi_{2i+1}-n\pi)\approx
E_{2i}+\sigma_02h\epsilon^{(low)}\sin(\varphi_{2i}),
\nonumber
\\
\varphi_{2i+2}&=&\varphi_{2i+1}+\frac{\omega_f}{\omega(E_{2i+2})}\pi\approx
\varphi_{2i}+2\pi n
+2\pi\frac{\omega_f-n\omega(E_{2i})}{\omega(E_{2i})}
\end{eqnarray}

\noindent (the second equality in the map for $E_{2i+2}$ takes into account
that $n$ is odd so that $\sin(\varphi-n\pi)=-\sin(\varphi)$.)

The quantity $\varphi_{2i+2}-\varphi_{2i}-2\pi n$ is small, so the map
$2i\rightarrow 2i+2$ (175) may be approximated by differential equations for
$E_{2i}$ and $\tilde{\varphi}_{2i}\equiv \varphi_{2i}- 2\pi ni $:

\begin{eqnarray}
&& \frac{{\rm d}E_{2i}}{{\rm
d}(2i)}=\sigma_0h\epsilon^{(low)}\sin(\tilde{\varphi}_{2i}),
\quad\quad
\frac{{\rm d}\tilde{\varphi}_{2i}}{{\rm d}(2i)}=
\frac{\pi}{\omega(E_{2i})}(\omega_f-n\omega(E_{2i})),
\\
&& \tilde{\varphi}_{2i}\equiv \varphi_{2i}- 2\pi ni.\nonumber
\end{eqnarray}

Let us (i) use for $\epsilon^{(low)}$ the asymptotic formula (171), (ii) take
into account that the increase of $i$ by 1 corresponds to an increase of time
by $\pi/\omega(E)$, and (iii) transform from the variables
$(E,\tilde{\varphi})$ to the variables $(I,\tilde{\psi}\equiv
n\pi(1-\sigma_0)/2-\tilde{\varphi})$. Equations (176) reduce then to:

\begin{eqnarray}
&& \frac{{\rm d}I}{{\rm
d}t}=-h\sqrt{2}(-1)^{\left[\frac{n}{4}\right]}\sin(\tilde{\psi}),
\quad\quad \frac{{\rm d}\tilde{\psi}}{{\rm d}t}= n\omega-\omega_f,
\\
&& \tilde{\psi}\equiv n\pi\frac{1-\sigma_0}{2}-\tilde{\varphi},
\quad\quad n\equiv 2j-1. \nonumber
\end{eqnarray}

Equations (177) are identical to the equations of motion of the system (87) in
their lowest-order approximation, i.e.\ to equations (98) where $q_n$ is
replaced by its asymptotic value (96) and the last term in the right-hand part
of the second equation is neglected, being of higher order in comparison with
the term $n\omega-\omega_f$.

Apart from the formal identity of Eqs.\ (177) and (98), $\tilde{\psi}$ in (177)
and $\tilde{\psi}$ in (98) are identical to. Necessarily $t_i^{\prime}$
corresponds to a turning point (see Fig.\ 20) while the corresponding $\psi$ is
equal to $2\pi i$ or $\pi+2\pi i$ for the right and left turning points
respectively (see (87)) i.e.\ $\psi=2\pi i+\pi(1-\sigma_i)/2$, so that
$\tilde{\psi}_{(98)}\equiv
n\psi-\omega_ft=n\pi(1-\sigma)/2-\tilde{\varphi}\equiv \tilde{\psi}_{(177)}$.

The relevant initial conditions for (177) follow from (173) and from the
relationship between $\tilde{\psi}$ and $\varphi$:

\begin{equation}
I(0)= I(E=E_b^{(1)}+h\sqrt{2}\pi|\sin(\tilde{\psi}(0))|),
\end{equation}

\noindent while $\tilde{\psi}(0)\equiv
n\pi(1-\sigma_0)/2-\varphi_{0}$ may be an arbitrary angle from the
ranges where

\begin{equation}
(-1)^{[n/4]}\sin(\tilde{\psi}(0))<0.
\end{equation}

For moderately small $\Phi$, it is better to use the  more accurate
dynamic equations (98) instead of (177) and the more accurate
initial value of action instead of (178):

\begin{equation}
I(0)=I(E=E_b^{(1)}+\delta_l|\sin(\tilde{\psi}(0))|), \quad\quad \delta_l\equiv
h|\epsilon^{(low)}|,
\end{equation}

\noindent with $\epsilon^{(low)}$ calculated by (172).

We name the quantity $\delta_l|\sin(\tilde{\psi})|$ the {\it generalized
separatrix split} (GSS) for the lower layer. Unlike the conventional separatrix
split $\delta_l$ \cite{zaslavsky:1998}, it is {\it angle-dependent}. The curve
$I(\tilde{\psi})=I(E=E_b^{(1)}+\delta_l|\sin(\tilde{\psi})|)$ may be called
then the GSS curve for the lower barrier and denoted as $I_{\rm
GSS}^{(l)}(\tilde{\psi})$.

Finally, let us investigate an important issue: whether the transformation from
the discrete separatrix map (i.e.\ (174) and (A14)) to the differential
equations (176) is valid for the very first step and, if it is so, for how long
it is valid after that. The transformation is valid as long as
$\omega(E_k)\approx n\omega_f$ i.e.\ as long as $E_k$ is not too close to the
barrier energy $E_b^{(1)}$. At the step $k=0$, the system stays at the GSS
curve, with a given (random) angle $\tilde{\psi}(0)$ from the range (179).
Thus, at this stage, the relation (168) is certainly valid (for the relevant
range of $h$ and for any angle except for the close vicinity of the multiples
of $\pi$). The change of energy at the next step is positive too:

\begin{eqnarray}
E_1-E_0 &&\equiv
\sigma_0h\epsilon^{(low)}\sin(\tilde{\varphi}_{0})\approx
-\sigma_{-1}h\epsilon^{(low)}\sin(\tilde{\varphi}_{-1}-n\pi)=
\nonumber\\
&& = \sigma_{-1}h\epsilon^{(low)}\sin(\tilde{\varphi}_{-1})\equiv
E_0-E_{-1}>0.\nonumber
\end{eqnarray}

\noindent This may  also be interpreted as a consequence of the
first equation in (177) and of the inequality (179).

Hence, (168) is valid at the step $k=1$ too. Similarly, one can show that
$E_2-E_1>0$, etc. Thus, the transformation (174,175)$\rightarrow$(176) is valid
at this initial stage indeed, and the evolution of $(E,\tilde{\varphi})$ does
reduce to the resonant trajectory (14) with an initial angle from the range
(179) and the initial action (180). This lasts until the resonant trajectory
meets the GSS curve in the adjacent $\pi$ range of $\tilde{\psi}$ i.e.\ at $t$
such that the state $(I(t),\tilde{\psi}(t))$ satisfies the conditions:

\begin{equation}
I(t)=I_{\rm GSS}^{(l)}(\tilde{\psi}(t)),\nonumber \quad\quad
[\tilde{\psi}(t)/\pi]-[\tilde{\psi}(0)/\pi]=1.
\end{equation}

\noindent At this instant, the absolute value of the change of energy $E_k$ in
the separatrix map (174) is equal to $E_k-E_b^{(1)}$ (just because the state
belongs to the GSS curve) but the sign of this change is negative because the
sign of $\sin(\varphi_k)$ is opposite to that of $\sin(\varphi_0)$. Therefore,
at the step $k+1$, the system gets to the separatrix itself, and the
regular-like evolution stops: at the next step of the map, the system may
either again get to the GSS curve but with a new (random) angle from the range
(179), and start a new regular-like evolution as described above; or it may get
to the similar GSS curve {\it below} the barrier and start an analogous
regular-like evolution in the energy range below the barrier, until it stops in
the same manner as described above, etc.

This approach makes it possible to describe all features of the transport
within the chaotic layer. In the present context, it is most important to
describe the {\it upper outer boundary} of the layer.

\subsubsection {Boundary of the layer}

We may now analyze the evolution of the boundary of the layer as $h$ grows.
Some of the stages of the evolution are illustrated by Figs.\ 13, 14 and 22.

\begin{figure}[tb]
\includegraphics[width = 5.8 cm]{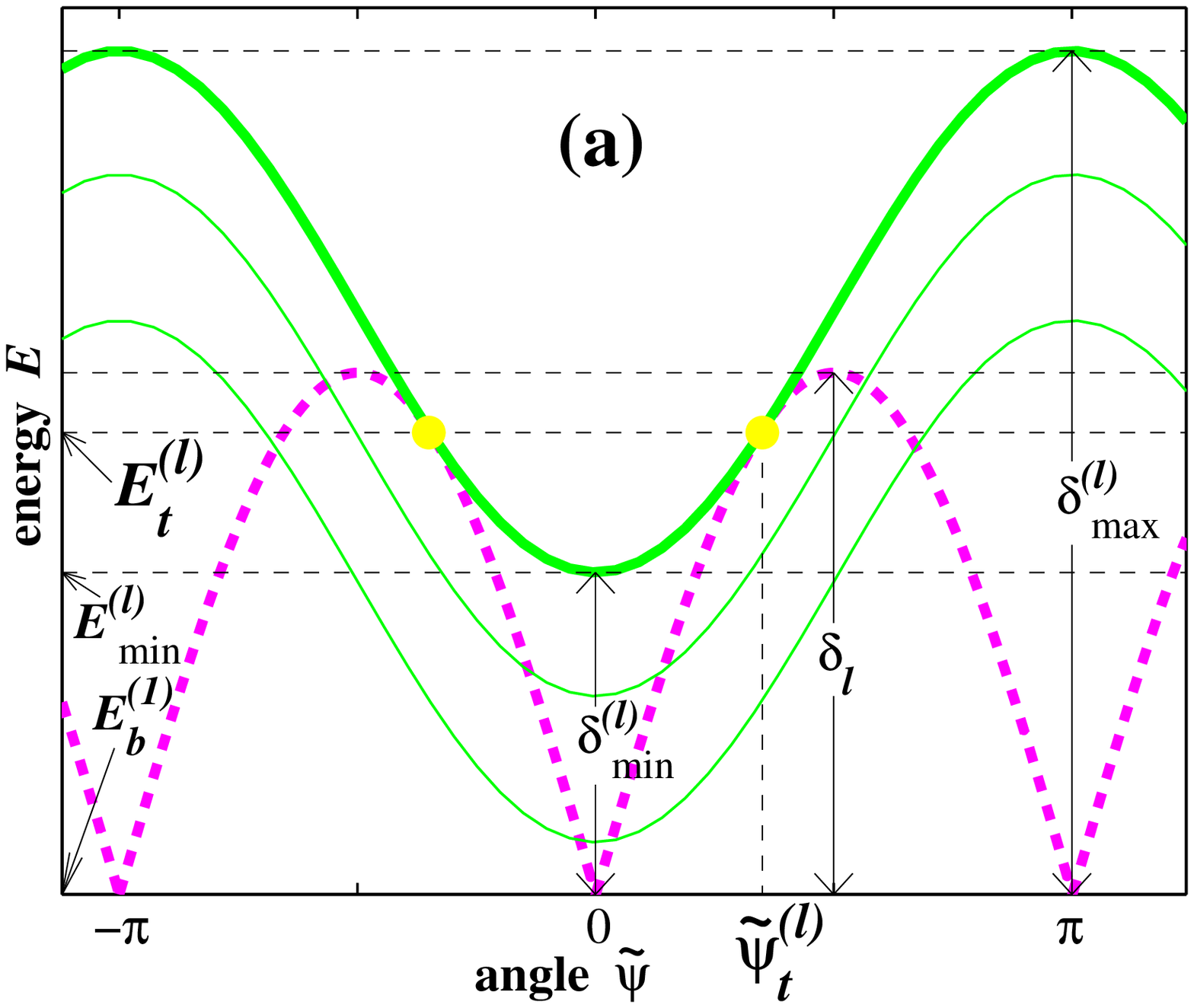}
\includegraphics[width = 5.8 cm]{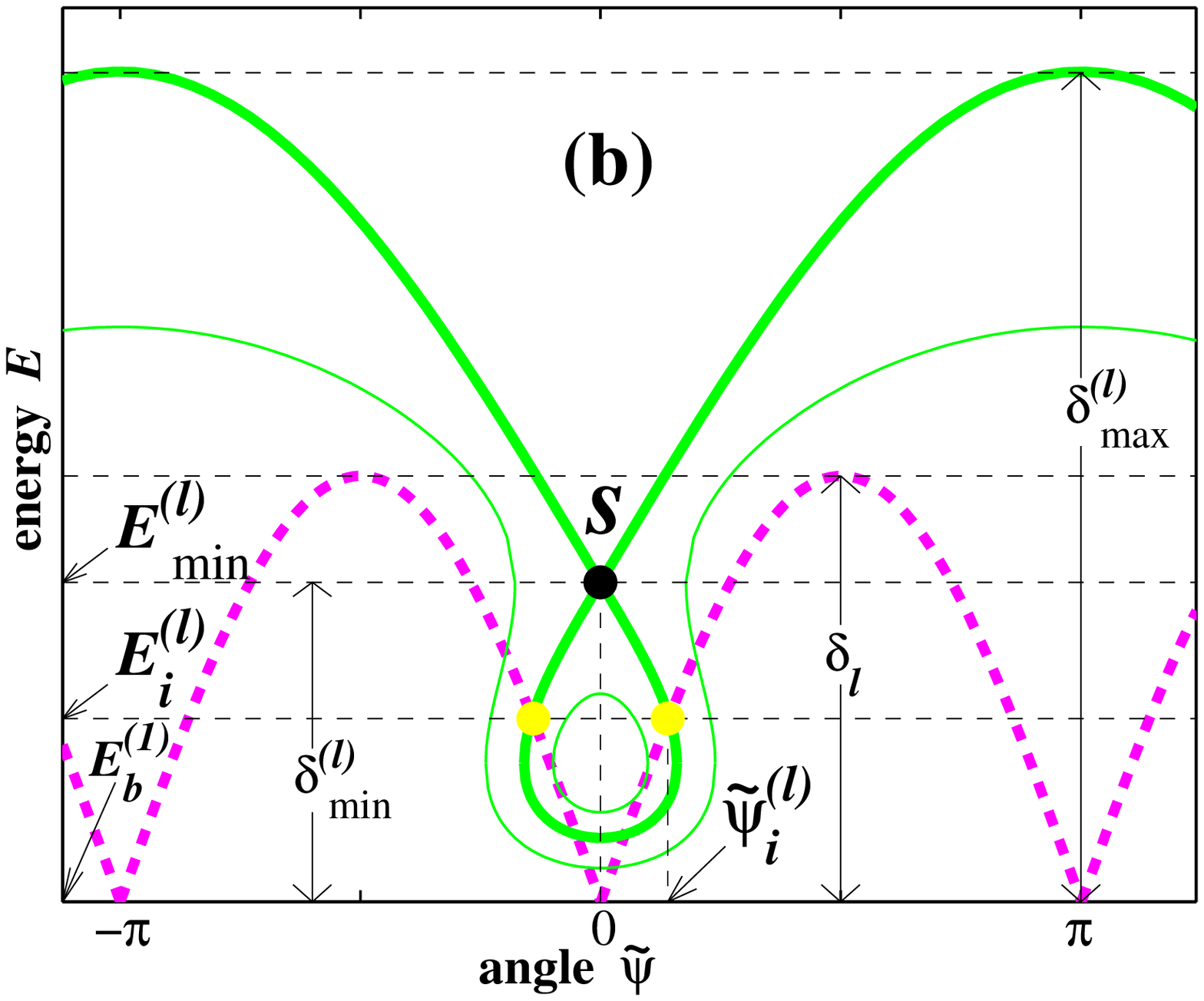}
\caption{(Color version may be found in the online version of \cite{pre2008} as
Fig.\ 16). A schematic figure illustrating the formation of the boundary of the
lower chaotic layer for $h<h_{cr}^{(l)}(\omega_f)$ in the ranges of $\omega_f$
relevant to (a) odd, and (b) even spikes. The dashed magenta line shows the GSS
curve in the energy-angle plane: $E(\tilde {\psi})=E_{\rm GSS}^{(l)}(\tilde
{\psi})\equiv E_b^{(1)}+\delta_l|\sin(\tilde {\psi})|$. Green lines show
examples of those trajectories (98) which have points in common with the GSS
curve. One of them (shown by the {\it thick} green line) relates to the
formation of the upper boundary of the lower chaotic layer: in (a), the
boundary is the trajectory {\it tangent} to the GSS curve; in (b), the boundary
is the upper part of the {\it separatrix} generated by the saddle \lq\lq $s$''.
Yellow dots indicate the relevant common points of the GSS curve and the thick
green line. They have angles $\pm\tilde {\psi}_t^{(l)}$ and energy $E_t^{(l)}$
in the case (a), and angles $\pm\tilde {\psi}_i^{(l)}$ and energy $E_i^{(l)}$
in the case (b). The minimum and maximum deviation of energy on the boundary
from the barrier energy are denoted as $\delta_{\min}^{(l)}$ and
$\delta_{\max}^{(l)}$ respectively. The maximum deviation on the GSS curve is
equal to $\delta_l$. }
\end{figure}

It follows from the analysis carried out in the previous subsection that {\it
any} state (in the $I-\tilde {\psi}$ plane) lying beyond the GSS curve but
belonging to any trajectory following the equations (98) which possesses common
points with the GSS curve belongs to the chaotic layer: the system starting
from such a state will, sooner or later, reach the separatrix where the
chaotization will necessarily occur. Therefore,  the {\it upper boundary} of
the chaotic layer coincides with the trajectory following equations (98) with
the initial action (180) and an initial angle $\tilde {\psi}(0)$ from the range
(179) such that the trajectory deviates from the barrier energy by more than
does a trajectory (98)-(179)-(180) with any other initial angle. There may be
only two topologically different options for such a trajectory: either it is
{\it tangent} to the GSS curve, or it is the separatrix trajectory which {\it
intersects} the GSS curve (some schematic examples are shown in Figs.\ 22(a)
and 22(b) respectively; some real calculations are shown in Figs.\ 13 and 14).

\paragraph{1. Relatively small $h$}

Consider first values of $h$ which are large enough for the condition (168) to
be satisfied (the explicit criterion will be given in (192)) but which are
smaller than the value $h_{cr}^{(l)}\equiv h_{cr}^{(l)}(\omega_f)$ determined
by Eq.\ (125) (its meaning is explained below). The further analysis within
this range of $h$ differs for the ranges of $\omega_f$ relevant to {\it odd}
and {\it even} spikes, and so we consider them separately.

\subparagraph{A. Odd spikes}

The relevant frequencies are:

\begin{equation}
\omega_f\approx n\omega_m, \quad\quad n\equiv 2j-1, \quad\quad j=1,3,5, \ldots
\end{equation}

Let us seek the state $\{I_t^{(l)},\tilde{\psi}_t^{(l)}\}$ (with
$\tilde{\psi}_t^{(l)}$ within the range $]0,\pi[$) where the resonant
trajectory is {\it tangent} to the GSS curve. With this aim, we equalise both
the actions and the derivatives of both curves. The equality of actions
immediately yields $I_t^{(l)}$ via $\tilde{\psi}_t^{(l)}$: $I_t^{(l)}\equiv
I(E=E_t^{(l)})=I_{\rm GSS}^{(l)}(\tilde{\psi}_t^{(l)})$. The derivative along
the GSS curve is obtained by  differentiation of $I_{\rm
GSS}^{(l)}(\tilde{\psi})$. The derivative along a resonant trajectory can be
found by dividing the first dynamic equation in (14) by the second one.
Substituting the expression of $I_t^{(l)}$ via $\tilde{\psi}_t^{(l)}$ into the
equality of the derivatives, we obtain a closed equation for
$\tilde{\psi}_t^{(l)}$, and its solution immediately gives us the relevant
$\tilde{\psi}(0)$:

\begin{eqnarray}
&&\left[
|\epsilon^{(low)}|\cos(\tilde{\psi}_{t}^{(l)})
\left(
1-\frac{\omega_f}{n\omega(E)}-h\frac{{\rm d}q_n(E)}{{\rm d}E}
\cos(\tilde{\psi}_{t}^{(l)})
\right)+q_n(E)\sin(\tilde{\psi}_{t}^{(l)})
\right]_{E=E_t^{(l)}}
\nonumber\\
&&=0,
\nonumber\\
&& E_t^{(l)}\equiv
E_b^{(1)}+h|\epsilon^{(low)}|\sin(\tilde{\psi}_{t}^{(l)}),
\quad\quad \tilde{\psi}_{t}^{(l)}\in [0,\pi],
\nonumber\\
&&
n\equiv 2j-1, \quad\quad j=1,3,5,\ldots,\quad\quad
\tilde{\psi}(0)=\tilde{\psi}_{t}^{(l)}.
\end{eqnarray}

A careful analysis of the phase space structure shows that, in the present case
(i.e.\ when $h<h_{cr}^{(l)}(\omega_f)$ while $j$ is odd), there is no
separatrix of the resonant Hamiltonian (4) which would both intersect the GSS
curve and possess points above the tangent trajectory\footnote{For odd numbers
$j\geq 3$, there are separatrices which lie in the range of $E$ where
$\omega(E)\ll \omega_m$ i.e.\ much closer to the barrier than the tangent
trajectory: these separatrices are therefore irrelevant.}. Thus, for this range
of $h$, the outer boundary of the chaotic layer is formed by the trajectory
following the dynamical equations (98) with the initial angle given by (183)
and the initial action by (180) (Fig.\ 22(a)).

Let us find the lowest-order solution of Eq.\ (183). We neglect the term
$1-\omega_f/(n\omega(E))$ (the result will justify this) and use the
lowest-order expression for the relevant quantities: namely, Eqs.\ (171) and
(96) for $\epsilon^{(low)}$ and $q_n$ respectively, and the lowest-order
expression for ${\rm d}q_n/{\rm d}E$ which can be derived from Eq.\ (95):

\begin{eqnarray}
&& \frac{{\rm d}q_n(E)}{{\rm d}E}=(-1)^{\left[\frac{n}{4}\right]}
\frac{\pi}{4\sqrt{2}\left(E-E_b^{(1)}\right)\ln\left(\Phi^{-1}\right)},
\nonumber
\\
&& n\equiv 2j-1,\quad\quad E-E_b^{(1)}\ll \Phi\rightarrow 0.
\end{eqnarray}

\noindent Then Eq.\ (183) reduces to the following equation

\begin{equation}
\tan^2(\tilde{\psi}_{t}^{(l)})=
\frac{n\pi}{8\ln\left(\Phi^{-1}\right)}.
\end{equation}

\noindent The lowest-order solution of (185) in the range $]0,\pi[$ is

\begin{equation}
\tilde{\psi}_{t}^{(l)}=(-1)^{\left[\frac{n}{4}\right]}\sqrt{\frac{n\pi}{8\ln(1/\Phi)}}
+\pi\frac{1-(-1)^{\left[\frac{n}{4}\right]}}{2}.
\end{equation}

\noindent It follows from the definition $E_t^{(l)}$ (183) and from (186) that
the lowest-order expression for $E_t^{(l)}-E_b^{(1)}$ is

\begin{equation}
E_t^{(l)}-E_b^{(1)} = \delta_l\sin(\tilde{\psi}_t^{(l)})  =
\frac{\pi^{3/2}} {2}\frac{h}{\sqrt{\ln\left(1/\Phi\right)/n}}.
\end{equation}

The next step is to find the {\it minimum} value of the energy on the boundary
of the layer, $E_{\min}^{(l)}$. It follows from the analysis of the dynamical
equations (98) that the corresponding angle $\tilde{\psi}_{\min}$ is equal to
$0$ if ${\rm sign} (q_{2j-1})>0$ (i.e.\ $j=1,5,9,\ldots$) or to $\pi$ if ${\rm
sign} (q_{2j-1})<0$ (i.e.\ $j=3,7,11,\ldots$): cf.\ Fig.\ 8(a). Given that the
Hamiltonian (87) is constant along any trajectory (98) while the boundary
coincides with one such trajectory, the values of the Hamiltonian (87) in the
states $\{I(E_{\min}^{(l)}),\tilde{\psi}=\tilde{\psi}_{\min}\}$ and
$\{I_{t}^{(l)},\tilde{\psi}_{t}^{(l)}\}$ should be equal to each other. In
explicit form, this equality may be written as

\begin{equation}
\int_{E_{\min}^{(l)}}^{E_t^{(l)}}{\rm d}E\left(1-\frac{\omega_f}{n\omega(E)}\right)
-h\left(q_n(E_t^{(l)})\cos(\tilde{\psi}_t^{(l)})-(-1)^{\left[\frac{n}{4}\right]}q_n(E_{\min}^{(l)})\right)=0.
\end{equation}

Let us find the lowest-order solution of Eq.\ (188). Assume that
$E_{\min}^{(l)}$ still belongs to the range of $E$ where
$\omega(E)\approx\omega_m$ (the result will confirm this assumption). Then the
integrand in (188) goes to zero in the asymptotic limit $\Phi\rightarrow 0$.
Hence the integral may be neglected (again, to be justified by the result). The
remaining terms in Eq.\ (188) should be treated very carefully. In particular,
it is insufficient to use the lowest-order value (96) for $q_n$ since it is the
difference between $q_n(E_t^{(l)})$ and $q_n(E_{\min}^{(l)})$ that matters.
Moreover, the approximate equality $q_n(E_t^{(l)})-q_n(E_{\min}^{(l)}) \approx
{\rm d}q_n(E_t^{(l)})/{\rm d}E_t^{(l)}(E_t^{(l)}-E_{\min}^{(l)})$ does not
apply here either since, as follows from Eq.\ (184), the derivative ${\rm
d}q_n(E)/{\rm d}E$ may vary strongly in the range
$[E_{\min}^{(l)},E_{t}^{(l)}]$ if
$(E_t^{(l)}-E_{\min}^{(l)})/(E_{\min}^{(l)}-E_b^{(1)})\stackrel{\sim}{>}1$
(again, to be justified by the result). That is why it is necessary to use the
more accurate expression (95) for $q_n$. Allowing for the asymptotic expression
(186) of $\tilde{\psi}_t^{(l)}$ and keeping only the lowest-order terms, one
can finally reduce Eq.\ (188) to the relation

\begin{equation}
\ln\left(\frac{E_t^{(l)}-E_b^{(1)}}{E_{\min}^{(l)}-E_b^{(1)}}\right)=
\frac{1}{2}.
\end{equation}

\noindent Substituting here the asymptotic value of $E_t^{(l)}$ (187), we
obtain the final lowest-order expression for the minimum deviation (along the
boundary) of the energy from the barrier:

\begin{equation}
\delta_{\min}^{(l)}\equiv
E_{\min}^{(l)}-E_b^{(1)}=(E_t^{(l)}-E_b^{(1)})/\sqrt{{\rm e}}
=\frac{\pi^{3/2}} {2{\rm
e}^{1/2}}\frac{h}{\sqrt{\ln\left(1/\Phi\right)/n}}.
\end{equation}

\noindent It is necessary and sufficient that the condition $\omega(E)\approx
\omega_m$ is satisfied at the {\it minimal} and {\it maximal} energies of the
boundary to ensure that the second equality in (168) holds true, i.e.\ that
$\omega(E)$ is close to $\omega_m$ for {\it all} points of the boundary.

At the {\it minimal} energy, this condition is

\begin{equation}
\omega_m-\omega(E_b^{(1)}+\delta_{\min}^{(l)})\ll\omega_m.
\end{equation}

\noindent Eq.\ (191) determines the lower limit of the relevant range of $h$.
The asymptotic form of (191) is:

\begin{equation}
\frac{\ln\left( \frac{\Phi\sqrt{\ln(1/\Phi)}}{h}\right)}{\ln\left(
1/\Phi \right)}\ll 1.
\end{equation}

\noindent We emphasize that any $h$ of the order of $h_{s0}$ (136) satisfies
this condition. In the asymptotic limit $\Phi\rightarrow 0$, the left-hand part
of Eq.\ (192) goes to zero.

As for the {\it maximal} energy, it may take values up to the energy of the
lower saddle \lq\lq $sl$'', i.e.\ $E_{sl}$ (102). Obviously, (168) is valid at
this saddle, too.

\subparagraph{B. Even spikes}

The relevant frequencies are:

\begin{equation}
\omega_f\approx n\omega_m, \quad\quad n\equiv 2j-1, \quad\quad
j=2,4,6, \ldots
\end{equation}

In this case, $q_n(E)$ and ${\rm d}q_n(E)/{\rm d}E$ have different signs for
all $E$ within the relevant range (i.e.\ where $\omega(E)\approx \omega_m$,
$q_n(E)\approx q_n(E_m)$): cf.\ (96) and (184). Then, in the asymptotic limit
$\Phi\rightarrow 0$, Eq.\ (183) for the tangency does not have any solution for
$\tilde{\psi}_t^{(l)}$ in the relevant range\footnote{In case of a {\it
moderately} small $\Phi$, tangency may exist in the relevant range of energies.
The boundary of the layer is then formed by the tangent trajectory.}. There may
only be solutions very close to some of $\pi$ integers, and the corresponding
energies $E_t^{(l)}$ are then very close to $E_b^{(1)}$ i.e.\
$\omega(E_t^{(l)})\ll \omega_m$: therefore they are irrelevant.

\noindent At the same time, unlike for the odd spikes, there exists a saddle
with an angle

\begin{equation}
\tilde{\psi}_s^{(l)}=\pi\frac{1-(-1)^{\left[\frac{n}{4}\right]}}{2},
\end{equation}

\noindent while the energy (which may be found as the appropriate solution of
Eq.\ (99)) lies in the relevant vicinity of the lower barrier (Fig.\ 22(b)). In
the lowest-order approximation, this saddle energy is:

\begin{equation}
E_s^{(l)}\equiv E_b^{(1)}+\delta_s,\quad\quad
\delta_s=\frac{\pi}{2\sqrt{2}}\frac{h}{\ln(\ln(4{\rm e}/\Phi))} \;
.
\end{equation}

This saddle (denoted in Fig.\ 22(b) as \lq\lq$s$'') generates a separatrix. Its
upper whiskers go to the similar adjacent saddles (shifted in $\tilde{\psi}$ by
$2\pi$). In the asymptotic limit $\Phi\rightarrow 0$, the upper whiskers are
much steeper than the GSS curve and hence they do not intersect it\footnote{In
case of a {\it moderately} small $\Phi$, they may intersect the GSS curve.
Then, the tangent trajectory lying above the separatrix necessarily
exists, so the boundary of the layer is formed by this tangent trajectory.}.
The lower whiskers do intersect the GSS curve and, moreover, two intersections
lie in the relevant energy range (Fig.\ 22(b)). Let us show this explicitly. We
write the expression for the Hamiltonian (87) in the relevant vicinity of the
barrier energy (i.e.\ where $\omega_m-\omega(E)\ll\omega_m)$, keeping in the
expression both the lowest-order terms and the terms of next order (in
particular, we use Eq.\ (95) for $q_n(E)$ and take into account that
$0<\sqrt{2}-nq_n(E)\ll \sqrt{2}$ for the relevant range of $E$):

\begin{eqnarray}
&&\tilde{H}(I=I(E=E_b^{(1)}+\delta),\tilde{\psi})=-\frac{n\delta\ln \left(\frac{2\Phi}{\delta}\right)}{2\ln
\left(\frac{4{\rm
e}}{\Phi}\right)}+\left(\omega_f-\frac{n\pi}{2\ln
\left(\frac{4{\rm e}}{\Phi}\right)}\right)\frac{2\Phi}{\pi}\ln
\left(\frac{4{\rm
e}}{\Phi}\right)- \nonumber\\
&&-(-1)^{\left[\frac{n}{4}\right]}h\sqrt{2}\left(1+\frac{n\pi\ln
\left(\frac{2\Phi}{\delta}\right)}{8\ln \left(\frac{4{\rm
e}}{\Phi}\right)}\right)\cos(\tilde{\psi}),\nonumber\\
&& \omega_m-\omega(E+\delta)\ll\omega_m.
\end{eqnarray}

The Hamiltonian $\tilde{H}$ should possess equal values at the
saddle \lq\lq $s$'' and at the intersections of the separatrix and
the GSS curve. Let us denote the angle of the intersection in the
range $]0,\pi[$ as $\tilde{\psi}_i^{(l)}$, and let us denote the
deviation of its energy $E_i^{(l)}$ from $E_b^{(1)}$ as
$\delta_i^{(l)}\equiv\delta_l\sin(\tilde{\psi}_i^{(l)})$.

Assuming that $|\tilde{\psi}_i^{(l)}-\tilde{\psi}_s^{(l)}|\ll 1$
(the result will confirm this) so that
$$\cos(\tilde{\psi}_i^{(l)})\approx
(-1)^{[n/4]}(1-(\tilde{\psi}_i^{(l)}-\tilde{\psi}_s^{(l)})^2/2)\approx$$
$$\approx(-1)^{[n/4]}(1-(\delta_i^{(l)}/\delta_l)^2/2)\approx
(-1)^{[n/4]}(1-(\delta_i^{(l)}/h)^2/4),$$ the equality of the values of
$\tilde{H}$ is:

\begin{equation}
\frac{n}{2\ln \left(\frac{4{\rm e}}{\Phi}\right)} \left(
\delta_s\ln \left(\frac{2\Phi}{\delta_s}\right)-\delta_i^{(l)}\ln
\left(\frac{2\Phi}{\delta_i^{(l)}}\right)\right)=  h\sqrt{2}\frac{n\pi}{8}\frac{\ln
\left(\frac{\delta_s}{\delta_i^{(l)}}\right)}{\ln
\left(\frac{4{\rm
e}}{\Phi}\right)}-\frac{(\delta_i^{(l)})^2}{2\sqrt{2}h}.
\end{equation}

Let us assume that, in the asymptotic limit $\Phi\rightarrow 0$,
$\delta_i^{(l)}\ll \delta_s$ (the result will confirm this). Then the left-hand
part is asymptotically smaller than the first term in the right-hand part. So,
Eq.\ (197) implies, in the asymptotic limit, that the right-hand side equals
zero. Expressing $h$ via $\delta_s$ from Eq.\ (195), we finally obtain a closed
transcendental equation for $\delta_s/\delta_i^{(l)}$:

\begin{equation}
\left(\frac{\delta_s}{\delta_i^{(l)}}\right)^2\ln\left(\frac{\delta_s}{\delta_i^{(l)}}\right)
=\frac{\pi\ln \left(\frac{4{\rm
e}}{\Phi}\right)}{n\left(\ln\left(\ln \left(\frac{4{\rm
e}}{\Phi}\right)\right)\right)^2}\equiv A.
\end{equation}

In the asymptotic limit  $\Phi\rightarrow 0$, the quantity $A$ diverges and,
hence, the lowest-order asymptotic solution of Eq.\ (198) reads as

\begin{equation}
\frac{\delta_s}{\delta_i^{(l)}}=\sqrt{\frac{2A}{\ln(A)}}.
\end{equation}

\noindent Substituting here the expression (195) for $\delta_s$ and
the expression (198) for $A$, we  obtain:

\begin{equation}
\delta_i^{(l)}=h\frac{1}{4}\sqrt {\frac{n\pi\ln\left(\ln
\left(\frac{4{\rm e}}{\Phi}\right)\right)}{\ln \left(\frac{4{\rm
e}}{\Phi}\right)}}.
\end{equation}

Thus, we have proved the following asymptotic properties of the separatrix
generated by the saddle \lq\lq$s$'': (i) the lower whiskers of the separatrix
do intersect the GSS curve in the relevant range of $E$ (i.e.\ where the
resonant approximation is valid); and (ii) the upper whiskers of the separatrix
do {\it not} intersect the GSS curve (there is no solution of Eq.\ (197) in the
range $\delta_i^{(l)}>\delta_s$). The former property confirms the
self-consistency of the asymptotic theory for even spikes; the latter property
means that the {\it upper outer boundary} of the lower chaotic layer is formed
by the {\it upper whiskers of the separatrix generated by the saddle
\lq\lq$s$''}.

Finally, we note explicitly that the minimal (along the boundary) deviation of
energy from the barrier energy occurs exactly at the saddle \lq\lq$s$'',
i.e.\

\begin{equation}
\delta_{\min}^{(l)}=\delta_s.
\end{equation}

\paragraph{2. Relatively large $h$.}

As $h$ grows, the boundary of the layer rises while the lower part of the
resonance separatrix, on the contrary, falls. They reconnect at the critical
value of $h$, $h_{cr}^{(l)}\equiv h_{cr}^{(l)}(\omega_f)$, determined by Eq.\
(125), which may be considered as the absorption of the resonance by the
chaotic layer. If $h$ grows further, then the GSS curve and the resonance
separatrix intersect. As a result, the trajectory starting from the state of
angle (183) and action (180), for odd spikes, or from the saddle \lq\lq
$s$\rq\rq, for even spikes, is {\it encompassed} by the resonance separatrix.
So, it no longer forms the outer boundary of the layer. Rather it forms the
inner boundary i.e.\ the boundary of the main island of stability inside the
layer, repeated periodically in $\tilde{\psi}$ with a period $2\pi$ (cf.\
analogous islands in the upper layer in Fig.\ 13). Unless the lower chaotic
layer reconnects with the upper one, the {\it outer} boundary of the lower
layer is formed by the upper part of the {\it resonance separatrix}. The
relevant initial angle $\tilde{\psi}(0)$ on the GSS curve corresponds to the
intersection of the GSS curve with the resonance separatrix (cf.\ the analogous
situation for the upper layer in Fig.\ 13).

\subsection {Upper chaotic layer}

The upper chaotic layer may be treated analogously\footnote{For any AC-driven
spatially periodic Hamiltonian system, the {\it upper} energy boundary of the
layer associated with the unbounded separatrix diverges in the adiabatic limit
$\omega_f\rightarrow 0$ \cite{prl2005}. However, this divergence is not
relevant for the present problem for the following reasons. The lower chaotic
layer relates to the {\it bounded} separatrix while, for the upper (unbounded)
layer, it is the {\it lower} boundary of the layer which is relevant for the
onset of global chaos in between the separatrices. Moreover, even for the upper
boundary of the upper layer, the divergence is not yet manifested for the
driving parameters $(h,\omega_f)$ in the vicinity of the spikes minima (cf.\
\cite{prl2005}).} to the lower layer. We present here only the results.

Similarly to the lower-layer case, one may consider the ranges of relatively
small $h$ (namely, smaller than $h_{cr}^{(u)}\equiv h_{cr}^{(u)}(\omega_f)$
determined by Eq.\ (126)) and relatively large $h$ (i.e.\ $h>h_{cr}^{(u)}$). In
the former range, the formation of the boundary occurs in a manner which is, in
a sense, opposite to that for the lower-layer case. For even spikes, the lower
outer boundary is formed by {\it tangency} while, for  odd spikes, it is formed
by the lower part of the {\it separatrix} generated by the saddle
\lq\lq$\tilde{s}$\rq\rq, analogous to the saddle \lq\lq$s$\rq\rq in the
lower-layer case\footnote{This tangency may exist for a {\it moderately} small
$\Phi$. The boundary is then formed by the tangent trajectory rather than by
the separatrix: see an example in Fig.\ 14(c).}.

So, for even spikes, the angle of tangency $\tilde{\psi}_t^{(u)}$ is determined
by:

\begin{eqnarray}
&&\left[ |\epsilon^{(up)}|\cos(\tilde{\psi}_{t}^{(u)}) \left(
1-\frac{\omega_f}{n\omega(E)}-h\frac{{\rm d}q_n(E)}{{\rm d}E}
\cos(\tilde{\psi}_{t}^{(u)}) \right) -q_n(E)\sin(\tilde{\psi}_{t}^{(u)})
\right]_{E=E_t^{(u)}}\nonumber
\\
&&  =0,\nonumber
\\
&& E_t^{(u)}\equiv
E_b^{(2)}-h|\epsilon^{(up)}|\sin(\tilde{\psi}_{t}^{(u)})\quad\quad \tilde{\psi}_{t}^{(u)}\in\left[0,\pi \right],\nonumber
\\
&&
n\equiv 2j-1, \quad\quad j=2,4,6,\ldots, \quad\quad\tilde{\psi}(0)=\tilde{\psi}_{t}^{(u)},
\end{eqnarray}

\noindent
and $\tilde{\psi}_t^{(u)}$ determines the tangency energy:

\begin{equation}
E_t^{(u)}= E_b^{(2)}-h|\epsilon^{(up)}|\sin(\tilde{\psi}_{t}^{(u)}),
\end{equation}

\noindent
where the quantity $\epsilon^{(up)}$ is described by the formula

\begin{equation}
\epsilon^{(up)}(\omega_f)=2\int_0^{\infty}{\rm
d}t\;\dot{q}_s^{(up)}(t)\cos(\omega_ft) \, ,
\end{equation}
where $\dot{q}_s^{(up)}(t)$ is the time dependence of the velocity along the
separatrix associated with the upper barrier and the instant $t=0$ is chosen so
that $q_s^{(up)}(t=0)$ is equal to the coordinate of the lower barrier while
$\dot{q}_s^{(up)}>0$ for $ t \in [ 0, \infty [$. The dependence
$\left|\epsilon^{(up)}(\omega_f)\right|$ in Eq.\ (204) is shown for $\Phi=0.2$
in Fig.\ 21(b).

\noindent The asymptotic form of Eq.\ (204) is

\begin{equation}
\epsilon^{(up)}\equiv
\epsilon^{(up)}(\omega_f)=2\pi\cos\left(\frac{\pi\omega_f}{4\omega_m}
\right) \, .
\end{equation}

\noindent For $\omega_f=\omega_s^{(j)}\approx (2j-1)\omega_m$, Eq.\ (204)
reduces to

\begin{eqnarray}
&& \epsilon^{(up)}(\omega_s^{(j)}) \approx 2 \pi \cos
\left((2j-1)\frac{\pi}{4} \right) =\sqrt{2}\pi (-1)^{ \left[
\frac{ 2j+1}{ 4 } \right] },\nonumber
\\
&& j =  1,2,3,...,\quad \Phi\rightarrow 0.
\end{eqnarray}

\noindent The lowest-order solution of (202) is given in Eq.\ (121), so that
$E_{t}^{(u)}$ is approximated by Eq.\ (122). The maximal energy on the lower
boundary of the layer corresponds to $\tilde{\psi}(t)=\pi$ if $ j=2,6,10,\ldots
$ or  $0$ if $ j=4,8,12,\ldots$ and is determined by Eq.\ (123). The asymptotic
value of the minimal deviation from the upper barrier of the energy at the
boundary, $\delta_{\min}^{(u)}$, is given in Eq.\ (124).

For odd spikes, the boundary is formed by the lower part of the separatrix
generated by the saddle \lq\lq $\tilde{s}$ \rq\rq. The angle of the saddle is
given in Eq.\ (117) while the deviation of its energy from the barrier is
approximated in lowest-order by Eq.\ (118).

As $h$ grows, the boundary of the layer falls while the upper part of the upper
resonance separatrix rises. They reconnect at $h=h_{cr}^{(u)}\equiv
h_{cr}^{(u)}(\omega_f)$, as determined by Eq.\ (126), which may be considered
as the absorption of the resonance by the layer.

For larger $h$, the boundary of the layer is formed by the lower part of the
upper resonance separatrix (Fig.\ 13), unless the latter intersects the lower
GSS curve (in which case, $h_{cr}^{(u)}$ marks the onset of global chaos).


\begin{thebibliography}{00}

\bibitem{abdullaev} Abdullaev S.S.: Construction of Mappings for Hamiltonian
    Systems and Their Applications. Springer, Berlin, Heidelberg (2006).

\bibitem{Abramovitz_Stegun} Abramovitz M., Stegun I.: Handbook of Mathematical
    Functions. Dover, New York (1970).

\bibitem{andronov} Andronov A.A., Vitt A.A., Khaikin S.E.: Theory of
    Oscillators. Pergamon, Oxford (1966).

\bibitem {arnold} Arnold V.I.: Instability of dynamical systems with several
    degrees of freedom. Sov. Math. Dokl. \textbf{5}, 581--585 (1964).
%\emph{Dokl. Acad. Nauk SSSR} {\bf 156}, 9 (1964).

\bibitem {bogmit} Bogolyubov N.N.,  Mitropolsky Yu.A.: Asymptotic Methods in
    the Theory of Nonlinear Oscillators. Gordon and Breach, New York (1961).

\bibitem{Oleg12} Carmona H.A. et al.: Two dimensional electrons in a lateral
    magnetic superlattice. Phys. Rev. Lett. {\bf 74}, 3009--3012 (1995).


\bibitem{A.A.} Chernikov A.A. et al.: Minimal chaos and stochastic webs. Nature
    {\bf 326}, 559--563 (1987).

\bibitem{chernikov:1987_} Chernikov A.A. et al.: Some peculiarities of
    stochastic layer and stochastic web formation. Phys. Lett. A {\bf 122},
    39--46 (1987).

\bibitem{chernikov:1988} Chernikov A.A. et al.: Strong changing of adiabatic
    invariants, KAM-tori and web-tori. Phys. Lett. A {\bf 129}, 377--380
    (1988).

\bibitem{Chirikov:79} Chirikov B.V.: A universal instability of
    many-dimensional oscillator systems. Phys. Rep. {\bf 52}, 263--379 (1979).

\bibitem{Diego} del-Castillo-Negrete D., Greene J.M., Morrison P.J.:
    Area-preserving non-twist maps: periodic orbits and transition to chaos.
    Physica D {\bf 61}, 1--23 (1996).

\bibitem{James} Dullin H.R., Meiss J.D., Sterling D.: Generic twistless
    bifurcations. Nonlinearity {\bf 13}, 203--224 (2000).

\bibitem{physica1985} Dykman M.I., Soskin S.M., Krivoglaz M.A.: Spectral
    distribution of a nonlinear oscillator performing Brownian motion in a
    double-well potential. Physica A {\bf 133}, 53--73 (1985).

%\bibitem{drsv} Dykman M.I., Rabitz H., Smelyanskiy V.N.,
%Vugmeister B.E., Phys.\ Rev.\ Lett.\ {\bf 79} 1178 (1997).

\bibitem{E&E:1991} Elskens Y. and Escande D.F.: Slowly pulsating separatrices
    sweep homoclinic tangles where islands must be small: an extension of
    classical adiabatic theory. Nonlinearity {\bf 4}, 615--667 (1991).

\bibitem{fromhold} Fromhold T.M. et al.: Effects of stochastic webs on chaotic
    electron transport in semiconductor superlattices. Phys. Rev. Lett. {\bf
    87}, 046803-1--046803-4 (2001).

\bibitem{fromhold_nature} Fromhold T.M. et al.: Chaotic electron diffusion
    through stochastic webs enhances current flow in superlattices. Nature {\bf
    428}, 726--730 (2004).

\bibitem{vasya} Gelfreich V., private communication.

\bibitem{gelfreich} Gelfreich V.G., Lazutkin V.F.: Splitting of separatrices:
    perturbation theory and exponential smallness. Russian Math. Surveys {\bf
    56}, 499--558 (2001).

%\bibitem{Gommers} Gommers R. et al. \emph{ Phys. Rev. Lett.} 94, 143001 (2005).

%\bibitem{Grynberg} Grynberg G. and Milliart-Robilliard C., \emph{Phys. Rep.} 355, 335 (2001).

\bibitem{Howard:84} Howard J.E. and Hohs S.M.: Stochasticity and reconnection
    in Hamiltonian systems. Phys. Rev. A {\bf 29}, 418--421 (1984).

\bibitem{Howard:95} Howard J.E. and Humpherys J.: Non-monotonic twist maps.
    Physica {\bf D 80}, 256--276 (1995).

%\bibitem{Jones} Jones P.H., Goonasekera M., and Renzoni F.,
%\emph{Phys. Rev. Lett.} 93, 073904 (2004).

\bibitem{Landau:76} Landau L.D. and Lifshitz E.M.: Mechanics. Pergamon, London
    (1976).

\bibitem {leonel} Leonel E.D.: Corrugated Waveguide under Scaling
    Investigation. Phys. Rev. Lett. {\bf 98}, 114102-1--114102-4 (2007).

\bibitem{lichtenberg_lieberman} Lichtenberg A.J. and Lieberman M.A.: Regular
    and Stochastic Motion. Springer, New York (1992).

\bibitem{luo2} Luo A.C.J.: Nonlinear dynamics theory of stochastic layers in
    Hamiltonian systems. Appl. Mech. Rev. {\bf 57}, 161--172 (2004).

\bibitem{luo1} Luo A.C.J., Gu K., Han R.P.S.: Resonant-Separatrix Webs in
    Stochastic Layers of the Twin-Well Duffing Oscillator. Nonlinear Dyn. {\bf
    19}, 37--48 (1999).

\bibitem{Albert} Morozov A.D.: Degenerate resonances in Hamiltonian systems
    with 3/2 degrees of freedom. Chaos {\bf 12}, 539--548 (2002).


\bibitem{Neishtadt:1986} Neishtadt A.I.: Change in adiabatic invariant at a
    separatrix. Sov. J. Plasma Phys. \textbf{12}, 568--573 (1986).

\bibitem{Neishtadt:1997} Neishtadt A.I., Sidorenko V.V., and Treschev D.V.:
    Stable periodic motions in the problem on passage trough a separatrix.
    Chaos \textbf{7}, 2--11 (1997).

\bibitem{treschev} Piftankin G.N., Treschev D.V.: Separatrix maps in
    Hamiltonian systems. Russian Math. Surveys {\bf 62}, 219--322 (2007).

\bibitem {prants} Prants S.V., Budyansky M.V., Uleysky M.Yu., Zaslavsky G.M.:
    Chaotic mixing and transport in a meandering jet flow. Chaos {\bf 16},
    033117-1--033117-8 (2006).

\bibitem{vered} Rom-Kedar V.: Transport rates of a class of two-dimensional
    maps and flows. Physica D {\bf 43}, 229--268 (1990).

\bibitem{vered1} Rom-Kedar V.: Homoclinic tangles -- classification and
    applications. Nonlinearity {\bf 7}, 441--473 (1994).

\bibitem{shepelyansky} Schmelcher P. and Shepelyansky D.L.: Chaotic and
    ballistic dynamics for two--dimensional electrons in periodic magnetic
    fields. Phys. Rev. B {\bf 49}, 7418--7423 (1994).

\bibitem {shevchenko:1998} Shevchenko I.I.: Marginal resonances and
    intermittent Behavious in the motion in the vicinity of a separatrix. Phys.
    Scr. {\bf 57}, 185--191 (1998).

\bibitem {shevchenko} Shevchenko I.I.: The width of a chaotic layer. Phys.
    Lett. A {\bf 372}, 808--816 (2008).

\bibitem{Shmidt:93} Schmidt G.J.O.: Deterministic diffusion and
    magnetotransport in periodically modulated magnetic fields. Phys. Rev. B
    {\bf 47}, 13007--13010 (1993).

\bibitem{webs} Soskin S.M., unpublished.

\bibitem{icnf_approach} Soskin S.M. and Mannella R.: New Approach To The
    Treatment Of Separatrix Chaos. In: Proceedings of the ICNF-2009. In press.

\bibitem{pre_submitted} Soskin S.M., Mannella R.: Maximal width of the
    separatrix chaotic layer. Submitted to Phys. Rev. E.

\bibitem {soskin2000} Soskin S.M., Mannella R., Array\'{a}s M. and Silchenko
    A.N.: Strong enhancement of noise-induced escape by transient chaos. Phys.
    Rev. E {\bf 63}, 051111-1--051111-6 (2001).

\bibitem {PR} Soskin S.M., Mannella R. and McClintock P.V.E.: Zero-Dispersion
    Phenomena in oscillatory systems. Phys. Rep. {\bf 373}, 247--409 (2003).

\bibitem{prl2005} Soskin S.M., Yevtushenko O.M., Mannella R.: Divergence of the
    Chaotic Layer Width and Strong Acceleration of the Spatial Chaotic
    Transport in Periodic Systems Driven by an Adiabatic ac Force. Phys. Rev.
    Lett. {\bf 95}, 224101-1--224101-4 (2005).

\bibitem {pre2008} Soskin S.M., Mannella R., Yevtushenko O.M.: Matching of
    separatrix map and resonant dynamics, with application to global chaos
    onset between separatrices. Phys. Rev. E {\bf 77}, 036221-1--036221-29
    (2008).

\bibitem {proceedings} Soskin S.M., Mannella R., Yevtushenko O.M.: Separatrix
    chaos: new approach to the theoretical treatment. In: Chandre C., Leoncini
    X., and Zaslavsky G. (eds.) Chaos, Complexity and Transport: Theory and
    Applications (Proceedings of the CCT-07), pp. 119-128. World Scientific,
    Singapore, (2008).

\bibitem {13_prime} Soskin S.M., Yevtushenko O.M., Mannella R.: Adiabatic
    divergence of the chaotic layer width and acceleration of chaotic and
    noise-induced transport. Commun. Nonlinear Sci. Numer. Simulat. In press,
    doi:10.1016/j.cnsns.2008.06.025.

\bibitem{icnf_enlargement} Soskin S.M., Khovanov I.A., Mannella R., McClintock
    P.V.E.: Enlargement of a low-dimensional stochastic web. In: Proceedings of
    the ICNF-2009. In press.

\bibitem {vecheslavov} Vecheslavov V.V.: Chaotic layer of a pendulum under
    low-and medium-frequency perturbations. Tech. Phys. {\bf 49}, 521--525
    (2004).

\bibitem{Oleg10} Ye P.D. et al.: Electrons in a periodic magnetic field induced
    by a regular array of micromagnets. Phys. Rev. Lett. {\bf 74}, 3013-3016
    (1995).

\bibitem{oleg98} Yevtushenko O.M. and Richter K.: Effect of an ac electric
    field on chaotic electronic transport in a magnetic superlattice. Phys.
    Rev. B {\bf 57}, 14839--14842 (1998).

\bibitem{oleg99} Yevtushenko O.M. and Richter K.: AC-driven anomalous
    stochastic diffusion and chaotic transport in magnetic superlattices.
    Physica {\bf E 4}, 256--276 (1999).

\bibitem{zaslavsky:1998} Zaslavsky G.M.: Physics of Chaos in Hamiltonian
    systems, 2nd edn. Imperial Colledge Press, London (2007).

\bibitem{zaslavsky:2005} Zaslavsky G.M.: Hamiltonian  Chaos and Fractional
    Dynamics. Oxford University Press, Oxford (2008).

\bibitem{ZF:1968} Zaslavsky G.M. and Filonenko N.N.: Stochastic instability of
    trapped particles and conditions of application of the quasi-linear
    approximation. Sov. Phys. JETP {\bf 27}, 851--857 (1968).

\bibitem{zaslavsky:1986} Zaslavsky G.M. et al.: Stochastic web and diffusion of
    particles in a magnetic field. Sov. Phys. JETP {\bf 64}, 294--303 (1986).

\bibitem {Zaslavsky:1991} Zaslavsky G.M., Sagdeev R.D., Usikov D.A. and
    Chernikov A.A.: Weak Chaos and Quasi-Regular Patterns. Cambridge University
    Press, Cambridge (1991).

\end{thebibliography}
\end{document}